\documentclass[aip,graphicx,jcp,twocolumn,amsmath,floatfix,10pt]{revtex4-1}
\usepackage{graphicx} 
\usepackage{algorithm2e}
\usepackage{algpseudocode}
\RestyleAlgo{ruled}
\usepackage{amsmath}
\usepackage{amssymb}
\usepackage{amsfonts}
\usepackage{amstext}
\usepackage{amsthm}
\usepackage[dvipsnames]{xcolor}
\usepackage{import}
\usepackage[hidelinks]{hyperref}

\begin{document}


\title{Shadow Molecular Dynamics for Flexible Multipole Models} 



\author{Rae A. Corrigan Grove}
\email{rcgrove@lanl.gov}
\affiliation{Theoretical Division, Los Alamos National Laboratory, Los Alamos, NM 87545, USA}

\author{Robert Stanton}
\affiliation{Theoretical Division, Los Alamos National Laboratory, Los Alamos, NM 87545, USA}


\author{Michael E. Wall}
\affiliation{Computing and Artificial Intelligence Division, Los Alamos National Laboratory, Los Alamos, NM 87545, USA}

\author{Anders M. N. Niklasson}
\email{amn@lanl.gov}
\affiliation{Theoretical Division, Los Alamos National Laboratory, Los Alamos, NM 87545, USA}

\date{\today}

\begin{abstract}
Shadow molecular dynamics provide an efficient and stable atomistic simulation framework for flexible charge models with long-range electrostatic interactions. While previous implementations have been limited to atomic monopole charge distributions, we extend this approach to flexible multipole models. We derive detailed expressions for the shadow energy functions, potentials, and force terms, explicitly incorporating monopole-monopole, dipole-monopole, and dipole-dipole interactions. In our formulation, both atomic monopoles and atomic dipoles are treated as extended dynamical variables alongside the propagation of the nuclear degrees of freedom. We demonstrate that introducing the additional dipole degrees of freedom preserves the stability and accuracy previously seen in monopole-only shadow molecular dynamics simulations. Additionally, we present a shadow molecular dynamics scheme where the monopole charges are held fixed while the dipoles remain flexible.  Our extended shadow dynamics provide a framework for stable, computationally efficient, and versatile molecular dynamics simulations involving long-range interactions between flexible multipoles. This is of particular interest in combination with modern artificial intelligence and machine learning techniques, which are increasingly used to develop physics-informed and data-driven foundation models for atomistic simulations. These models aim to provide transferable, high-accuracy representations of atomic interactions that are applicable across diverse sets of molecular systems, which requires accurate treatment of long-range charge interactions.
\end{abstract}

\pacs{}

\maketitle 

\section{Introduction}

Molecular dynamics (MD) simulations provide a powerful computational framework for investigating chemical and biological systems at the atomistic level \cite{MAllen90,DawBaskes93,MarxHutter,DFrenkel02,KarplusMcCammon-2002,AVoter02,MTuckerman02,Tuckerman2010,Schlick21,Wu22}. Classical MD methods that use charge-independent, short-range potentials can give direct and valuable insights into the molecular properties and behaviors of many interesting materials. Long-range electrostatic interactions are typically modeled using fixed atom-centered monopole charges \cite{Levitt69,MAllen90,Ponder03,Riniker18}. A limitation of such models, however, is that they are not able to capture more complex phenomena such as charge transfer, polarization, or chemical reactions involving significant changes in the electronic structure.

Quantum-mechanical MD methods address the limitations of fixed charges by explicitly representing the electronic structure  using molecular orbitals \cite{MKarplus73,CLeforestier78, RCar85,MCPayne92,BLandman1993,MarxHutter,GaoTruhlar-2002,KarplusMcCammon-2002, JHerbert06, Tuckerman2010, Brunk-2015, Mouvet22}. This enables modeling of polarization and charge transfer, often with high accuracy. 
Unfortunately, the high computational cost of these methods limits their applicability to smaller systems and shorter simulations times. 

Polarizable force field models using atomic dipoles offer a low-cost  alternative to orbital-based quantum-mechanical formulations for enhancing chemical detail and physical accuracy of simulations \cite{Vesley77, Liu98, Dang00, Halgren01, Koneshan01, Dang02, Kaminski02, Dang03, Lamoureax03, Kaminski04,  Wick05, Wick07, GreshPiquemal07, WickDang07, Cieplak09, Lopes09,  Baranyai10, Ponder10, Wang11-1, Wang11-2, WangLuo12, Chang14, Li14, JingRen19}. However, polarizable force field models generally do not account for net charge transfer between atoms such as in covalent bond formation or dissociation. Charge equilibration models \cite{Vesley77, Mortier86, MSprik88, MSprik90, RappeGoddard91, Rick94, NaserifarGoddard18, TVerstraelen13, TVerstraelen14, RStanton25} address this limitation by dynamically adjusting the atomic charges as the atoms move.  Combining charge equilibration with the flexible dipoles used in polarizable force fields leads to flexible multipole models, bridging the cost and accuracy gap between classical fixed-charge and quantum-mechanical approaches \cite{Stern99,Lopes09,Jensen22}. Flexible multipole models, which are the main focus of this work, efficiently capture long-range Coulomb interactions arising from charge redistributions occurring both between and within atoms at only a fraction of the cost of quantum-based methods. 

Flexible multipole models can be derived from first-principle density functional theory (DFT) \cite{HohenbergKohn64, Parr89, Jones89, Haunschild19, Clark21} using a coarse-grained representation of the electron density \cite{DMYork96, GTabacchi02, TVerstraelen13, Atlas_2021, Niklasson21, Goff23, Vuong23} and are naturally connected to the framework of conceptual DFT \cite{RParr78,Parr89,Geerlings03,Geerlings2020}, including atom-projected properties such as electronegativities and chemical hardness parameters. While direct computation of these atom-projected properties is possible, artificial intelligence (AI) or statistical machine learning (ML) methods offer a more practical approach by fitting them to large reference datasets generated from first-principles calculations using, for example, neural networks, atomic cluster expansions, or kernel ridge regression, which can capture dependencies on the local atomic environments \cite{SGoedecker15, TWKo20,Goff23,LiNiklasson25,Rinaldi25,Venkatraman2025}. 
These parameterized and optimized (or trained) flexible multipole models can then be used for MD simulations. 

\begin{figure}
    \centering
    \includegraphics[width=\columnwidth]{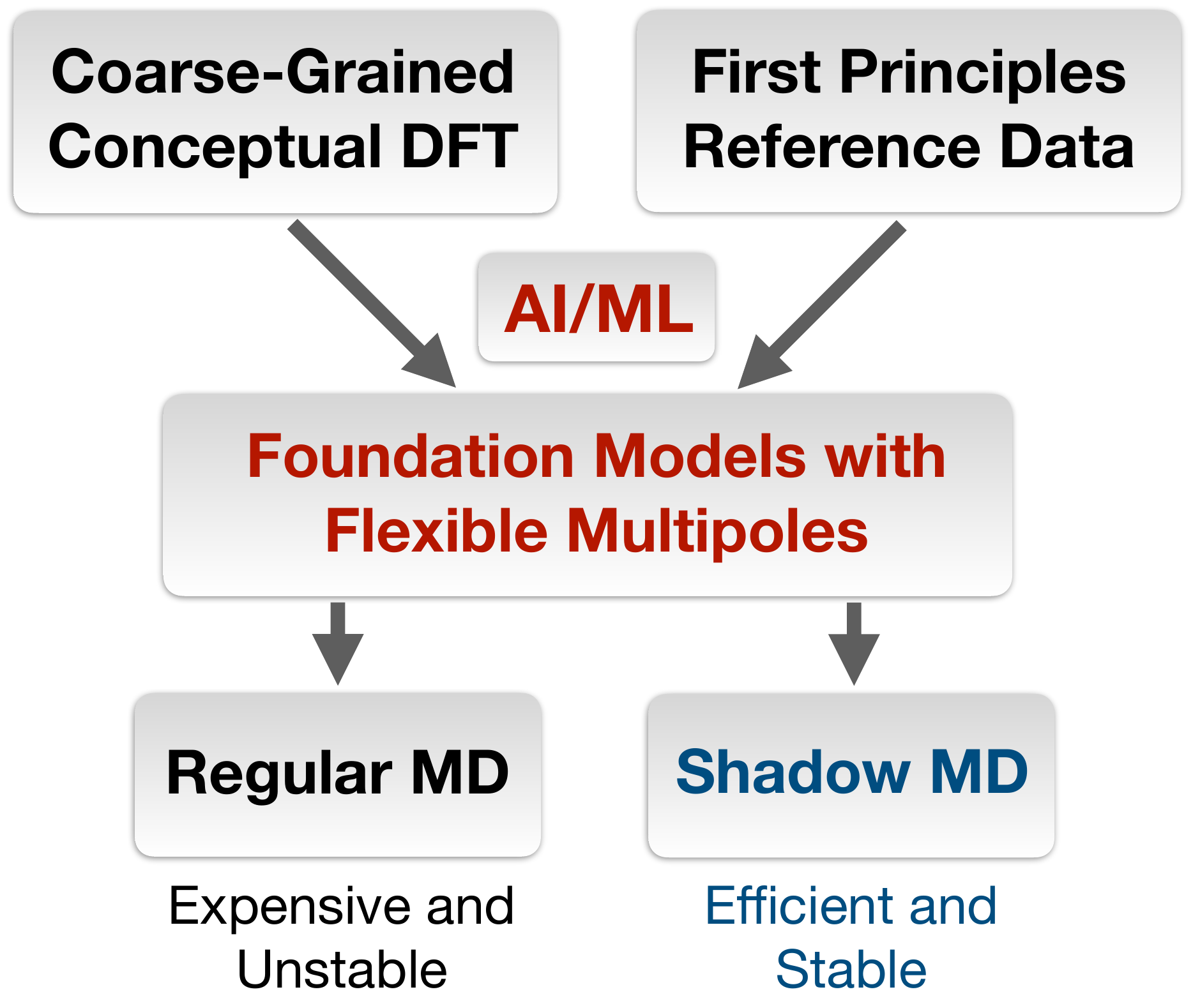}
    \caption{Conceptual picture of how the shadow MD framework can provide accurate and stable atomistic simulations using physics-informed and data-driven foundation models, including long-range flexible multipole interactions, where the interatomic potential is based on coarse-grained conceptual DFT and is parameterized using AI/ML trained on first-principles reference data.}
    \label{fig:concept}
\end{figure}

The integration of AI/ML with MD simulations is rapidly transforming computational chemistry and materials science \cite{JBehler07,MRupp12,Anatole15, JBehler16,Smith17,NSchutt17,Botu17,BNebgen18,Zhang18,ASifain18,Deringer19,Smith19, Devereux20, RZubatyuk21,Drautz2020,Friederich2021, Christensen21, Rai22, Behler21, Anstine23, WangRen24,GWang24, Olexandr25}. Because of their broad applicability and physical fidelity compared to short-ranged charge-independent force fields, flexible multipole models are of particular interest in this AI-driven development. 

Despite their advantages, flexible multipole models have some notable limitations. Although substantially faster than quantum-based MD methods, multipole equilibration is needed for each MD time step. This requires solving non-local, all-to-all systems of equations to determine the equilibrated multipoles. This equilibration is based on the Born–Oppenheimer approximation \cite{Born27}, which assumes instantaneous relaxation of the charge distribution for each new atomic configuration, and requires iterative solvers with tight convergence, which can be computationally demanding. Insufficient convergence can lead to instabilities and unphysical MD trajectories driven by non-conservative forces that invalidate the simulation results.

To address these challenges, charge equilibration models have been combined with extended Lagrangian Car–Parrinello MD (XL-CPMD) \cite{RCar85, MSprik88, MarxHutter, Sprik91, VanBelle92, Rick94}. However, while XL-CPMD reduces the computational costs per MD step, it often requires significantly shorter integration time steps compared to conventional Born–Oppenheimer MD for stable and reliable results. More recently, extended Lagrangian Born–Oppenheimer MD (XL-BOMD) has emerged as a robust and efficient alternative, overcoming several limitations of both XL-CPMD and direct Born-Oppenheimer MD methods \cite{ANiklasson08,KNomura15,Albaugh15,AAlbaugh17,AAlbaugh18,Leven19,YLi19,Niklasson21b}. In particular, the most recent ``shadow MD'' formulation of XL-BOMD replaces the exact Born-Oppenheimer potential with a carefully designed approximate ``shadow potential'', enabling accurate and stable MD simulations without iterative solvers for the equilibration problem \cite{Niklasson21, Goff23, LiNiklasson25, RStanton25, Niklasson2023}. This technique, which initially was only focused on first-principles quantum-mechanical MD, is based on ideas from backward error analysis and shadow Hamiltonian dynamics, originally developed for classical integrators \cite{HYoshida90, CGrebogi90, SToxvaerd94, JGans00, BJLeimkuhler07, RDEngel05, SToxvaerd12}, but applied to self-consistent, nonlinear charge-dependent MD. 

The shadow MD framework for flexible multipole models, which is the focus of our presented work, can speed up and stabilize MD simulations for physics-informed and data-driven foundation models, as shown conceptually in Fig.\ \ref{fig:concept}. Such foundation models aim to provide transferable, high-accuracy representations of atomic interactions that are applicable across diverse sets of molecular systems, which requires long-range charge interactions between flexible charges \cite{SGoedecker15, MCeriotti19, TWKo20, LZhang22, Goff23, Anstine23, LiNiklasson25, Olexandr25}. Apart from the physics-based flexible multipole models, there are several alternative AI-approaches that can be applied to capture long-range interactions, e.g.\ using iterative message passing networks or multiscale techniques \cite{Gao2022, AGrisafi21, Anstine23}. Here we will not consider such techniques.

Existing implementations and simulations of shadow Born–Oppenheimer MD have so far included only atomic monopole charges \cite{Niklasson21b, Niklasson21b, Goff23, LiNiklasson25, RStanton25, Kaymak25}. In this work, we extend this theoretical framework to include multipolar electrostatic interactions up to dipole order. Our goal is not to demonstrate practical applications with optimized potentials using advanced neural networks or high-performance large-scale simulations. Applying ML to capture environment-dependent atom-projected properties, such as electronegativities and chemical hardness parameters will be performed elsewhere. Instead, in this article we focus on the underlying mathematical and theoretical framework of shadow MD for stable and accurate Born–Oppenheimer MD simulations involving flexible multipolar electrostatics. 

We present both regular Born-Oppenheimer and shadow MD formulations for the monopole-only and monopole–dipole flexible charge equilibration models. We also discuss straightforward generalizations to higher-order multipoles. The extension from charge monopoles to multipoles significantly increases the number of electrostatic degrees of freedom per atom, making stability and the computational efficiency particularly important. Explicit analytical expressions for energies and forces are provided in Appendix \ref{ap: energyAndForce}. Some prototype code and additional theory is supplied in the Supplementary Information to provide clarity and facilitate implementation. Additionally, we introduce and explore a shadow MD scheme with fixed atomic monopoles and flexible dipoles. Our shadow MD framework for flexible multipole models is evaluated using simulations of solvated molecular systems, demonstrating both numerical stability and accuracy.

We conclude the paper with a summary of our findings and a brief discussion of future opportunities and challenges.

\section{Flexible Monopole Model}

First we present a Born-Oppenheimer MD for a regular flexible monopole-only charge equilibration model. Thereafter we discuss the corresponding shadow Born-Oppenheimer MD. This theory then serves as the backbone of our multipole generalization.   

\subsection{Regular Born-Oppenheimer MD}
A regular flexible monopole-only charge equilibration model can be defined by a charge-dependent energy function, $E(\mathbf{R, q)}$, which is based on a second-order expansion of the atomic energies in terms of the net partial charges, ${\bf q} \in {\mathbb{R}^N}$, for $N$ atoms at coordinates, ${\bf R} \in \mathbb{R}^{3N}$. In this energy function, 
\begin{equation} \label{eq: monoEnergy}
    E(\mathbf{R, q)} = \mathbf{q}^{\rm T}{\boldsymbol \chi} + \frac{1}{2} \mathbf{q}^{\rm T} \mathbf{C} \mathbf{q},
\end{equation}
${\boldsymbol \chi} \equiv [\chi_1,\chi_2, \ldots, \chi_N]^{\rm T}\in \mathbb{R}^N$ are the atomic electronegativities and $\mathbf{C} \in \mathbb{R}^{N \times N}$ is the Coulomb interaction matrix, 
\begin{equation}
    \mathbf{C} \equiv \begin{bmatrix}
        u_1 & C_{12} & C_{13} &\ldots & C_{1N}\\
        C_{21} & u_2 & C_{23} & \ldots & C_{2N}\\
        C_{31} & C_{32} & \ddots & \ddots & \vdots\\
        \vdots & \vdots & \ddots & u_{N-1} & C_{N-1 N}\\
         C_{N1} & C_{N2} & \ldots & C_{NN-1} & u_N\\
    \end{bmatrix}.
\end{equation}
Here $u_i \equiv C_{ii}$ is the chemical hardness (or Hubbard-U) parameter for atom $i$ and $C_{ij} \equiv f(r_{ij}) ={\operatorname{erf}(r_{ij}  u {\sqrt \pi}/2)}\big /{r_{ij}}$ 
is the Coulomb interaction potential between atom $i$ and $j$ with two atom-centered spherical monopole Gaussian charge distribution, where $r_{ij} = ||\mathbf{R}_i - \mathbf{R}_j ||_2$ is the distance between atoms $i$ and $j$, and
\begin{equation}
    u = \dfrac{2u_iu_j}{(u_i + u_j)}.
\end{equation} 

In principle, the Coulomb interaction potential can be defined as $f(r_{ij}) = r_{ij}^{-1}$, corresponding to point-charge interactions, but this may lead to divergence and instabilities in the equilibration, such as the polarization catastrophe in generalizations to dipole interactions. These problems can be avoided if we assume finite atomic charge distributions. Several functional forms of $f(r_{ij})$, corresponding to various finite shapes of the atomic charge distributions are possible, provided they have continuous derivatives up to third order.
Here we assume Gaussian shaped atom-centered charge distributions. The $i=j$ terms then correspond to the on-site Coulomb repulsion energies of overlapping Gaussian-shaped charge densities.  Accounting for this screened overlap between penetrating atomic charge distributions with widths determined by ${\bf u} \equiv \{u_i\}$, is thus not only more physically accurate, but also provides better numerical stability compared to simple point charge models.

From the energy function, $E(\mathbf{R, q})$, we can define the exact regular Born-Oppenheimer potential, $U_{\rm BO}(\mathbf{R})$, which assumes an instantaneous equilibration or relaxation to the lowest energy ground state of the charge distribution. The Born-Oppenheimer potential is given by
\begin{equation}\label{eq: monoPotential}
    \begin{aligned}
    &U_{\rm BO}(\mathbf{R}) = V(\mathbf{R}) +\\
    &~~~~\Big \{ E(\mathbf{R, q}) ~\Big | ~\frac{\partial E({\bf R, q})}{\partial \mathbf{q}} = 0,  \sum_i q_i = Q_{\rm tot} \Big \},
    \end{aligned}
\end{equation}
where $V(\mathbf{R})$ is the charge-independent part of the potential, e.g.\ including short-range repulsion or Van der Waal's interactions, and $Q_{\rm tot}$ is the total net charge of the system. The charge-dependent potential energy is determined by the value of $E({\bf R},{\bf q})$ under the constraint that $\partial E({\bf R},{\bf q})\big / \partial {\bf q} = {\bf 0}$ and $\sum_i q_i = Q_{\rm tot}$. This corresponds to an equilibrated relaxed ground state with a predefined net total charge.

To enforce the total net charge constraint, we can use a Lagrange multiplier, $\mu$. The equilibrated ground state solution, ${\bf c}_0 \equiv [{q_0}_1, {q_0}_2, \ldots, {q_0}_N,\mu]^{\rm T} \in {\mathbb R}^{N+1}$, is then determined by an all-to-all system of equations, 
\begin{equation} \label{eq: matrixqcalc}
    \mathbf{A}{\bf c}_0 = \mathbf{b}. 
\end{equation}
For a simple three-atom system ($N = 3$), the matrices and vectors are
\begin{equation}\label{matrix: Aqb}
\mathbf{A} = \begin{bmatrix}
u_1 & C_{12} & C_{13} & 1\\
C_{21} & u_2 & C_{23} & 1\\
C_{31} & C_{32} & u_3 & 1\\
1 & 1 & 1 & 0
\end{bmatrix},~
\mathbf{c}_0 = 
\begin{bmatrix}
{q_0}_1\\
{q_0}_2\\
{q_0}_3\\
\mu
\end{bmatrix},~
\mathbf{b} =
\begin{bmatrix}
-\chi_1\\
-\chi_2\\
-\chi_3\\
Q_{\rm tot}
\end{bmatrix},
\end{equation}
where $\mu$ is the chemical potential. The resulting equilibrated charges are given by, ${\bf q}_0 = \{{{c}_0}_i\}_{i = 1}^{N}$, from which the exact regular Born-Oppenheimer potential is evaluated as
\begin{equation}
    U_{\rm BO}({\bf R}) = V({\bf R}) + E({\bf R},\mathbf{q}_0).
\end{equation}

The equations of motion used to generate the MD trajectories with the exact regular Born-Oppenheimer potential in Eq.\ (\ref{eq: monoPotential}) are then given by
\begin{equation}\label{eq: eqofMotion-bo}
    m_i {\bf \ddot R}_i = -\frac{\partial {U}_{\rm BO}({\bf R})}{\partial {\bf R}_i},
\end{equation}
where $m_i$ is the mass of atom $i$ and ${\mathbf{ \ddot R}}_i$ is its acceleration. From here, integration of the equations of motion can be performed time step by time step, for example, using a velocity-Verlet integrator as described in Alg.\ \ref{alg: bomd}. In this way we generate the molecular trajectories, ${\bf R}(t)$, and velocities, ${\bf \dot R}(t)$.

\begin{algorithm}
\caption{Born-Oppenheimer Molecular Dynamics using a Leapfrog Velocity-Verlet Integration Scheme}\label{alg: bomd}
    \SetAlgoNoLine
    \tcp{Input Coordinates and Parameters}
    ${\bf R} = {\bf R}_{\rm in}$, ~
    $\boldsymbol{\chi} = \boldsymbol{\chi}_{\rm in}$, ~
    ${\bf u} = {\bf u}_{\rm in}$, ~
    ${\bf m} = {\bf m}_{\rm in}$
    \vspace{1mm}
    
    \tcp{Initialize Velocities (e.g. ${\bf v}_0 = {\bf 0}$)} 
    ${\bf v}(t_0) = \bf{v}_0$ 
    \vspace{3mm}

    \tcp{Initialize Energy and Forces}
    Calculate using Eqs.\ (\ref{eq: matrixqcalc}-\ref{eq: eqofMotion-bo})
    \vspace{3mm}

    \tcp{Main MD Loop ($t = t_0$)}
    \While{$t \leq {\rm MaxTime}$}{
        \vspace{1mm}
        
        ~~~~\tcp{Update Velocity, First Half-Step} 
        ~~~~${\bf v}(t+\frac{1}{2}\delta t) = {\bf v}(t) + \frac{1}{2}\delta t {\bf \ddot R}(t)$ 
        \vspace{1mm}
        
        ~~~~\tcp{Update Position}
        ~~~~${\bf R}(t + \delta t) = {\bf R}(t) + \delta t  * {\bf v}(t + \frac{1}{2} \delta t)$
        \vspace{3mm}

        ~~~~\tcp{Update Energy and Forces}
        ~~~~Calculate using Eqs.\ (\ref{eq: matrixqcalc}-\ref{eq: eqofMotion-bo})\\
        \vspace{3mm}
        
        ~~~~\tcp{Update Velocity, Second Half-Step}
        ~~~~${\bf v}(t + \delta t) = {\bf v}(t + \frac{1}{2} \delta t) + \frac{1}{2}\delta t {\bf \ddot R}(t + \delta t)$
        \vspace{1mm}
        
        ~~~~\tcp{Increase Time}
        ~~~~$t = t + \delta t$
    }
\end{algorithm}

\subsection{Shadow Born-Oppenheimer MD}

Using the exact regular Born-Oppenheimer potential energy, Eq.\ (\ref{eq: monoPotential}), in MD simulations is expensive due to the non-linear dependency of the energy on ${\bf q}$, leading to the system of linear equations in Eq.\ (\ref{eq: matrixqcalc}), which needs to be solved for the equilibrated ground state, ${\bf c}_0$, in each time step. For large systems, we can use iterative solvers to reduce the computational cost, which is of particular importance for systems with periodic boundary conditions. However, the iterative solutions must be tightly converged to provide conservative forces. Inadequate equilibration may cause instabilities and produce unphysical molecular trajectories, which invalidate the simulation results. Using thermostats does not solve the problem \cite{EMartinez15}.

To avoid the stability problems and the high cost of tight convergence, we can use the concept of shadow MD \cite{Niklasson21, Niklasson21b, Goff23, LiNiklasson25}, which is based on the idea of backward error analysis or a shadow Hamiltonian approach \cite{HYoshida90,CGrebogi90,SToxvaerd94,JGans00,BJLeimkuhler04,RDEngel05,SToxvaerd12, BLeimkuhler15}. Instead of calculating approximate forces, using an expensive iterative procedure for an underlying exact regular Born-Oppenheimer potential, we do the opposite. We calculate exact forces in a cheap and direct way, but for an underlying approximate {\it shadow} Born-Oppenheimer potential. The original development of shadow MD for non-linear charge models was performed for self-consistent, first-principles, quantum-mechanical MD simulations \cite{Niklasson21b} in combination with an extended Lagrangian formalism, in the spirit of Car-Parrinello MD \cite{RCar85}.

There is no unique or exact way to define the shadow potential. Here we construct a shadow Born-Oppenheimer potential by replacing the charge-dependent energy function, ${E}({\bf R},{\bf q})$, in Eq.\ (\ref{eq: monoEnergy}) with an approximate shadow energy function, $\mathcal{E}(\mathbf{R, q, n})$, which is partially linearized around an approximate ground-state solution, ${\bf n} \approx {\bf q}_0$. We carry out this approximation by first splitting the Coulomb matrix into short-range (S) and long-range (L) parts and performing a partial linearization of ${\bf E}({\bf R},{\bf q})$ around ${\bf q} \approx{\bf n}$  such that
\begin{equation}\label{eq: shadowMono}
    \mathcal{E}(\mathbf{R, q, n}) = \mathbf{q}^{\rm T}{\boldsymbol \chi} + 
    \frac{1}{2} {\bf q}^{\rm T}{\bf C}_{\rm S} {\bf q} + \frac{1}{2}\big(2\mathbf{q} - {\bf n}\big)^{\rm T}\mathbf{C}_{\rm L}\mathbf{n}.
\end{equation}
Here the Coulomb matrix is decomposed as ${\bf C} = {\bf C}_{\rm S} + {\bf C}_{\rm L}$, where ${\bf C}_{\rm S} \equiv \big \{u_i \delta_{ij}\big \}$ contains the diagonal (short-range) contributions, and ${\bf C}_{\rm L} \equiv \big\{ C_{ij}(1-\delta_{ij}) \big \}$ includes the complementary off-diagonal (long-range) contributions.
The particular division between short- and long-range interactions is flexible and can be modified, for example, by using diagonal (intramolecular) blocks in ${\bf C}_{\rm S}$ instead of the purely diagonal terms. Because of the linearized term, the leading error of $\mathcal{E}(\mathbf{R, q, n})$ compared to ${E}({\bf R},{\bf q})$ scales as $\propto |{\bf q}-{\bf n}|^2$.

Using the shadow energy function, $\mathcal{E}(\mathbf{R, q, n})$, we then define the shadow Born-Oppenheimer potential as
\begin{equation}\label{eq: monoShadowPoten}
    \begin{aligned}
        &\mathcal{U}_{\rm BO}(\mathbf{R}, \mathbf{n}) = V(\mathbf{R}) + \\
        &~~~~\Big\{ \mathcal{E}(\mathbf{R, q, n}) ~\Big | ~\frac{\partial \mathcal{E}(\mathbf{R, q, n})}{\partial \mathbf{q}} = 0, \sum_i q_i = Q_{\rm tot} \Big\}.
    \end{aligned}
\end{equation}
The charge-dependent shadow potential energy is determined by the value of $E({\bf R},{\bf q}, {\bf n})$ under the constraint that $\partial {\cal E}({\bf R},{\bf q},{\bf n})\big / \partial {\bf q} = {\bf 0}$ and $\sum_i q_i = Q_{\rm tot}$. This corresponds to an equilibrated relaxed ground state with a predefined net total charge. Thanks to the linearization, this equilibration can be performed by solving a quasi-diagonal system of linear equations,
\begin{equation} \label{eq: matrixqcalcshadow}
    \mathbf{A_{\rm S}c[x]} = \mathbf{b} -{\bf A}_{\rm L}{\bf x} ~\Rightarrow~ \mathbf{c}[{\bf x}] = \mathbf{A}^{-1}_{\rm S}\big(\mathbf{b} -{\bf A}_{\rm L}{\bf x}\big).
\end{equation}
The equilibrated ground state solution, ${\bf c}_0$, which defines the regular Born-Oppenheimer potential, is here replaced by an ${\bf x}$-dependent ground state, ${\bf c}[{\bf x}]$, with ${\bf x} \equiv [n_1, n_2,\ldots, n_N, 0]^T$.
For a three-atom system we have that ${\bf x} = [n_1,n_2,n_3,0]^{\rm T}$ and
\begin{equation}\label{matrix: As}
\mathbf{A}_{\rm S} = \begin{bmatrix}
u_1 & 0 & 0 & 1\\
0 & u_2 & 0 & 1\\
0 & 0 & u_3 & 1\\
1 & 1 & 1 & 0
\end{bmatrix},~
\mathbf{A}_{\rm L} = 
\begin{bmatrix}
0 & C_{12} & C_{13} & 0\\
C_{21} & 0 & C_{23} & 0\\
C_{31} & C_{32} & 0 & 0\\
0 & 0 & 0 & 0
\end{bmatrix}.
\end{equation} 
The ${\bf b}$ matrix remain the same as in Eq.\ (\ref{matrix: Aqb}).

The inversion of  the quasi-diagonal matrix ${\bf A}_{\rm S}$ is cheap. It can be performed using, for example, the Woodbury formula, based on the fact that ${\bf A}_{\rm S}$ is a rank-2 update of a diagonal matrix with a trivial matrix inverse \cite{LiNiklasson25}. The main cost is instead dominated by the Coulomb summation associated with the calculation of ${\bf A}_{\rm L} {\bf x}$ on the right-hand side of Eq.\ (\ref{eq: matrixqcalcshadow}). This summation can be performed, for example, using an Ewald summation technique for periodic boundary conditions \cite{Kaymak25}. Since the summation only needs to be performed once, no iterative solver is needed and convergence problems are avoided. This significantly decreases the cost of calculating ${\bf c}[{\bf x}]$ compared to solving for ${\bf c}_0$ in Eq.\ (\ref{eq: matrixqcalc}).

From the solution, ${\bf c}[{\bf x}]$, the equilibrated ${\bf n}$-dependent charges, are then given by ${\bf q}[{\bf n}] = \{{{\bf c}[{\bf x}]}_i\}_{i = 1}^{N}$, from which the ${\bf n}$-dependent shadow Born-Oppenheimer potential is evaluated as
\begin{equation}
    {\cal U}_{\rm BO}({\bf R}, {\bf n}) = V({\bf R}) + {\cal E}({\bf R},\mathbf{q}[{\bf n}],{\bf n}).
\end{equation}

The error between the shadow Born-Oppenheimer potential and the exact regular Born-Oppenheimer potential will depend on how far ${\bf n}$ is from the exact regular ground state solution, ${\bf q}_0$, or from the equilibrated charges of the shadow potential, ${\bf q}[{\bf n}]$. In the initial configuration of an MD simulation, we may chose ${\bf n}$ as the exact regular ground state, ${\bf q}_0$. However, as the atoms move, we need to update ${\bf n}$. We do this by introducing ${\bf n}(t) \equiv {\bf n}$ and its time derivative ${\bf \dot n}(t) \equiv {\bf \dot n}$ as extended dynamical vector variables that are driven by a harmonic oscillator centered around the exact ground state charges, ${\bf q}_0(t)$, or the equilibrated ground state charges, ${\bf q}[{\bf n}]$, of the shadow energy function (to highlight the time dependence of the charge degrees of freedom we often include an explicit (t) - i.e. ${\bf n}(t)$). We define this extended dynamics through the Lagrangian,
\begin{align}\label{extendedLagr}
    &{\cal L}({\bf R},{\bf \dot R},{\bf n},{\bf \dot n}) = \frac{1}{2}\sum_i m_i \vert {\bf \dot R}_i\vert^2 - {\cal U}_{\rm BO}({\bf R},{\bf n}) \\
    &~~~~~~~+ \frac{1}{2} \mu_{\rm e} \sum_i {\dot n}_i^2 - \frac{1}{2} \mu_{\rm e} \omega^2 \big({\bf q}[{\bf n}] - {\bf n}\big)^{\rm T}{\bf K}^{\rm T}{\bf K} \big({\bf q}[{\bf n}]-{\bf n}\big). \nonumber
\end{align}
Here $\mu_{\rm e}$ is a fictitious mass parameter and $\omega$ the frequency of the harmonic oscillator. 
In the harmonic potential, ${\bf K}^{\rm T}{\bf K}$ is a symmetric positive definite metric tensor, where ${\bf K}$ is a kernel defined as the inverse of the Jacobian of the residual function, ${\bf f}({\bf n}) =  {\bf q}[{\bf n}]-{\bf n}$, i.e.\
\begin{align}
    & {\bf K} = {\bf J}^{-1},~~~J_{ij} = \frac{\partial \big(q_i[{\bf n}] - n_i \big)}{\partial n_j}. 
\end{align}
The Euler-Lagrange equations,
\begin{align}
    & \frac{\rm d}{\rm dt}\left(\frac{\partial {\cal L}}{\partial {\bf \dot R}_i}\right) = \frac{\partial {\cal L}}{\partial {\bf R}_i},\\
    & \frac{\rm d}{\rm dt}\left(\frac{\partial {\cal L}}{\partial {\bf \dot n}}\right) = \frac{\partial {\cal L}}{\partial {\bf n}} , 
\end{align}
then generate the equations of motion and are derived assuming an adiabatic separation between the fast extended charge degrees of freedom governed by $\omega$ and $\mu_{\rm e}$ and the slower nuclear motion. This can be seen as a classical analogue to the quantum-mechanical Born-Oppenheimer approximation. In this adiabatic limit, where $\omega \rightarrow \infty$, and $\mu_{\rm e} \rightarrow 0$ such that $\mu_{\rm e} \omega = {\rm constant}$, we get the coupled equations of motion,
\begin{align}\label{eq:shadowEqMotion}
    & m_i {\bf \ddot R}_i = -\frac{\partial {\cal U}_{\rm BO}({\bf R},{\bf n})}{\partial {\bf R}_i}\Big \vert_{\bf n}\\
    & {\bf \ddot n} = - \omega^2{\bf K}\big({\bf q}[{\bf n}]-{\bf n}\big). \label{eq:shadowEqMotion2}
\end{align}
This formulation is exact in continuous time, meaning that if one would integrate with an infinitesimally small integration time step, $\delta t$, the equations would be exact for the dynamics defined by ${\cal L}({\bf R},{\bf \dot R},{\bf n},{\bf \dot n})$ in Eq.\ (\ref{extendedLagr}). Related mass-zero limits have also been applied successfully in connection with Car-Parrinello MD by Bonella {\it et al.}\ \cite{SBonella20,SBonella21}.

\begin{algorithm}
\caption{Shadow Born-Oppenheimer Molecular Dynamics using a Velocity-Verlet Integration Scheme. The coefficients, $\{c_k\}$, in the dissipative term for the modified Verlet schemes are given in Ref.\ \cite{ANiklasson09}} \label{alg: shadowbomd}
    \SetAlgoNoLine
    \tcp{Input Coordinates and Parameters}
    ${\bf R} = {\bf R}_{\rm in}$, ~
    $\boldsymbol{\chi} = \boldsymbol{\chi}_{\rm in}$, ~
    ${\bf u} = {\bf u}_{\rm in}$, ~
    ${\bf m} = {\bf m}_{\rm in}$
    \vspace{1mm}

    \tcp{Initialize Velocities (e.g., ${\bf v}_0 = {\bf 0}$)} 
    ${\bf v}(t_0) = \bf{v}_0$ 
    \vspace{1mm}

    \tcp{Initialize ${\bf x}$ as Ground State Multipoles, ${\bf c}_0$}
    ${\bf x}(t_0) = {\bf c}_0(t_0)$ 
    \vspace{1mm}

    \tcp{Determine a Preconditioner}
    $\mathbf{K}_0 = \mathbf{J}^{-1}(t_0)$
    \vspace{1mm}

    \tcp{Initialize ${ \bf \ddot x}$}
    ${\bf \ddot x}(t_0) = {\bf 0}$
    \vspace{3mm}

    \tcp{Initialize Energy and Forces}
    Calculate Using Eqs.\ (\ref{eq: matrixqcalcshadow}-\ref{eq:shadowEqMotion})
    \vspace{3mm}
    
    \tcp{Main MD Loop ($t = t_0$)}
    \While{$t \leq {\rm MaxTime}$}{
        \vspace{1mm}
        
        ~~~~\tcp{Update Velocity, First Half-Step} 
        ~~~~${\bf v}(t+\frac{1}{2}\delta t) = {\bf v}(t) + \frac{1}{2}\delta t{\bf \ddot{R}}(t)$ 
        \vspace{1mm}
        
        ~~~~\tcp{Update Position}
        ~~~~${\bf R}(t + \delta t) = {\bf R}(t) + \delta t * {\bf v}(t + \frac{1}{2} \delta t)$
        \vspace{1mm}

        ~~~~\tcp{Multipole Integration using Verlet}
        ~~~~$\mathbf{n}(t+\delta t) = 2\mathbf{n}(t) - \mathbf{n}(t-\delta t) + \delta t^2\ddot{{\bf n}}(t)$
        \vspace{1mm}
        
        ~~~~\tcp{Add Weak Dissipative Term}
        ~~~~$\mathbf{n}(t+\delta t) = \mathbf{n}(t + \delta t) + \alpha \sum_{k=0}^{k_{\rm max}} c_k \mathbf{n}(t-k\delta t)$
        \vspace{1mm}

        ~~~~\tcp{Determine Relaxed Multipoles}
        ~~~~ ${\bf x} = {\bf c}[{\bf x}]$
        \vspace{3mm}
        
        ~~~~\tcp{Update Energy and Forces }
        ~~~~Calculate Using Eqs.\ (\ref{eq: matrixqcalcshadow}-\ref{eq:shadowEqMotion})\\
        \vspace{3mm}

        ~~~~\tcp{Update $\ddot{{\bf n}}$ using Low-Rank Update}
        ~~~~\tcp{(Appendix \ref{ap:lrupdate}, Including ${\bf K}_0$)}
        ~~~~$\ddot{{\bf n}} = -\omega^2\mathbf{K}(\mathbf{q}[\mathbf{n}] - \mathbf{n})$
        
        ~~~~\tcp{Update Velocity, Second Half-Step}
        ~~~~${\bf v}(t + \delta t) = {\bf v}(t + \frac{1}{2} \delta t) + \frac{1}{2}\delta t{\bf \ddot{R}}(t + \delta t)$
        \vspace{1mm}
        
        ~~~~\tcp{Increase Time}
        ~~~~$t = t + \delta t$
    }
\end{algorithm}

The coupled equations of motion in Eqs.\ (\ref{eq:shadowEqMotion}) and (\ref{eq:shadowEqMotion2}) can be integrated using a leapfrog velocity-Verlet scheme for the nuclear coordinates and velocities, combined with a modified Verlet scheme for the net partial charges as described in Alg.\ \ref{alg: shadowbomd}. The Verlet scheme is modifed to include a weak dissipative term to remove the accumulation of numerical noise that otherwise would build up in a perfectly time-reversible integration \cite{ANiklasson09,PSteneteg10,GZheng11}. In the initial time step, we solve the the full regular charge equilibration problem and set ${\bf n} = {\bf q}_0$. A full equilibration is thus necessary, but only in the very first time step. The kernel, ${\bf K}$, acting on the residual function in Eq.\ (\ref{eq:shadowEqMotion2}), can be approximated using the preconditioned, low-rank Krylov subspace approach \cite{ANiklasson20,Niklasson21b} as described in Appendix \ref{ap:lrupdate}. 
 
The kernel, ${\bf K}$, acting on the residual function ${\bf q}[{\bf n}] - {\bf n}$, resembles a Newton optimization step, causing the extended charge degrees of freedom, ${\bf n}(t)$, to oscillate more closely around the exact ground-state partial charges, ${\bf q}_0$, than around the approximate ground state defined by the shadow potential, ${\bf q}[{\bf n}]$.This further improves the stability and the accuracy of the shadow MD.

The total computational cost of the shadow MD is dominated by the additional kernel approximation step, which is not included in regular Born-Oppenheimer MD. However, the propagation of the extended charge degrees of freedom, ${\bf n}(t)$, does not need to be highly accurate. As long as ${\bf n}(t)$ stays reasonably close to the `exact' ground state, the shadow MD remains very accurate. The cost of the low-rank kernel approximation can therefore be kept low.

Generalizations to higher-level shadow potentials for monopole-charge equilibration models are also possible \cite{Niklasson2023}. These formulations can provide higher-order accuracy at the cost of one extra Coulomb potential calculation. Other generalizations also include the use of higher-order geometric or symplectic integration schemes that previosuly have been used for quantum-mechanical XL-BOMD simulations \cite{ANiklasson08,AOdell09,AOdell11}. None of the these higher-order generalizations will be considered in this article.

\begin{figure}
    \centering
    \includegraphics[width=\columnwidth]{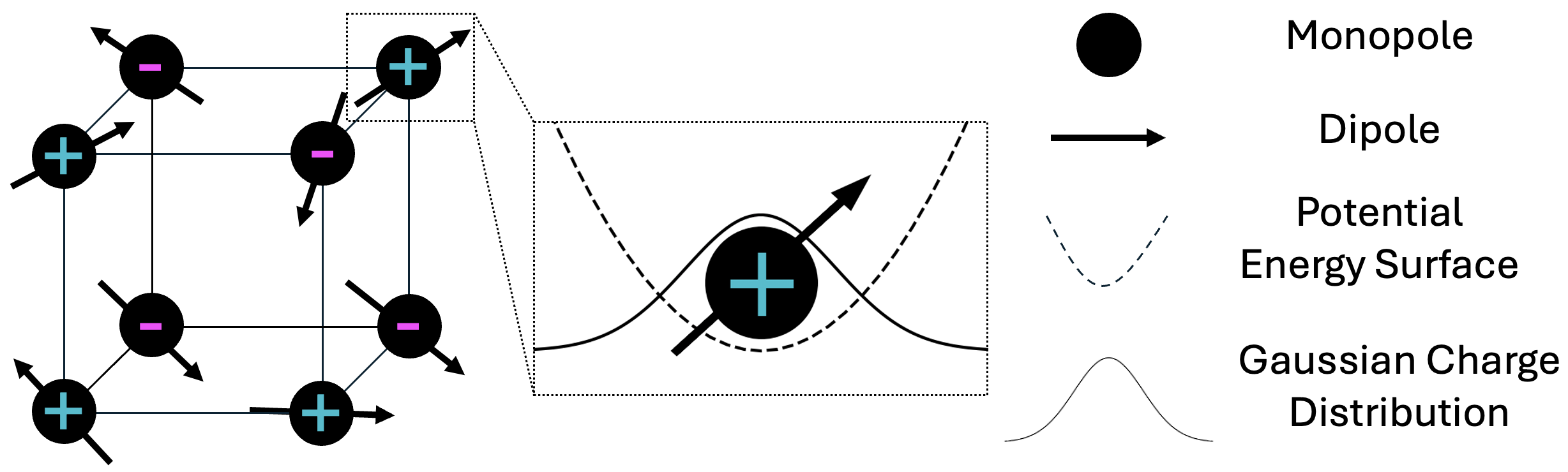}
    \caption{Model atomic system of eight atoms on the corners of a box with a single atom zoomed in to show details of the model. Blue "${\bf +}$" and magenta "${\bf -}$" symbols represent positive and negative monopole partial charges and arrows represent dipoles. The dotted curve illustrates the charge-independent potential energy surface, $V({\bf R})$, and the solid curve illustrates the Gaussian monopole charge distribution.}
    \label{fig:eightAtom_exampleAtom}
\end{figure}

\section{Flexible Multipole Model} \label{FlexMultipole}

For the expanded flexible multipole model, we follow the same general format of presentation as for the regular flexible monopole model, i.e.\ by first describing regular Born-Oppenheimer MD using the expanded multipole model and then discussing the corresponding shadow Born-Oppenheimer MD. The theory is formulated up to dipole-dipole interactions. However, generalizations to higher-order multipoles should be straightforward, at least in principle.

\subsection{Regular Born-Oppenheimer MD}

In a flexible multipole model we need to go beyond the monopole-monopole interactions in the energy expression in Eq.\ (\ref{eq: monoEnergy}) to include additional monopole-dipole and dipole-dipole terms. The corresponding flexible multipole energy function is given by
\begin{equation}\label{eq:dipole_energy}
    E({\bf R,q,p}) = {\bf q}^{\rm T}{\boldsymbol \chi} + \frac{1}{2} [{\bf q}^{\rm T}{\bf p}^{\rm T}] 
        \begin{bmatrix}
        {\bf C}& {\bf W}^{\rm T}\\
        {\bf W} & {\boldsymbol \Lambda}
        \end{bmatrix}
        \begin{bmatrix}
        {\bf q}\\
        {\bf p}
        \end{bmatrix} ,
\end{equation}
where
\begin{align}
&{\bf q} \in {\mathbb R}^{N} ~~~ \{{\rm Monopoles\}},\\
&{\bf p} \in {\mathbb R}^{3N} ~~\{{\rm Dipoles\}},
\end{align}
are the monopoles and dipoles, respectively. An example system of eight atoms is shown in Fig.\ \ref{fig:eightAtom_exampleAtom} which displays the monopole and dipole components as well as the charge-independent potential energy surface and the atom-centered Gaussian charge distributions that are used in our model.
In Eq.\ (\ref{eq:dipole_energy}), the electronegativities are given by ${\boldsymbol \chi} \in {\mathbb R}^{N}$ and the interaction matrices are given by
\begin{align}
&{\bf C} \in {\mathbb R}^{N\times N} ~~~ \{{\rm monopole - monopole\}},\\
&{\bf W} \in {\mathbb R}^{3N \times N} ~~\{{\rm dipole-monopole\}},\\
&{\boldsymbol \Lambda }\in {\mathbb R}^{3N \times 3N} ~~\{{\rm dipole-dipole}\}.
\end{align}
Here ${\bf C}$ is the monopole-monopole Coulomb interaction matrix, ${\bf W}$ is the dipole-monopole interaction matrix, and ${\boldsymbol \Lambda }$ is the dipole-dipole interaction matrix. The explicit expressions for the interaction matrices are given in Appendix \ref{ap: interact}.

The exact regular Born-Oppenheimer potential energy surface, $U_{\rm BO}(\mathbf{R})$, is then given by the energy for the equilibrated multipoles plus a charge-independent potential, $V(\mathbf{R})$, where
\begin{equation} \label{eq:dipole_bopoten}
    \begin{aligned}
        &U_{\rm BO}(\mathbf{R}) = V(\mathbf{R}) + \\
        &~~~\left\{ E(\mathbf{R, q, p}) \Big | \frac{\partial E}{\partial \mathbf{q}} = \mathbf{0}, \frac{\partial E}{\partial \mathbf{p}} = \mathbf{0}, \sum_i q_i = Q_{\rm tot} \right\}.
    \end{aligned}
\end{equation}
The constrained multipole equilibration is determined by the constraints,
\begin{equation}
\left\{ \begin{array}{l}
{\partial E}/{\partial \mathbf{q}} = \mathbf{0},\\
{\partial E}/{\partial \mathbf{p}} = \mathbf{0}, \\
 \sum_i q_i = Q_{\rm tot},  
\end{array}\right.
\end{equation}
which give the equilibrated monopoles and dipoles, ${\bf q}_0$ and ${\bf p}_0$, from the solution of the set of linear equations,
\begin{equation} \label{eq:RegularMultipoleEquilibration}
\begin{bmatrix}
\bf C & \bf W^{\rm T} & \bf 1 \\
\bf W & \boldsymbol \Lambda & \bf 0 \\
\bf 1^{\rm T} & \bf 0^{\rm T} & 0
\end{bmatrix}
\begin{bmatrix}
{\bf q}_0 \\
{\bf p}_0 \\
{\mu}_0
\end{bmatrix}
=
\begin{bmatrix}
-\boldsymbol \chi \\
\bf 0 \\
Q_{\rm tot}
\end{bmatrix}.
\end{equation}
The solution gives us the Born-Oppenheimer potential,
\begin{equation}
    U_{\rm BO}({\bf R}) = V({\bf R}) + E({\bf R},{\bf q}_0,{\bf p}_0).
\end{equation}
This potential can then be used to define our dynamical system through the Lagrangian,
\begin{equation}
    \mathcal L({\bf R}, {{\bf \dot R}}) = \frac{1}{2} \sum_i m_i |{\bf \dot R}_i|^2 - U_{\rm BO}({\bf R}).
\end{equation}
The Euler-Lagrange equations for the stationary action functional, 
\begin{align}
    & \frac{\rm d}{\rm dt}\left(\frac{\partial {\cal L}}{\partial {\bf \dot R}_i}\right) = \frac{\partial {\cal L}}{\partial {\bf R}_i}
\end{align}
then give us Newton's equations of motion,
\begin{equation}
    m_i {\mathbf{\ddot R}}_i = -\frac{\partial U_{\rm BO}(\mathbf{R})}{\partial {\bf R}_i}.
\end{equation}
The integration of the equations of motion can be performed using a velocity-Verlet integrator as described in Alg.\ \ref{alg: bomd}, which then generates the trajectories of an MD simulation. 

This approach to Born-Oppenheimer MD is thus very closely aligned with the atomic monopole-only equilibration model and is easily extendable to higher-order multipoles. The same holds for shadow Born-Oppenheimer MD discussed in the next section. 

In the same way as for the monopole-only charge equilibration model, the expense of solving for the relaxed constrained monopoles and dipoles in Eq.\ (\ref{eq:RegularMultipoleEquilibration}) dominates the overall cost. For larger systems or systems with periodic boundary conditions, iterative solvers can be used to reduce this cost. However, unless sufficiently converged (which likely is harder with the increased number of degrees of freedom for multipole models) the dynamics are unstable and may yield an unphysical results. Here we will show how shadow MD for multipole models alleviates these problems.

\subsection{Shadow Born-Oppenheimer MD}

The shadow MD for the flexible multipole model is designed similarly to the monopole-only charge equilibration model. To construct the approximate shadow energy function based on $E({\bf R},{\bf q}, {\bf p})$ in Eq.\ (\ref{eq:dipole_energy}), we start by splitting the interaction matrix into short-range (S) and long-range (L) parts, where
\begin{equation}\label{eq:Range_Separation}
     \begin{bmatrix}
        {\bf C}& {\bf W}^{\rm T}\\
        {\bf W} & {\boldsymbol \Lambda}
        \end{bmatrix} = 
        \begin{bmatrix}
    {\bf C}_{\rm S}& {\bf W}^{\rm T}_{\rm S}\\
     {\bf W}_{\rm S} & {\boldsymbol \Lambda}_{\rm S}
    \end{bmatrix}
    +
    \begin{bmatrix}
    {\bf C}_{\rm L}& {\bf W}^{\rm T}_{\rm L}\\
    {\bf W}_{\rm L} & {\boldsymbol \Lambda}_{\rm L}
    \end{bmatrix} .
\end{equation}
In the same way as for the shadow monopole model, we chose ${\bf C}_{\rm S}$ as the diagonal of the Coulomb matrix given by the chemical hardness parameters, ${\bf u} \in {\mathbb R}^{N}$. We set the short-range monopole-dipole interaction matrix to zero, i.e.\ $ {\bf W}_{\rm S} \equiv {\bf 0}$, and choose ${\boldsymbol \Lambda}_{\rm S}$ as the diagonal, onsite, dipole-dipole interaction elements given by inverse polarizability parameters, $\boldsymbol{\alpha}^{-1}$ (See Appendix \ref{ap: interact} for details). However, other choices are possible, e.g.\ choosing ${\bf C}_{\rm S}$ and ${\boldsymbol \Lambda}_{\rm S}$ as block-diagonal parts corresponding to some  intramolecular Coulomb interactions. We then take the long-range interaction matrices as the remaining parts of the interaction matrices, i.e.\ such that Eq.\ (\ref{eq:Range_Separation}) holds.
The shadow energy function is then defined by a partial linearization of $E({\bf R},{\bf q}, {\bf p})$ in Eq.\ (\ref{eq:dipole_energy}) around some approximate equilibrated solution,
\begin{equation}
    \begin{bmatrix}
        {\bf n}\\
        {\bf p}
    \end{bmatrix}
    \approx 
    \begin{bmatrix}
        {\bf q}_0\\
        {\bf p}_0
    \end{bmatrix}.
\end{equation}
Our approximate shadow energy function is then defined by 
\begin{equation} \label{eq:shadowMultipole-energy}
    \begin{array}{l}
        {\cal E}({\bf R}, {\bf q,p},{\bf n,d}) = {\bf q}^{\rm T}{\boldsymbol \chi} + \frac{1}{2} [{\bf q^{\rm T}}~{\bf p^{\rm T}}] 
    \begin{bmatrix}
    {\bf C}_{\rm S}& {\bf W}^{\rm T}_{\rm S}\\
    {\bf W}_{\rm S} & {\boldsymbol \Lambda}_{\rm S}
    \end{bmatrix}
    \begin{bmatrix}
    {\bf q}\\
    {\bf p}
    \end{bmatrix}+\\
     ~~\\
    ~~~~~~~+ \frac{1}{2} \left(2[{\bf q^{\rm T}}~{\bf p^{\rm T}}] - [{\bf n^{\rm T}}~{\bf d^{\rm T}}] \right)
    \begin{bmatrix}
    {\bf C}_{\rm L}& {\bf W}^{\rm T}_{\rm L}\\
    {\bf W}_{\rm L} & {\boldsymbol \Lambda}_{\rm L}
    \end{bmatrix}
    \begin{bmatrix}
    {\bf n}\\
    {\bf d}
    \end{bmatrix}.
     \end{array}
\end{equation}
Only the long-range part (L) of the energy term is linearized around $[{\bf n}^{\rm T},{\bf d}^{\rm T}]$, whereas the short-range part (S) is kept to second order.

To simplify the notation and to highlight the straightforward generalization to higher-order multipole models, we can use the composite notation
\begin{align}
    &\mathbf{x} \equiv 
    \begin{bmatrix}
        \mathbf{n} \\
        \mathbf{d}
    \end{bmatrix}, ~\mathbf{c}[\mathbf{x}] \equiv 
    \begin{bmatrix}
        \mathbf{q}[\mathbf{x}]\\
        \mathbf{p}[\mathbf{x}]
    \end{bmatrix}, ~~
    {\bf b} \equiv  
    \begin{bmatrix}
        {-\boldsymbol \chi}\\
        \mathbf{0}
    \end{bmatrix}
\end{align}
and
\begin{equation}
     {\bf G} \equiv  \begin{bmatrix}
        {\bf C}& {\bf W}^{\rm T}\\
        {\bf W} & {\boldsymbol \Lambda}
        \end{bmatrix} = {\bf G}_{\rm S} + {\bf G}_{\rm L}
\end{equation}
for the short-range and long-range interaction separation. Here we assume that ${\bf G}_{\rm S}$ consists of the diagonal (or short-ranged) parts of ${\bf C}$ and ${\boldsymbol \Lambda}$.
In this way our shadow energy function is given by
\begin{equation} \label{eq:shadowFlexEnergy}
        {\cal E}({\bf R},{\bf c},{\bf x}) = {\bf c}^{\rm T}{\bf b}+ \frac{1}{2} {\bf c}^{\rm T} {\bf G}_{\rm S} {\bf c} + \frac{1}{2} \big(2{\bf c}^{\rm T} - {\bf x}^{\rm T}\big) {\bf G}_{\rm L} {\bf x}.
\end{equation}
The equilibrated constrained monopoles and dipoles then give us the shadow Born-Oppenheimer potential,
\begin{equation}\label{eq:shadowMultipole_bopoten}
    \begin{aligned}
        &\mathcal{U}_{\rm BO}(\mathbf{R}, \mathbf{x}) = V(\mathbf{R}) +\\
        &~~~~+\Big\{ \mathcal{E}(\mathbf{R}, {\bf c},{\bf x}) ~\Big | ~\frac{\partial \mathcal{E}({\bf R, c, x})}{\partial \mathbf{c}} = \mathbf{0},~ \sum_i q_i = Q_{\rm tot} \Big \}.
    \end{aligned}
    \end{equation}
The constrained equilibration is determined from 
\begin{equation}
\left \{\begin{array}{l}
{\partial {\cal E}}/{\partial \mathbf{c}} = \mathbf{0},\\
 \sum_i q_i = Q_{\rm tot},  
\end{array}\right.
\end{equation}
which gives us the ${\bf x}$-dependent equilibrated monopoles and dipoles, ${\bf c}[{\bf x}]$, from the system of linear equations,
\begin{equation} \label{eq:MultipoleShadowEquilibration}
\begin{bmatrix}
{\bf G}_{\rm S}&  {\bf 1}\\
{\bf 1}^{\rm T} &  0
\end{bmatrix}
\begin{bmatrix}
{\bf c}[{\bf x}]\\
\mu
\end{bmatrix}
= 
\begin{bmatrix}
{\bf b}\\
Q_{\rm tot}
\end{bmatrix} - 
\begin{bmatrix}
    {\bf G}_{\rm L} & {\bf 0}\\
    {\bf 0} & {\bf 0}
\end{bmatrix}
\begin{bmatrix}
    {\bf x}\\
    {\bf 0}
\end{bmatrix}
.
\end{equation}
The corresponding shadow Born-Oppenheimer potential is then given by
\begin{equation}
    {\cal U}_{\rm BO}({\bf R}, {\bf x}) = V({\bf R}) + {\cal E}({\bf R},{\bf c}[{\bf x}],{\bf x}).
\end{equation}
The linear set of equations in Eq.\ (\ref{eq:MultipoleShadowEquilibration}) is easy to solve directly by rewriting the left-hand side system matrix as a rank-2 update of a diagonal (or block diagonal) matrix and then use using the Woodbury formula for the matrix inverse. Only a single electrostatic vector potential calculation from the monopoles and dipoles is necessary on the right-hand side, which can be performed, for example, with linear scaling complexity using the Ewald summation method for periodic boundary conditions, taking advantage of the fast Fourier transform \cite{Kaymak25}. This represents a significant speed-up compared to the regular multipole model in Eq.\ (\ref{eq:RegularMultipoleEquilibration}), which in general needs to be solved iteratively and with tight convergence, requiring the construction of a new electrostatic potential in each iteration.

It is easy to see that we could replace ${\bf c}$ and ${\bf x}$ and the corresponding interaction matrices ${\bf G}_{\rm S}$ and ${\bf G}_{\rm L}$ by vectors of higher-order multipoles and their multipole interaction matrices. The theory presented here is thus straightforward to generalize to any order in the flexible multipole expansion.

In the same way as for the shadow Born-Oppenheimer MD with the flexible monopole-only model, we can now define our dynamics with the extended Lagrangian,
\begin{equation}
   \begin{aligned} 
    &\mathcal{L}(\mathbf{R, \dot R, x, \dot x}) = \frac{1}{2} \sum_i m_i |\mathbf{\dot R}_i|^2 -\mathcal{U}_{\rm{BO}}(\mathbf{R, x}) +\\
    &~~~~~+ \frac{\mu_{\rm e}}{2}\sum_i \dot x_i^2- \frac{\mu_{\rm e} \omega^2}{2}({\bf c}^{\rm T}[{\bf x}] - {\bf x}^{\rm T}) {\bf K}^{\rm T}{\bf K}({\bf c}[{\bf x}] - {\bf x}).
\end{aligned} 
\end{equation}
Here we have included ${\bf x}(t) \equiv {\bf x}$ and ${\bf \dot x}(t) \equiv {\bf \dot x}$ as extended time-dependent dynamical variables that are driven by a harmonic oscillator, which is centered around the equilibrated multipoles, ${\bf c}[{\bf x}]$. In this way ${\bf x}(t)$ will follow the optimized shadow ground state ${\bf c}[{\bf x}]$. ${\bf K}^{\rm T}{\bf K}$ is a symmetric positive definite metric tensor, where the kernel is defined by the inverse Jacobian of the residual function, i.e.\
\begin{align}
&{\bf K} = {\bf J}^{-1},~~
J_{ij} = \frac{\partial \left(  c_i[{\bf x}]-{ x_i}\right)}{\partial x_j}.
\end{align}
As in the extended Lagrangian using the monopole shadow Born-Oppenheimer potential, $\mu_{\rm e}$ is a fictitious charge mass parameter and $\omega$ is the frequency of the harmonic oscillator. 

The equations of motion are derived in an adiabatic limit from Euler-Lagrange equations, 
\begin{align}
    & \frac{{\rm d}}{{\rm d}t}\left( \frac{\partial {\cal L}}{\partial {\bf \dot R}_i} \right) = \frac{\partial {\cal L}}{\partial {\bf R}_i},\\
    & \frac{{\rm d}}{{\rm d}t}\left( \frac{\partial {\cal L}}{\partial {\bf \dot x}} \right) = \frac{\partial {\cal L}}{\partial {\bf x}},
\end{align}
where we assume that the nuclear motion is slow compared to the extended multipole degrees of freedom. We capture this Born-Oppenheimer-like assumption in the derivation of the equations of motion by asserting an adiabatic limit, where $\mu_{\rm e} \rightarrow 0$, $\omega \rightarrow \infty$, with $\mu_{\rm e} \omega = {\rm constant}$. In this mass-zero limit we get the equations of motion as 
\begin{align} 
&m_i {\bf \ddot R}_i = -\frac{\partial {\cal U}_{\rm BO}({\bf R},{\bf x})}{\partial {\bf R}_i}\big \vert _{{\bf x}} \label{eq:shadowMultipole_eqofmotion_1},\\
&{\bf \ddot x} = 
- \omega^2 {\bf K} \left(
{\bf c}[{\bf x}]-{\bf x}
 \right) \label{eq:shadowMultipole_eqofmotion_2},
\end{align} 
which are exact in continuous time. The kernel, ${\bf K}$, acts on the residual function, similar to a Newton step, such that the extended electronic dynamical variables, ${\bf x}$, oscillate around a close approximation to the exact ground state, $[{\bf q}_0,{\bf p}_0]$, of the exact regular Born-Oppenheimer multipoles. This improves the stability and accuracy of the shadow potential and keeps ${\bf x}$ and {\bf c}[{\bf x}] close to the exact ground state, $[{\bf q}_0,{\bf p}_0]$.
Integration of these equations of motion can be performed using the velocity-Verlet scheme for the atomic coordinates and velocities and with the modified Verlet schemes for the extended charge degrees of freedom, as described in Algorithm \ref{alg: shadowbomd}, with ${\bf q}$ replaced by ${\bf c}$ and ${\bf n}$ by ${\bf x}$. The modified Verlet scheme includes a weak dissipative force term that avoids accumulation of numerical noise that would occur in an exactly time-reversible propagation \cite{ANiklasson09,Niklasson21b}.

The kernel, ${\bf K}$, makes ${\bf x}(t)$ follow an even closer approximation of the exact regular equilibrated ground state than ${\bf c}[{\bf x}]$ \cite{Niklasson21b}.
We use the preconditioned, low-rank Krylov subspace approach to approximate the kernel, ${\bf K}$, from the inverse Jacobian of the residual function, ${\bf f}({\bf x})=({\bf c}[{\bf x}] - {\bf x})$, as described in Appendix \ref{ap:lrupdate}. This approach is also straightforward to extend to 
arbitrary multipole order. 
Detailed expressions for the multipole interaction terms and forces are given in Appendix \ref{ap: energyAndForce}.

\section{Fixed Monopole/Flexible Dipole Model} \label{fixedmono}

Charge equilibration models can experience problems with unphysical net partial charge transfer. This appears, for example, in dissociated unattached atomic fragments that attain fractional equilibrated charges, or when the polarization increases non-linearly with system size in non-conjugated molecular systems, such as in a metal where the charge is free to move \cite{RStanton25}. A way to address these problems is to keep the atomic partial monopole charges fixed (or determined by some separate local environment-dependent ML approach) while letting only the dipoles fluctuate. This approach is commonly used in polarizable force fields \cite{Liu98, Dang00, Halgren01, Koneshan01, Dang02, Kaminski02, Dang03, GreshPiquemal07, Wick05, Wick07, WickDang07, Baranyai10, Ponder10, Wang11-1, Wang11-2, WangLuo12, Chang14, Li14, JingRen19}. Here we will present a shadow MD formulation for this fixed monopole/flexible dipole model. First we show the regular Born-Oppenheimer model and then the corresponding shadow MD formulation.

\subsection{Regular Born-Oppenheimer MD}

In the regular Born-Oppenheimer MD formulation of the fixed monopole/flexible dipole model, we start by defining our energy function as for the flexible multipoles in Eq.\ (\ref{eq:dipole_energy}), but with fixed monopole charges, ${\bf q}_0$. In this case 
\begin{equation}
    E_{{\bf q}_0}({\bf R},{\bf p}) = {\bf q}_0^{\rm T}{\boldsymbol \chi} + \frac{1}{2} [{\bf q}_0^{\rm T}{\bf p}^{\rm T}] 
    \begin{bmatrix}
    {\bf C}& {\bf W}^{\rm T}\\
    {\bf W} & {\boldsymbol \Lambda}
    \end{bmatrix}
    \begin{bmatrix}
    {\bf q}_0\\
    {\bf p}
    \end{bmatrix},
\end{equation}
where
\begin{equation}    
\left\{\begin{array}{l}
{\bf q}_0\in {\mathbb R}^N  ~~~{\rm fixed~atomic ~monopole~charges},\\
{\bf p} \in {\mathbb R}^{3N} ~~{\rm ~flexible~ atomic~dipoles}.
\end{array}
\right.
\end{equation}
Because the atomic monopole charges are fixed, we do not need to include a net charge constraint with the corresponding chemical potential in the construction of the Born-Oppenheimer potential. The exact regular Born-Oppenheimer potential is then given by
\begin{equation}\label{eq: multiPotential_fc}
    U_{\rm BO}(\mathbf{R}) = V(\mathbf{R}) +  \left\{ E_{{\bf q}_0}({\bf R},\mathbf{p}) \Big | \frac{\partial E({\bf R},\mathbf{p})}{\partial \mathbf{p}} = \mathbf{0} \right\}.
\end{equation}
From the system of linear equations, $\partial E_{{\bf q}_0}({\bf R},{\bf p}) \big/\partial {\bf p} = {\bf 0}$, i.e.\
\begin{equation} \label{eq:DipoleEquil-fc}
     {\boldsymbol \Lambda}{\bf p}  = - {\bf W}{\bf q}_0 
\end{equation}
we get the equilibrated dipoles,
\begin{equation} \label{eq:DipoleEquil-fc2}
 {\bf p}_0 = -{\boldsymbol \Lambda }^{-1} {\bf W} {\bf q}_0
\end{equation}
that determine the Born-Oppenheimer potential
\begin{equation}
    { U}_{\rm BO}({\bf R}) = V({\bf R}) +  {E}_{{\bf q}_0}({\bf R},{\bf p}_0).
\end{equation}
This potential can then be used to define the dynamics with the Lagrangian,
\begin{equation}
    \mathcal L({\bf R}, \dot{{\bf R}}) = \frac{1}{2} \sum_i m_i |{\bf \dot R}_i|^2 - U_{\rm BO}({\bf R}).
\end{equation}
The Euler-Lagrange equations, 
\begin{align}
    & \frac{\rm d}{\rm dt}\left(\frac{\partial {\cal L}}{\partial {\bf \dot R}_i}\right) = \frac{\partial {\cal L}}{\partial {\bf R}_i},\\
\end{align}
then give us Newton's equations of motion,
\begin{equation}
    m_i \ddot{\mathbf{R}}_i = -\frac{\partial U_{\rm BO}(\mathbf{R})}{\partial {\bf R}_i} \Big |_{\bf p}.
\end{equation} 
These equations of motion can be integrated in the same way as the regular Born-Oppenheimer MD scheme, Alg.\ \ref{alg: bomd}. Detailed expressions for the multipole interaction energy terms and forces are given in Appendix \ref{ap: energyAndForce}.

The main cost of this fixed-monopole/flexible dipole model, is the dipole equilibration, which requires the solution to Eq.\ (\ref{eq:DipoleEquil-fc}). Solving this system of linear equations is in general most easily done with an iterative method. To generate accurate and stable dynamics, the solutions need to be tightly converged, requiring repeated dipole summations. With the shadow MD formulation this costly iterative approach and potential instabilities can be avoided.

\subsection{Shadow Born-Oppenheimer MD} 

The shadow MD and the shadow Born-Oppenheimer potential can be constructed from a partially linearized shadow energy function in the same way as before. 
Thanks to the fixed monopole net charges, ${\bf q}_0$, the shadow energy function is simplified.
In the shadow energy function, we only need to separate the dipole-dipole interaction matrix, $\boldsymbol{\Lambda}$, into a short-range (S) and a long-range (L) part where
\begin{equation}
{\boldsymbol \Lambda} 
 \equiv 
 {\boldsymbol \Lambda}_{\rm S} 
+
{\boldsymbol \Lambda}_{\rm L} .
\end{equation}
Here, we choose $\boldsymbol{\Lambda}_{\rm S}$ to be the diagonal components of ${\boldsymbol{\Lambda}}$, given by the inverse polarizability parameters, ${\boldsymbol{\alpha}}^{-1}$, where ${\boldsymbol{\Lambda}}_{\rm L}$ contain the remaining off-diagonal entries of the ${\boldsymbol{\Lambda}}$ matrix. However, other options are possible, for example, choosing some block diagonal form of $\boldsymbol{\Lambda}_{\rm S}$. We then construct the shadow energy function by performing a linearization for the long-range dipole-dipole interaction term around some approximate solution, ${\bf d}$, to the equilibrated dipoles of the regular fixed-monopole/flexible dipole Born-Oppenheimer model, i.e.\ where ${\bf d} \approx {\bf p}_0$. This give us the shadow energy function,
\begin{equation}\label{eq: fc-shadowMulti-energymatform}
    \begin{aligned}
    {\cal E}_{{\bf q}_0}({\bf R}, {\bf p},{\bf d}) = {\bf q}_0^{\rm T}{\boldsymbol \chi} + \frac{1}{2} {\bf q}_0^{\rm T} \mathbf{C} \mathbf{q}_0 + \mathbf{q}_0^{\rm T} \mathbf{W} \mathbf{p} + \\
    \frac{1}{2} \mathbf{p}^{\rm T} \mathbf{\Lambda}_{\rm S} \mathbf{p} + \frac{1}{2} \left( 2 \mathbf{p}^{\rm T} - \mathbf{d}^{\rm T}\right) \mathbf{\Lambda}_{\rm L} \mathbf{d}.
    \end{aligned}
\end{equation}
The shadow Born-Oppenheimer potential is then given from the constrained optimization of this energy function, where
\begin{equation}\label{eq: fc-multiShadowPoten}
  \begin{aligned}
      &\mathcal{U}_{\rm BO}(\mathbf{R}, \mathbf{d}) = V(\mathbf{R}) + \\
      &~~~~\left\{ \mathcal{E}_{{\bf q}_0}(\mathbf{R, p, d}) \Big| \frac{\partial \mathcal{E}_{{\bf q}_0}({\bf R,p,d})}{\partial \mathbf{p}} = \mathbf{0}\right\}.
  \end{aligned}
\end{equation}
This dipole-only equilibration for the shadow Born-Oppenheimer potential is simplified, because no monopole net-charge constraint is required. 
The ${\bf d}$-dependent equilibrated dipoles, ${\bf p}[{\bf d}]$, are determined from the linear system of equations,
\begin{equation}\label{eq:DipoleEquil_a}
\frac{\partial {\cal E}_{{\bf q}_0} ({\bf R,p,d})}{ \partial {\bf p}} = {\bf 0}, 
\end{equation}
or 
\begin{equation}\label{eq:DipoleEquil}
{\boldsymbol \Lambda }_{\rm S}{\bf p}[{\bf d}] = -\big( {\bf W} {\bf q}_0 + {\boldsymbol \Lambda}_{\rm L}{\bf d}\big).
\end{equation}
The ${\bf d}$-dependent solution, ${\bf p}[{\bf d}]$, then gives us the shadow Born-Oppenheimer potential,
\begin{equation}
    \mathcal{U}_{\rm BO}(\mathbf{R}, \mathbf{d}) = V(\mathbf{R}) +
       \mathcal{E}_{{\bf q}_0}(\mathbf{R}, {\bf p}[{\bf d}], {\bf d}).
\end{equation}
Because the system matrix ${\boldsymbol \Lambda}_{\rm S}$ is diagonal we get the exact solution in Eq.\ (\ref{eq:DipoleEquil}) directly, i.e.\ in contrast to the regular equilibration problem in Eq.\ (\ref{eq:DipoleEquil-fc}). The computational cost is dominated by a single electrostatic vector potential calculation from the monopoles and dipoles. No iterative solver is required. 

\begin{algorithm}
\caption{Conjugate Gradient Algorithm to Update $\mathbf{\ddot d}$ for the Fixed Charge/Flexible Dipole Model.}\label{alg: lrupdate-fc}
    \SetAlgoNoLine
    \SetAlgoNoEnd
    \tcp{Set initial guess $\mathbf{z}_0$}
    $\mathbf{z} = \mathbf{0}~~{\rm or} ~~\mathbf{z} =\mathbf{d} - \mathbf{p}[\mathbf{d}]$
    \vspace{1mm}

    \tcp{Calculate Residual ${\bf r}$}
    ${\bf r} = \left(\mathbf{Wq} + \mathbf{\Lambda_{\rm S} d} + \mathbf{\Lambda_{\rm L} d} \right) - \mathbf{\Lambda z}$ 
    \vspace{1mm}

    \tcp{Set Max Rank and Convergence Tolerance}
    $k_{max} = max\_rank$, ~~
    $\varepsilon = error\_threshold$
    \vspace{1mm}
        
    \tcp{Initialize Rank and Error}
    $k = 0$, ~~
    ${\rm Err} = ||{\bf r}||_2$
    \vspace{3mm}
    
    \tcp{Conjugate Gradient Loop to Update $\mathbf{z}$}
    \While{${\rm Err} > \varepsilon$ {\bf and} $k \leq k_{max}$}{

        \vspace{1mm}
         
        ~~~~\tcp{Increase Rank}
        ~~~~$k = k + 1$
        
        \vspace{1mm}
        
        ~~~~\tcp{Preconditioned Residual Function} 
        ~~~~$\mathbf{y} = \mathbf{\Lambda}_{\rm S}^{-1} {\bf r}$ 
        \vspace{1mm}
        
        ~~~~$v = {\bf r}^{\rm T} \mathbf{y}$
        \vspace{2mm}

        ~~~~\tcp{Update $\mathbf{p}$ Based on Current Rank}
        ~~~~\uIf{$k == 1$}{
            ~~~~~~~~$\mathbf{p} = \mathbf{y}$
        ~~~~}
        ~~~~\ElseIf{$k > 1$}{
            ~~~~~~~~$\mathbf{p} = \mathbf{y} + (v/v_0)\mathbf{p}$
        ~~~~}
        ~~~~$\mathbf{end}$
        \vspace{3mm}

        ~~~~\tcp{Update potential, $\mathbf{w}$ (Main Cost!)}
        ~~~~$\mathbf{w} = \mathbf{\Lambda p}$ \\
        ~~~~$\alpha = v/(\mathbf{p}^{\rm T}\mathbf{w})$
        ~~~~\vspace{1mm}

        ~~~~\tcp{Update solution $\mathbf{z}$}
        ~~~~$\mathbf{z} = \mathbf{z} + \alpha \mathbf{p}$
        \vspace{1mm}

        ~~~~\tcp{Update Residual}
        ~~~~${\bf r} = {\bf r} - \alpha \mathbf{w}$
        \vspace{1mm}

        ~~~~\tcp{Recalculate Error}
        ~~~~${\rm Err} = ||{\bf r}||_2$
        \vspace{1mm}

        ~~~~$v_0 = v$
        \vspace{1mm}
        
    }
    $\mathbf{end}$
    \vspace{3mm}

    \tcp{For $k_{max}=0$}
    \uIf{$k_{max} == 0$}{~~~~$z=\boldsymbol{\Lambda}_{\rm S}{\bf r}$}
    $\mathbf{end}$
    \vspace{3mm}
    
    \tcp{Final answer at convergence}
    $\mathbf{\ddot d} = - \omega^2 \mathbf{z}$
\end{algorithm}

To keep the shadow Born-Oppenheimer potential in close agreement with the regular exact Born-Oppenheimer potential in Eq.\ (\ref{eq: multiPotential_fc}), we need to update the approximate dipoles, ${\bf d}$, around which we did the expansion of our shadow energy function. We can do this by including ${\bf d}$ and its time derivative, ${\bf \dot d}$, as extended dynamical vector variables, ${\bf d(t)}$ and ${\bf \dot d}(t)$, which follow the relaxed ground state solution (or our best available approximation of the relaxed ground state) through a harmonic oscillator. This can be achieved, in the same way as before, within  extended Lagrangian dynamics, which we here define by the Langrangian,
\begin{equation}
    \begin{aligned}
        &\mathcal{L}(\mathbf{R, \dot R, d, \dot d}) = \frac{1}{2} \sum_i m_i |\mathbf{\dot R}_i|^2 \\
        & -\mathcal{U}_{\rm{BO}}(\mathbf{R, d}) + \frac{1}{2} \mu_{\rm e} \sum_i |{\bf \dot d}_i |^2 \\
        &-\frac{1}{2} \mu_{\rm e} \omega^2 \left( \mathbf{p}^{\rm T}[\mathbf{d}] - \mathbf{d}^{\rm T} \right) \mathbf{K}^{\rm T} \mathbf{K} \left( \mathbf{p}[\mathbf{d}] - \mathbf{d} \right).
    \end{aligned}
\end{equation}
As before, $\mu_{\rm e}$ is a fictitious mass parameter, $\omega$ is the frequency of the extended harmonic oscillator, and $\mathbf{K}^{\rm T} \mathbf{K}$ is a symmetric positive definite metric tensor. 
In the same way a before, we define the kernel, ${\bf K}$, as the inverse Jacobian of the residual function, i.e.\
\begin{equation}
    {\bf K} = {\bf J}^{-1}, ~~J_{i\alpha,j \beta} = \frac{\partial (p_{i\alpha}[\mathbf{d}] - d_{i \alpha})}{\partial d_{j\beta}},
\end{equation}
where $\alpha$ and $\beta$ are the $[x,y,z]$ components of the dipole vectors for each atom $i$ or $j$.
In this case it is easy to see from Eq.\ (\ref{eq:DipoleEquil}) that 
\begin{equation} \label{eq:fc_jacobian}
    \mathbf{J} = - \mathbf{\Lambda}_{\rm S}^{-1} \mathbf{\Lambda}_{\rm L} - \mathbf{I}. 
\end{equation}

The equations of motion for the shadow fixed-monopole/flexible dipole model with the shadow potential are derived from Euler-Lagrange's equations in the adiabatic Born-Oppenheimer-like limit, in the same way as before, where $\mu_{\rm e} \rightarrow  0$, and $\omega \rightarrow \infty$ such that $\mu_{\rm e} \omega = {\rm constant}$. This gives us the coupled equations of motion,
\begin{align} 
&m_i {\mathbf {\ddot R}}_i = - \dfrac{\partial {\cal U}_{\rm BO}({\mathbf R}, {\mathbf d})}{\partial {\mathbf R}_i} \Big |_{\mathbf {q}_0,{\bf d}} \label{eq:eqofmotion_fc_1},\\
&{\mathbf {\ddot d}} = - \omega^2 {\mathbf K}(\mathbf{p}[\mathbf{d}] - \mathbf{d})\label{eq:eqofmotion_fc_2},
\end{align} 
which are exact in continuous time.

For this fixed-monopole/ flexible dipole model, we can reformulate the equations of motion for ${\bf \ddot{d}}$ in Eq.\ (\ref{eq:eqofmotion_fc_2}) by using the explicit expression for the Jacobian in Eq.\ (\ref{eq:fc_jacobian}). After some manipulations we get the equivalent but computationally convenient form,
\begin{align} 
&m_i {\mathbf {\ddot R}}_i = - \dfrac{\partial {\cal U}_{\rm BO}({\mathbf R}, {\mathbf d})}{\partial {\mathbf R}_i} \Big |_{\mathbf {q}_0,{\bf d}} \label{eq:eqofmotion_fc_1_B},\\
&    {\boldsymbol \Lambda}{\bf \ddot d} = -\omega^2  \left({\bf W} {\bf q}_0 + {\boldsymbol \Lambda}{\bf d} \right).\label{eq:eqofmotion_fc_2_B}
\end{align}

These equations of motion provide an alternative form to Eqs.\ (\ref{eq:eqofmotion_fc_1}) and (\ref{eq:eqofmotion_fc_2}).
This reformulation is also possible for the monopole-only and the multipole models if we include chemical potential as an extra dynamical variable. This reformulation is useful because it allows us to use established Krylov subspace solvers, such as the generalized minimum residual (GMRES) method \cite{Saad86}, or the conjugate gradient scheme \cite{Nazareth09}.
In place of the low-rank update scheme described for the flexible charge and dipole model described in Appendix \ref{ap:lrupdate}, we can use the preconditioned conjugate gradient algorithm for the fixed monopole/flexible dipole model to integrate the equations of motion in Eq.\ (\ref{eq:eqofmotion_fc_2_B}), as described in Alg.\ \ref{alg: lrupdate-fc}. 

The conjugate gradient method in Alg.\ \ref{alg: lrupdate-fc} can be used because the $\mathbf{\Lambda}$ dipole interaction matrix is symmetric positive definite. The conjugate gradient algorithm starts with an initial guess, $\mathbf{z}$, which typically is set to zero, and we use the diagonal ${\boldsymbol \Lambda}_{\rm S}^{-1}$ matrix as a preconditioner.

The integration of Eqs.\ (\ref{eq:eqofmotion_fc_1_B}) and (\ref{eq:eqofmotion_fc_2_B}) can be performed with the same mixed Verlet scheme as described in Alg.\ \ref{alg: shadowbomd}, but with {\bf n} replaced by {\bf d} and ${\bf q}[{\bf n}]$ by ${\bf p}[{\bf d}]$.
Detailed expressions for the multipole interaction energy terms and forces are given in Appendix \ref{ap: energyAndForce}.

Equation (\ref{eq:eqofmotion_fc_2_B}) is, in principle, of the same form and complexity as the original problem in the standard Born-Oppenheimer formulation given by Eq.\ (\ref{eq:DipoleEquil-fc}). At first glance, it may therefore seem that our shadow formulation offers no advantage. However, this is not the case.

The solution, ${\bf \ddot d}$, to Eq.\ (\ref{eq:eqofmotion_fc_2_B}) enables time-reversible propagation of the extended dynamical dipole moments. Crucially, the propagated dipole moments, ${\bf d}(t)$, do not need to be exact. The extended dynamical variables ${\bf d}(t)$ serve only as an approximate solution to the exact ground state, ${\bf p}_0$, around which we expand our shadow energy function.  As a result, Eq.\ (\ref{eq:eqofmotion_fc_2_B}) can be solved with a loose convergence tolerance, and a good initial guess is also available. This is in contrast to the original problem in Eq.\ (\ref{eq:DipoleEquil-fc}) which must be solved to high precision, requiring tight convergence to prevent non-conservative forces and ensure long-term stability. An example is demonstrated in subsection \ref{Precond}.

\begin{figure}
    \centering
    \includegraphics[width=\columnwidth]{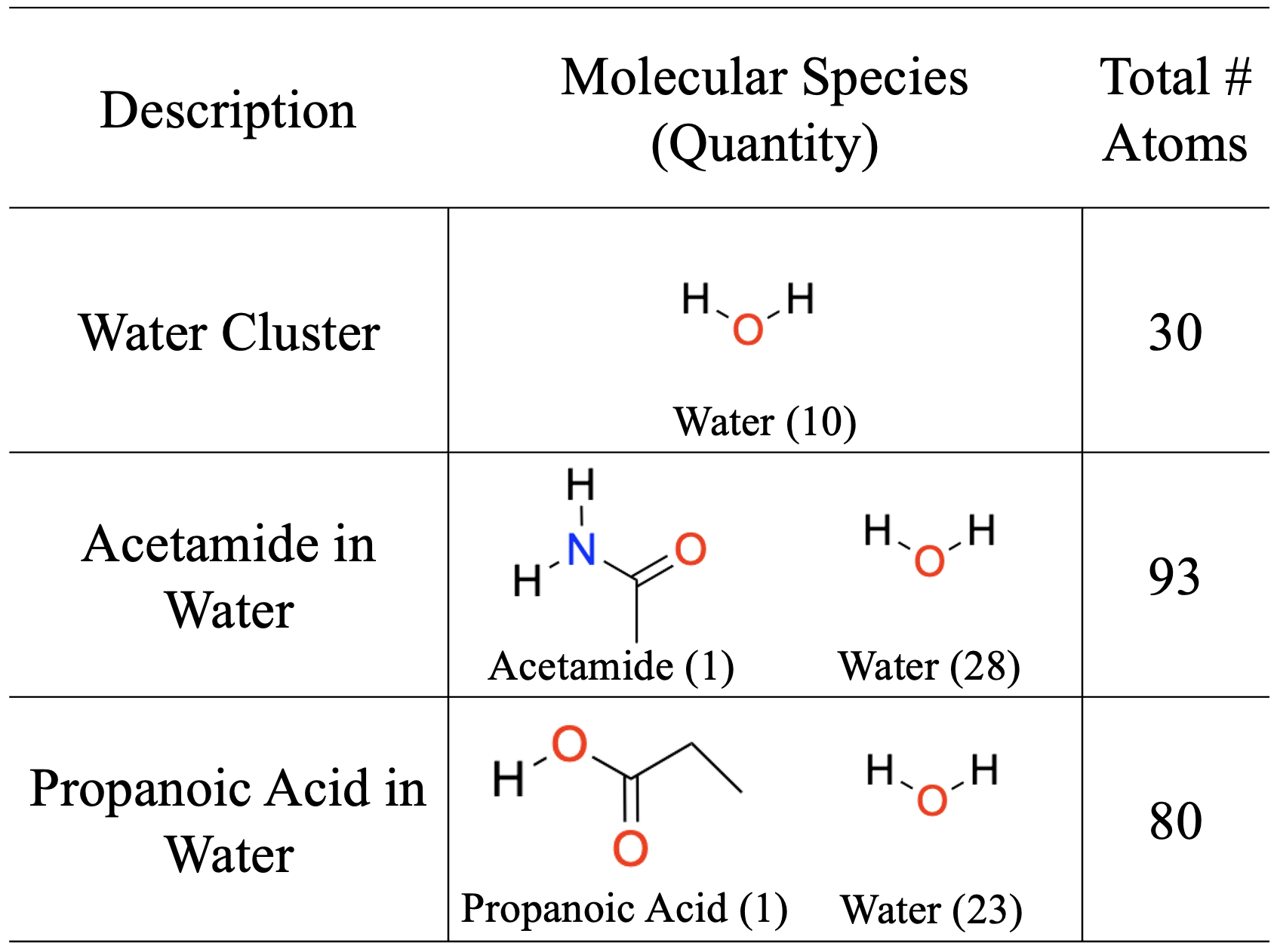}
    \caption{Description of the three tested molecular systems including the total number of atoms and molecules in each system.}
    \label{fig:molec_table}
\end{figure}

\begin{figure*}
    \centering
    \includegraphics[width=\textwidth]{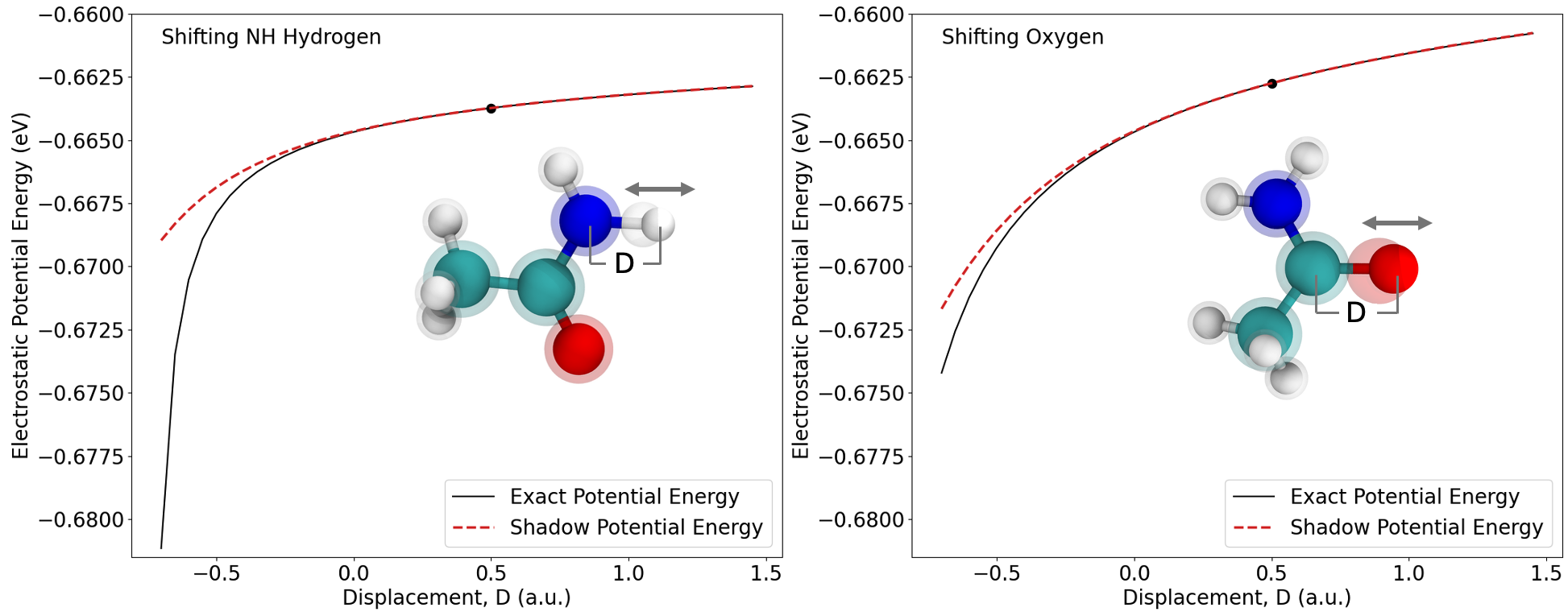}
    \caption{Comparison of electrostatic potential energy values using exact and shadow potentials as a single atom of acetamide in vacuum is displaced (hydrogen in the left panel, oxygen in the right panel).  The shadow energy function was expanded around the exact solution at a displacement $D = 0.5$ a.u. (black dot) from the equilibrium bond distance. Around the expansion point ($D \in [0,1.5]$ a.u.) the exact and shadow potentials agree closely, but begin to diverge as the displacement changes ($D < 0$).}
    \label{fig:aceshifting}
\end{figure*}

\section{Demonstration and Evaluation}

To demonstrate and evaluate the shadow multipole MD schemes, we will use three different test systems, each integrated up to 100 ps of MD simulation time, and compare the results with the `exact' regular Born-Oppenheimer schemes. The three molecular test systems are are described in Fig.\ \ref{fig:molec_table} and consists of: 1) a cluster of waters; 2) acetamide in water; and 3) propanoic acid in water. 

In our flexible multipole models (both with flexible and fixed monopoles), the chemical hardness parameters, ${\bf u}$, and the electronegativities, ${\boldsymbol \chi}$, were taken from the work by Rappe and Goddard \cite{RappeGoddard91} and the atomic polarizabilities, ${\boldsymbol \alpha}$, were taken from the AMOEBA Bio 2018 force field \cite{Ponder10, ZhangRen18}. 
For the charge-independent potential, $V({\bf R})$, we used the charge-independent tight-binding forces from the xTB method \cite{Grimme17, Grimme19xtb}.
All parameters (converted to the same shared units) are chosen as constants, one for each atomic type. They are thus not optimized for our particular test systems and they are not environment dependent. The goal of this work is not to demonstrate performance or high accuracy with respect to first-principles theory or experiments, but to demonstrate stability and to show that our shadow multipole MD is in close agreement with the regular and tightly converged `exact' regular Born-Oppenheimer schemes. Alternative high-performance and AI-driven formulations with environment-dependent parameterizations, for example, using neural networks, the atomic cluster expansion, or kernel ridge-regression, are straightforward, at least in principle \cite{TWKo20, Goff23, LiNiklasson25, Rinaldi25, Venkatraman2025}. 
We first discuss the flexible multipole model presented in Section \ref{FlexMultipole} and then the fixed monopole/flexible dipole model presented in Section \ref{fixedmono} .

\subsection{Flexible Multipole Model}

\subsubsection{Shadow Potential}

To demonstrate the accuracy of the shadow potential in the flexible multipole model relative to the exact regular Born-Oppenheimer potential, we performed single-atom displacement on an isolated acetamide molecule in vacuum. In these tests, we displaced a single atom from its equilibrium position along one dimension. This displacement is denoted as ``D'' in Fig.\ \ref{fig:aceshifting}.

The shadow energy function, ${\cal E}({\bf R},{\bf q},{\bf p},{\bf n},{\bf d})$ in Eq.\ (\ref{eq:shadowMultipole-energy}), was expanded around constant reference values for ${\bf n}$ and ${\bf d}$, set equal to the regular solutions ${\bf q}_0$ and ${\bf p}_0$, which correspond to the fully equilibrated net charges and dipoles at a bond displacement of $D = 0.5$ a.u. At this reference displacement, the optimized electrostatic shadow Born-Oppenheimer potential and the corresponding exact regular and shadow Born-Oppenheimer potentials are therefore equal.

By varying the atomic displacement, $D$, we then compute the optimized shadow potential energy (dashed line) using the same fixed values of ${\bf n}$ and ${\bf d}$, and compared it to the corresponding fully optimized exact Born–Oppenheimer potential (solid line), as shown in Fig.\ \ref{fig:aceshifting}. The left panel shows how the electrostatic potential energy changes as a function of the displacement of a hydrogen atom, while the right panel shows the same for an oxygen atom. In both cases, the shadow potential energy closely follows the exact Born–Oppenheimer potential energy in a fairly wide region ($D \in [0,1.5]$ a.u.).

The gradual divergence between the shadow and exact potential energies as the displacement changes from $D = 0.5$ a.u. demonstrates both the local accuracy of the shadow potential approximation and the necessity for periodic updates of ${\bf n}$ and ${\bf d}$ as the atomic configurations evolve. Notably, the expansion points, ${\bf n}$ and ${\bf d}$, do not need to coincide exactly with the exact ground-state regular values, ${\bf q}_0$ and ${\bf p}_0$. As long as ${\bf n}$ and ${\bf d}$ are not too far away from these ground-state values, the shadow potential provides a highly accurate representation of the exact reference potential.

\begin{figure}
    \centering
    \includegraphics[width=\columnwidth]{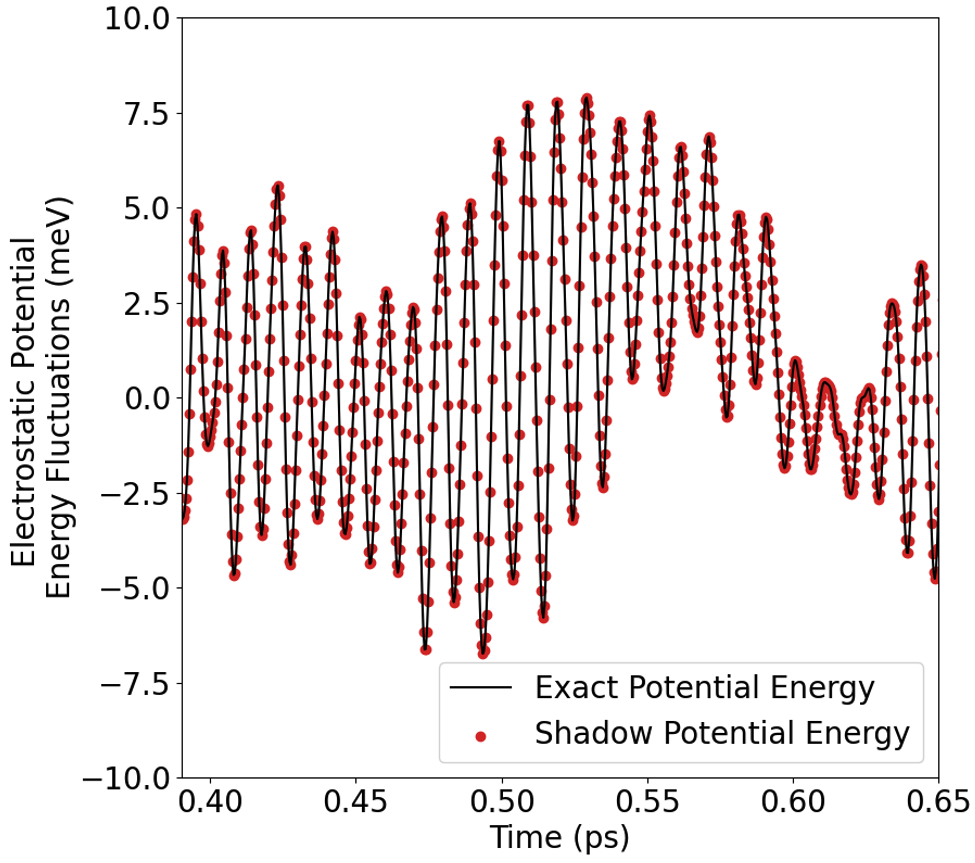}
    \caption{The electrostatic potential energy fluctuations (the Born-Oppenheimer potential without $V({\bf R})$) for acetamide in water along a shadow MD trajectory generated with the flexible multipole model. The figure compares the exact regular electrostatic potential (black line) with the shadow potential (red dots). Simulations were performed using a time step of $\delta t = 0.4$ fs.}
    \label{fig:acePotenEFluc}
\end{figure}

In our extended Lagrangian shadow MD, ${\bf n}(t) \equiv {\bf n}$ and ${\bf d}(t) \equiv {\bf d}$ appear as dynamical time-dependent variables propagated through extended harmonic oscillators that follow the exact ground state solutions. This Car-Parrinello-like approach provides a shadow potential energy curve that closely matches the exact Born-Oppenheimer potential energy, but at significantly lower cost. Fig. \ref{fig:acePotenEFluc} shows the electrostatic potential energy fluctuations across a brief section of simulation time for a MD simulation of the acetamide in water system using both the exact regular electrostatic potential and the shadow potential. The two potential energy curves match nearly perfectly, demonstrating the high accuracy of the shadow potential generated with propagated ${\bf n}(t)$ and ${\bf d}(t)$ values.

\begin{figure*}
    \centering
    \includegraphics[width=\textwidth]{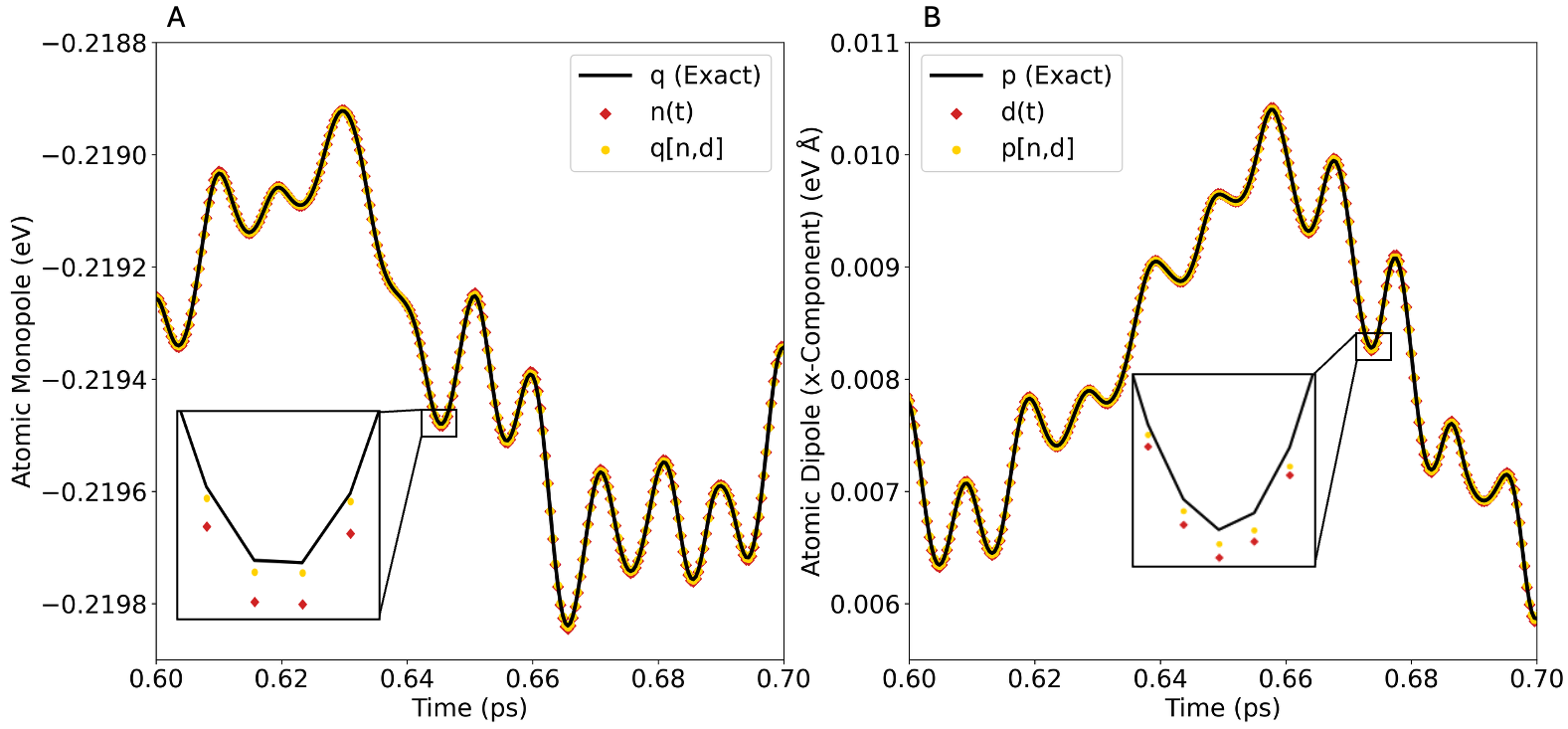}
    \caption{(A) Comparison of the exact charge, $q$, with the approximate charge, $q[{\bf n}, {\bf d}]$, and the dynamical charge variable value, $n(t)$, for the oxygen atom in the acetamide molecule within the acetamide in water system. All three curves are superimposed across simulation time. The call-out shows a zoomed in view of a subsection of the plot that highlights the slight offsets among the curves. (B) Comparison of the exact x-direction dipole component, $p_{x}$, with the equilibrated shadow dipole component, $p_x[{\bf n}, {\bf d}]$, and the dynamical dipole component variable value, $d_{x}(t)$, for the oxygen atom in the acetamide molecule within the acetamide in water system. All three curves are superimposed across simulation time. The call-out shows a zoomed in view of a subsection of the plot that highlights the slight offsets among the curves.}
    \label{fig:monodicomp-meq}
\end{figure*}

\begin{figure}
    \centering
    \includegraphics[width=\columnwidth]{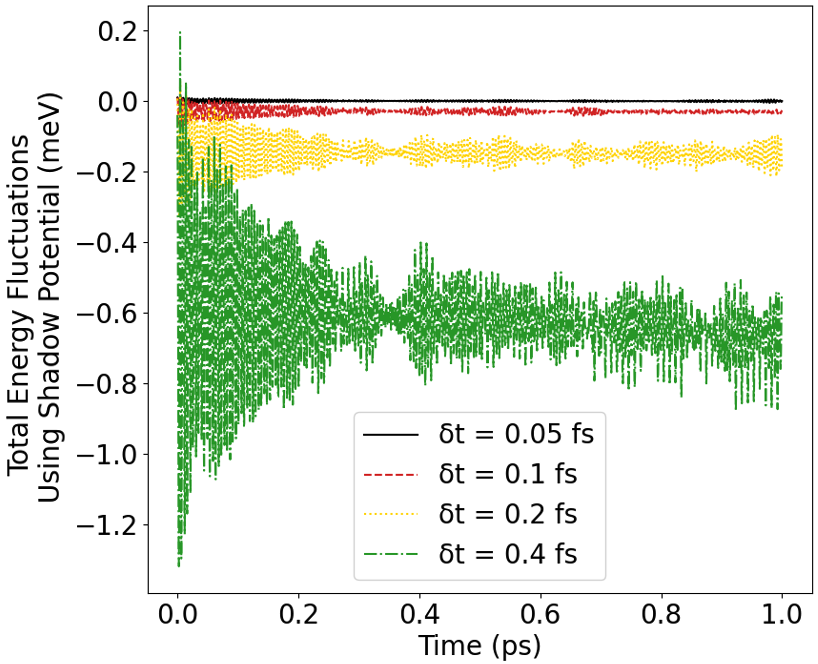}
    \caption{Fluctuations in total energy (kinetic energy + potential energy) along a shadow MD trajectory across 1 ps of simulation time for the acetamide in water system using four different integration time steps (in units of femtoseconds). Fluctuations are centered around the average of the total energies sampled using $\delta t = 0.05$ fs. All four trajectories were simulated using the flexible multipole model and used a maximum rank 4 update. Fluctuations scale with $\delta t^2$, as expected.}
    \label{fig:dtScaling-flexqd}
\end{figure}

\subsubsection{Dynamical Multipoles}

In Fig.\ \ref{fig:monodicomp-meq} we track the propagated monopole and dipole values, ${\bf n}(t)$ and ${\bf d}(t)$, as well as their optimized shadow values, ${\bf q}[{\bf n}, {\bf d}]$ and ${\bf p}[{\bf n},{\bf d}]$, and we find that they closely follow the exact regular Born-Oppenheimer values, ${\bf q}$ and ${\bf p}$, across the simulation time. Fig. \ref{fig:monodicomp-meq} shows the monopole values (left panel) for the oxygen atom in the acetamide molecule and the x-component of the dipole (right panel) for the same atom for the 93 atom test system (acetamide in water) across a brief section of simulation time. The cutouts in both plots show magnified sections, highlighting how close the dynamical variables, ${\bf n}(t)$, and the optimized values ${\bf q}[{\bf n},{\bf d}]$, are to the exact reference values, ${\bf q}$, in the left panel, and how close the dynamical variables, ${\bf d}(t)$, and optimized values ${\bf p}[{\bf n}, {\bf d}]$ are to the exact reference values, ${\bf p}$, in the right panel. The dynamical variables, ${\bf n}(t)$ and ${\bf d}(t)$, show a slightly larger deviation from ${\bf q}$ and ${\bf p}$, respectively, compared to the optimized values. This demonstrates the accuracy of the model in propagating ${\bf n}(t)$ and ${\bf d}(t)$, keeping them close to the exact relaxed ground state across the simulation time. In this way the shadow Born-Oppenheimer potential will closely follow the corresponding exact Born-Oppenheimer potential.

\subsubsection{Energy Stability}

A sensitive gauge of the accuracy of our MD simulations is the scaling of the amplitude of the local fluctuations in the total energy as a function of the integration time step and the long-term stability of the total energy. For the Verlet integration scheme the amplitude should scale as $\propto\delta t^2$ ~\cite{Niklasson21b}.
The MD simulations with our chosen test systems (displayed in Fig.\ \ref{fig:molec_table}) all show this expected $\delta t^2$ scaling of the total energy fluctuations. The magnitude of the total energy fluctuations approximately quadruples when the the integration time step $\delta t$ doubles in size. Fig.\ \ref{fig:dtScaling-flexqd} shows the total energy fluctuations (kinetic energy + potential energy) for four different 1 ps simulations of the acetamide in water system using four different sizes of the integration time steps, $\delta t$ (in units of femtoseconds). This comparison demonstrates the expected $\delta t^2$-scaling behavior.  

\begin{figure}
    \centering
    \includegraphics[width=\columnwidth]{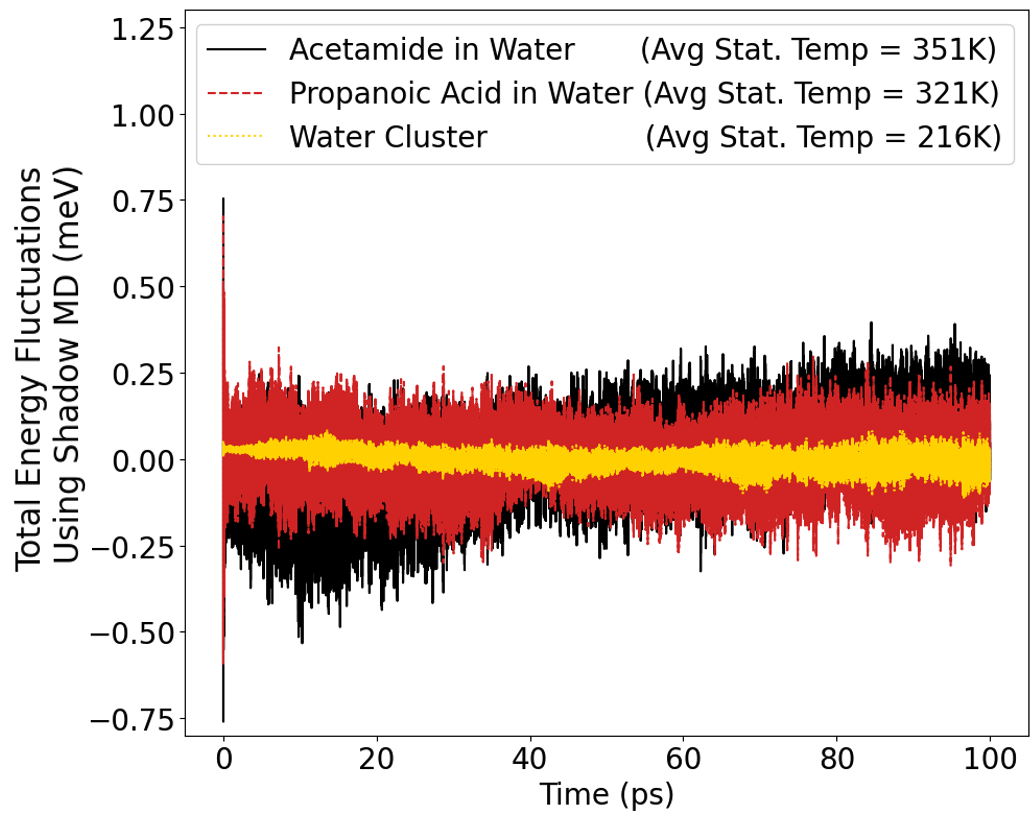}
    \caption{Fluctuations in total energy (kinetic energy + potential) over 100 ps shadow MD simulations of our three stest ystems using an integration time step of $\delta t = 0.4$ fs with the flexible multipole model. Average statistical temperatures across each simulation are given in the legend. No significant long-term drift in the total energy fluctuations is visible.}
    \label{fig:dt4100ps}
\end{figure}

To evaluate the important long-term energy stability we performed longer runs over 100 ps of shadow MD simulation time for our test systems with average statistical temperatures of around 200-350 K using an integration time step of $\delta t = 0.4$ fs. For these simulations, we used a simple diagonal Jacobi preconditioner, $\mathbf{\Lambda}_{\rm S}^{-1}$, for the adaptive kernel approximation with a maximum of rank-4 (in almost all time steps, only a rank-2 or rank-3 update was used).  The fluctuations in the total energy (kinetic + potential) are shown in Fig.\ \ref{fig:dt4100ps}. All simulations demonstrate excellent long-term stability. No significant systematic long-term energy drift is visible and the behavior is the same as seen for previous monopole-only shadow MD simulations \cite{Goff23, LiNiklasson25}. This long-term stability test is an important gauge on the accuracy of the shadow model and its implementation. Notice that if we have problems with the long-term energy conservation, we cannot use a thermostat to avoid the unphysical behavior. If the underlying microcanoncial (NVE) ensemble has an energy drift, the corresponding canonical (NVT) ensemble will have unphysical thermal fluctuations, which can lead to a number of problems \cite{EMartinez15}. An energy-conserving dynamics is thus not only of importance for microcanical studies, for example of exothermic reactions, but is a general quality measure for a wide range of properties that can be derived from MD simulations.

\subsubsection{IR Spectra}

\begin{figure}
    \centering
    \includegraphics[width=\columnwidth]{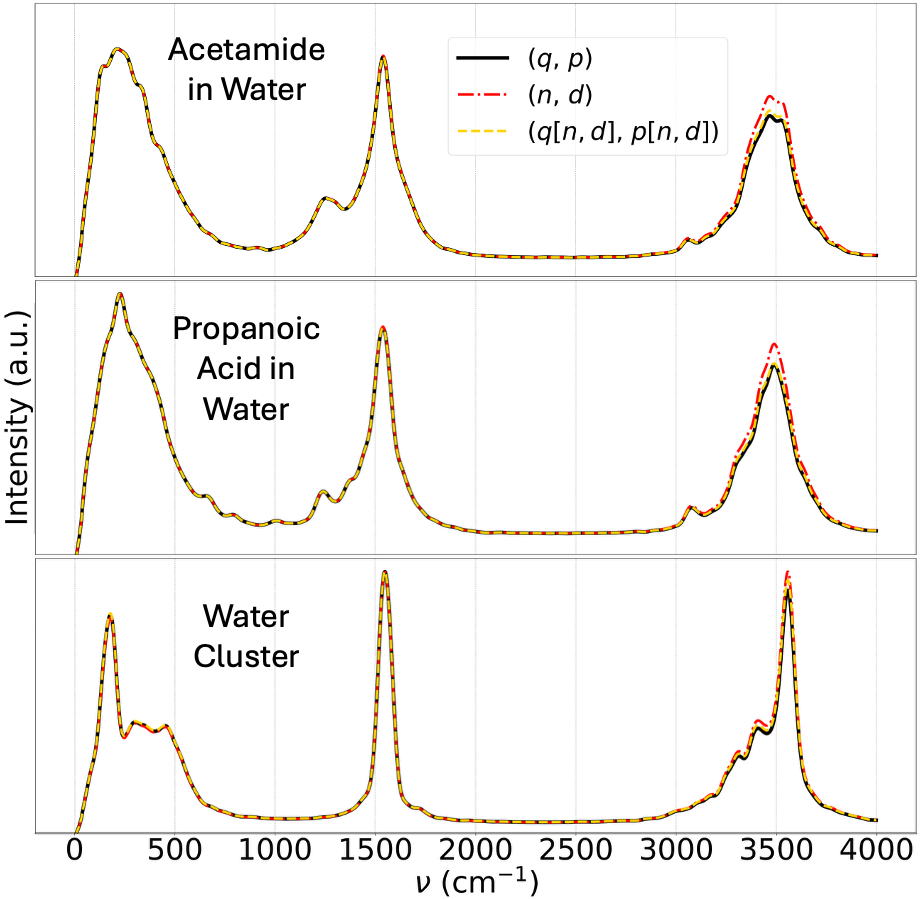}
    \caption{Calculated IR spectra from the net dipole auto-correlation function sampled from shadow MD trajectories with the flexible multipole model for our three three systems (See Fig.\ \ref{fig:molec_table}). The net molecular dipole values determined from the atomic monopoles and dipoles were calculated either exactly (solid black line), using the exact regular Born-Oppenheimer monopoles, ${\bf q}$, and atomic dipoles, ${\bf p}$, (dot-dashed red line) or using either the propagated dynamical parameters, ${\bf n}(t)$ and ${\bf d}(t)$, or the equilibrated monopoles and dipoles for the shadow potential, ${\bf q}[{\bf n},{\bf d}]$ and ${\bf p}[{\bf n},{\bf d}]$ (dashed yellow line). All three lines are virtually on top of each other. Only at high frequencies ($>3000~ \rm{cm}^{-1}$) is it possible to see small differences in the amplitude, but no frequency shifts in the peak positions are visible.}
    \label{fig:irspectra}
\end{figure}

By sampling the net dipole autocorrelation function given by the atomic monopole charges and dipoles over 100 ps of MD simulation time, we can calculate the infrared (IR) spectra for the three different test systems \cite{SMukamel95, CRBaiz20, MGaigeot03, MGaigeot07, LiNiklasson25,RStanton25}. 
The net molecular dipole values were calculated either exactly, using the fully optimized exact regular Born-Oppenheimer monopoles, ${\bf q}$, and atomic dipoles, ${\bf p}$, or using either the propagated dynamical parameters, ${\bf n}(t)$ and ${\bf d}(t)$, or the relaxed multipoles for the shadow potential, ${\bf q}[{\bf n},{\bf d}]$ and ${\bf p}[{\bf n},{\bf d}]$. The samples MD trajectories were determined by the shadow Born-Oppenheimer potential. The different calculated spectra for our three test systems are shown in Fig.\ \ref{fig:irspectra}. For each test system the three different IR spectra are virtually on top of each other. However, while the frequency positioning of the peaks remain the same, we find a small gradual deviation in their amplitude between the three different molecular dipole approximations at high frequencies ($\nu > 3000$ cm$^{-1}$). At higher frequencies, the effective time sampling is reduced. The gradual deviation at high frequencies may therefore be understood from an expected increase in the difference in the propagated monopoles and dipoles compared to the `exact' regular reference values, which scales as $\delta t^2~$ \cite{Niklasson21}, i.e.\ in the same way as the total energy fluctuations shown in Fig.\ \ref{fig:dtScaling-flexqd}.

\subsubsection{Local Dipole Behavior}

\begin{figure}
    \centering
    \includegraphics[width=\columnwidth]{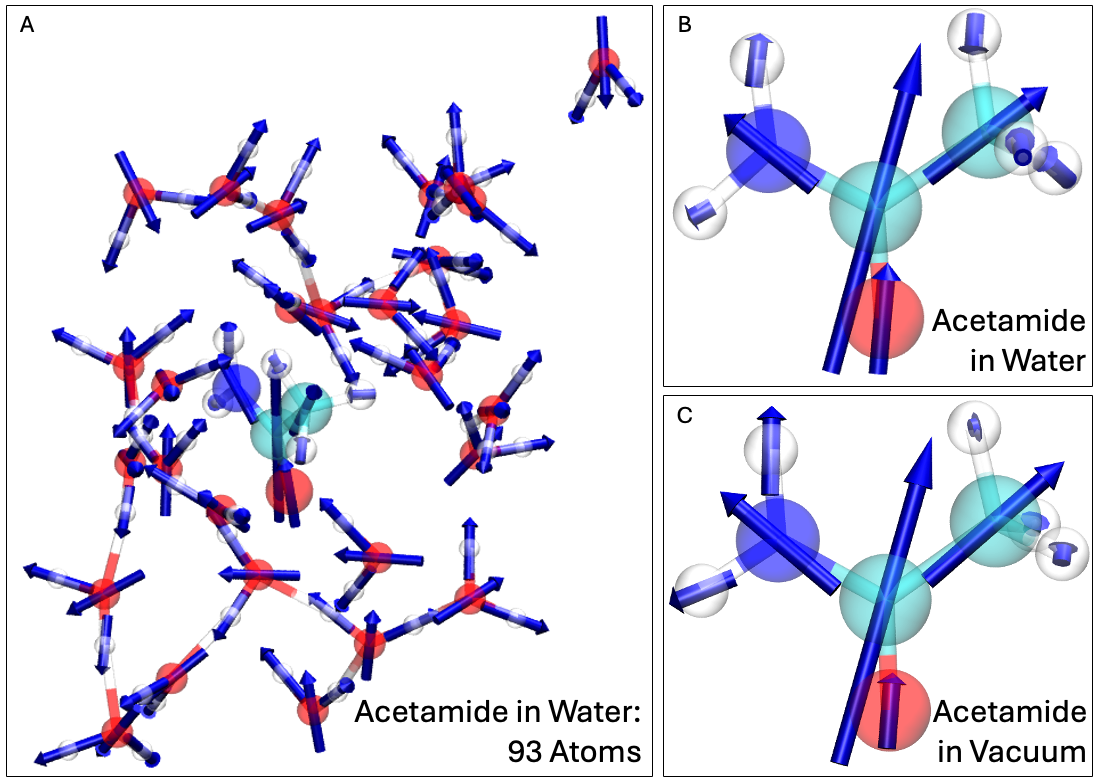}
    \caption{Representation of the atomic dipoles for acetamide with the flexible multipole model. Hydrogen atoms are white, carbon atoms are teal, oxygen atoms are red, and nitrogen atoms are blue. Left panel A) shows the molecular representation of the acetamide in water system including on-site dipoles (blue arrows). Upper right panel B) shows the acetamide molecule with on-site dipoles in condensed phase (in water). Lower right panel C) shows the corresponding acetamide molecule with on-site dipoles in vacuum. Changes in dipole magnitude and direction between panels B) and C) illustrates how the flexible dipoles are affected by the solvent.}
    \label{fig:acevmddipoles}
\end{figure}

 It is interesting to see how the atomic dipoles are affected by their local environments. Using the visualization tool VMD\cite{Humphrey96}, we can plot the dipole vectors on each atom for a given test system configuration. Fig.\ \ref{fig:acevmddipoles} panel (A) shows the acetamide in water system with atomic dipoles. Panels (B) and (C) of Fig.\ \ref{fig:acevmddipoles} show the on-site dipoles for the acetamide molecule, either in water (panel (B) upper right) or in vacuum (panel (C) lower right). This highlights the subtle but distinct differences observed between the two phases and how the atomic dipoles are affected by the field generated by nearby atoms. The strength of this field is dominated by the nearby atomic charges that are driven by hte interatomic charge transfer caused by differences in the electronegativities. An atom that is surrounded by atoms that have large differences in electronegativity (i.e., the central acetamide carbon positioned among oxygen, nitrogen, and the methyl carbon) will therefore have a larger atomic dipole magnitude than an atom of the same type positioned among atoms that have small to no differences in electronegativity (i.e., the acetamide methyl carbon which is mainly surrounded by hydrogen atoms). This effect dominates over differences in the atomic polarizabilities and depends strongly on the geometry, which changes as the molecules move and interact, causing dipole fluctuations. The atomic dipoles add important flexibility and fidelity to monopole-only models, since the atomic dipoles capture environment-dependent behavior. This is also of significance for the long-range electrostatic interactions driven by the molecular dipole moments.

 Heuristic analysis indicates that variations in atomic electronegativities, ${\boldsymbol \chi}$, exert a stronger influence on monopole and dipole magnitudes than variations in the atomic polarizabilities, ${\boldsymbol \alpha}$. Thus, monopole modifications substantially affect dipoles, whereas dipole variations have comparatively minor effects on the monopoles. This result is particularly relevant for our fixed-monopole/flexible-dipole model,  where we keep the monopoles fixed and only allow the atomic dipoles to be flexible.

\subsection{Fixed Monopole/Flexible Dipole Model}

The fixed monopole/flexible dipole model presented in Section \ref{fixedmono} is commonly used in polarizable force fields for biomolecular simulations \cite{Liu98, Dang00, Halgren01, Koneshan01, Dang02, Kaminski02, Dang03, GreshPiquemal07, Wick05, Wick07, WickDang07, Baranyai10, Ponder10, Wang11-1, Wang11-2, WangLuo12, Chang14, Li14, JingRen19} and an efficient and stable shadow extended Lagrangian Born-Oppenheimer MD formulation of this approach is therefore of interest. In our simulations below, the fixed monopole charges where chosen at a snapshot (after 1 ps) of an MD simulation with acetamide in water using the flexible multipole model. Other choices can be made, for example, where the fixed monopoles are predicted using AI/ML with a parameterization that depends on the local atomic environments for each atom. 

\subsubsection{Shadow Potential}

\begin{figure}
    \centering
    \includegraphics[width=\columnwidth]{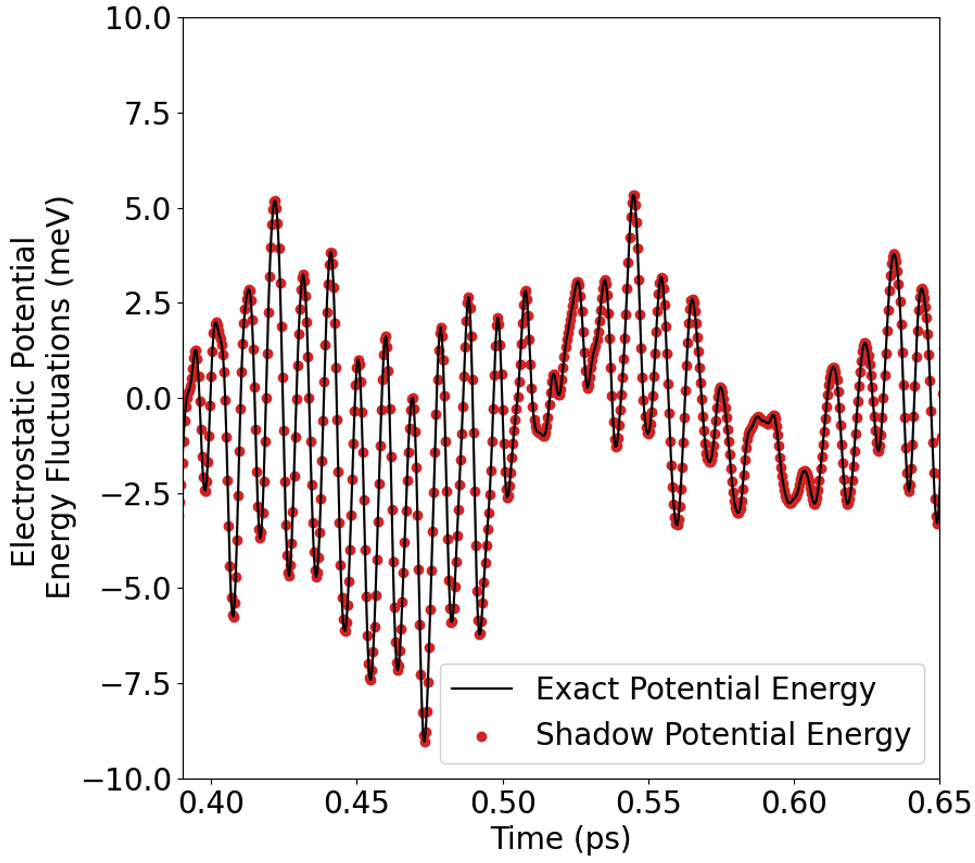}
    \caption{The electrostatic shadow potential energy (red dots) for acetamide in water along a shadow MD trajectory in comparison to the `exact' reference values for the regular Born-Oppenheimer electrostatic energy (black line). The MD trajectory was generated using the shadow fixed monopole/flexible dipole model with a time step of $\delta t = 0.4$ fs.}
    \label{fig:acePotenEFluc-fc}
\end{figure}

Fig. \ref{fig:acePotenEFluc-fc} shows the electrostatic potential energy surface along a shadow MD trajectory of acetamide in water using the fixed monopole/flexible dipole model. 
Here we compare the fluctuations in the shadow electrostatic potential (red dots) with 
the exact reference values for the regular Born-Oppenheimer electrostatic energy (black line). In the same way as for the flexible multipole model, we find that the shadow potential energy closely follows the exact regular potential energy, as expected.

\subsubsection{Dynamical Dipoles}

\begin{figure}
    \centering
    \includegraphics[width=\columnwidth]{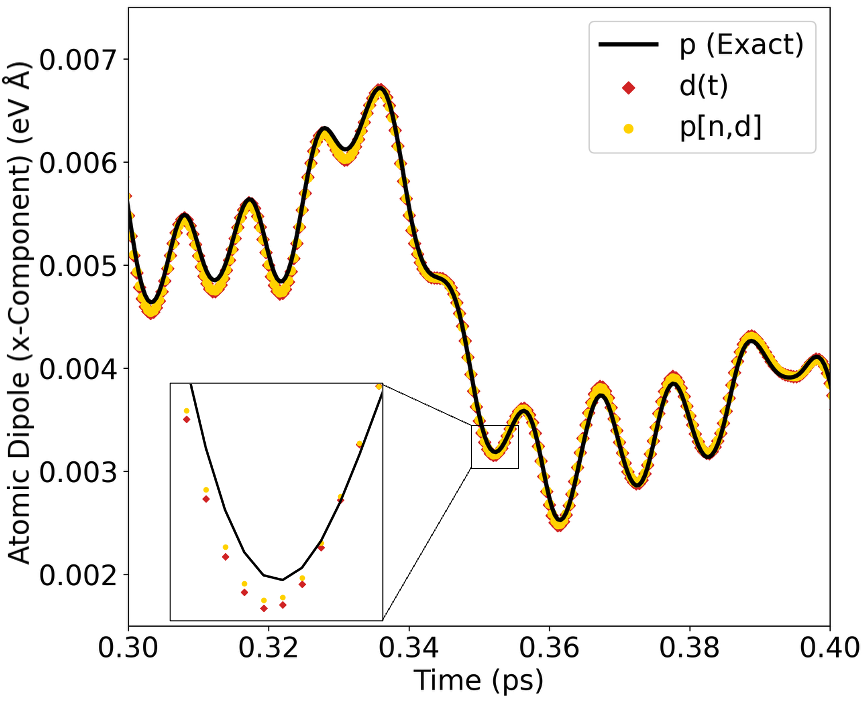}
    \caption{Comparison of ${\bf d}(t)$, and $\bf p[\bf n, d]$ for a shadow MD simulations with the exact reference values of $\bf p$ using the fixed monopole/flexible dipole model. Only the x-component of the atomic dipoles for the oxygen atom in the acetamide molecule within the acetamide in water system is shown along its trajectory. The zoomed in detailed view of the curves highlights the differences among the exact and approximate dipole values. Simulations were performed using the fixed monopole/flexible dipole model with a time step of $\delta t = 0.4$ fs.}
    \label{fig:fcdipoleComp}
\end{figure}

Fig. \ref{fig:fcdipoleComp} shows the dynamical dipole, ${\bf d}(t)$, in the x-direction of the oxygen atom in the acetamide molecule within the acetamide in water system for the fixed monopole/flexible dipole model in comparison to the corresponding optimized shadow potential values, ${\bf p}[{\bf d}]$, and the exact reference values, ${\bf p}$, from a direct regular Born-Oppenheimer optimization along the same shadow MD trajectory. The results are very similar to the fully flexible multipole model in Fig.\ \ref{fig:monodicomp-meq}. All three values for the dipoles closely follow each other, demonstrating the accuracy of the shadow MD for the fixed monopole/flexible dipole model. 

\subsubsection{Energy Stability}

\begin{figure}
    \centering
    \includegraphics[width=\columnwidth]{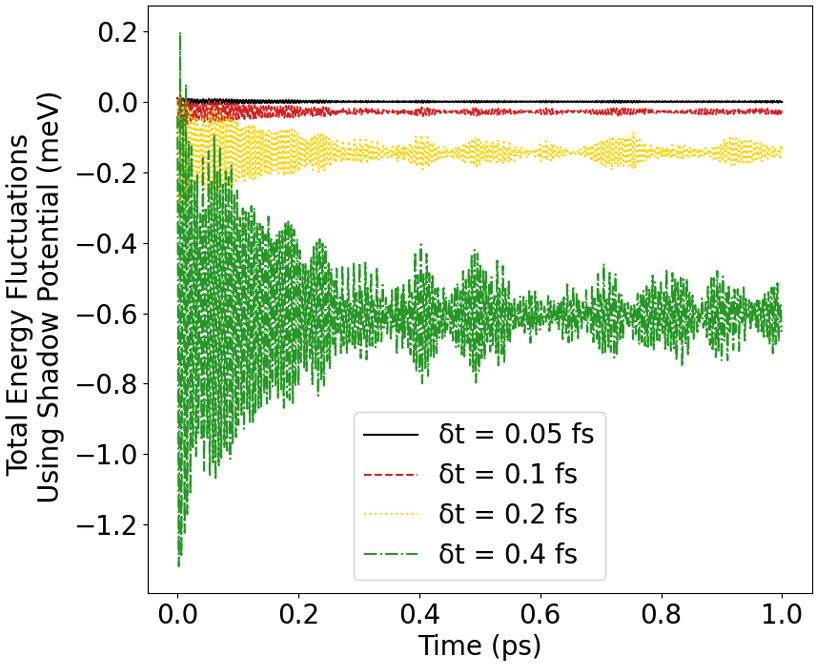}
    \caption{Fluctuations in the total energy (kinetic energy + potential energy) for the fixed monopole/flexible dipole model using the shadow potential across 1 ps of simulation time for the acetamide in water system using four different integration time steps (in units of femtoseconds). Fluctuations are centered around the average of the total energies sampled using $\delta t = 0.05$ fs. All four trajectories were simulated using the fixed monopole/flexible dipole model (Sec.\ \ref{fixedmono}) with Alg.\ \ref{alg: lrupdate-fc} which uses the conjugate gradient algorithm to approximate ${\bf \ddot d}$ with $k_{\rm max} = 4$ and a diagonal Jacobian preconditioner, $\boldsymbol{\Lambda}_{\rm S}^{-1}$. Fluctuations in the amplitude of the total energy scale approximately with $\delta t^2$, as expected.}
    \label{fig:dtscalingfc}
\end{figure}

In Fig.\ \ref{fig:dtscalingfc} we show the total energy fluctuations (kinetic energy + potential energy) for the shadow MD simulations using the fixed monopole/flexible dipole model across 1 ps of simulation time for the acetamide in water system with four different integration time steps (in units of femtoseconds). 
These simulations show very similar behavior and scaling to the flexible multipole model (Fig.\ \ref{fig:dtScaling-flexqd}), e.g.\ the $\delta t^2$ scaling of the amplitude of the total energy fluctuations. This is the same behavior as for monopole-only shadow MD as well \cite{Niklasson21, Niklasson21b, Goff23}.

\begin{figure}
    \centering
    \includegraphics[width=\columnwidth]{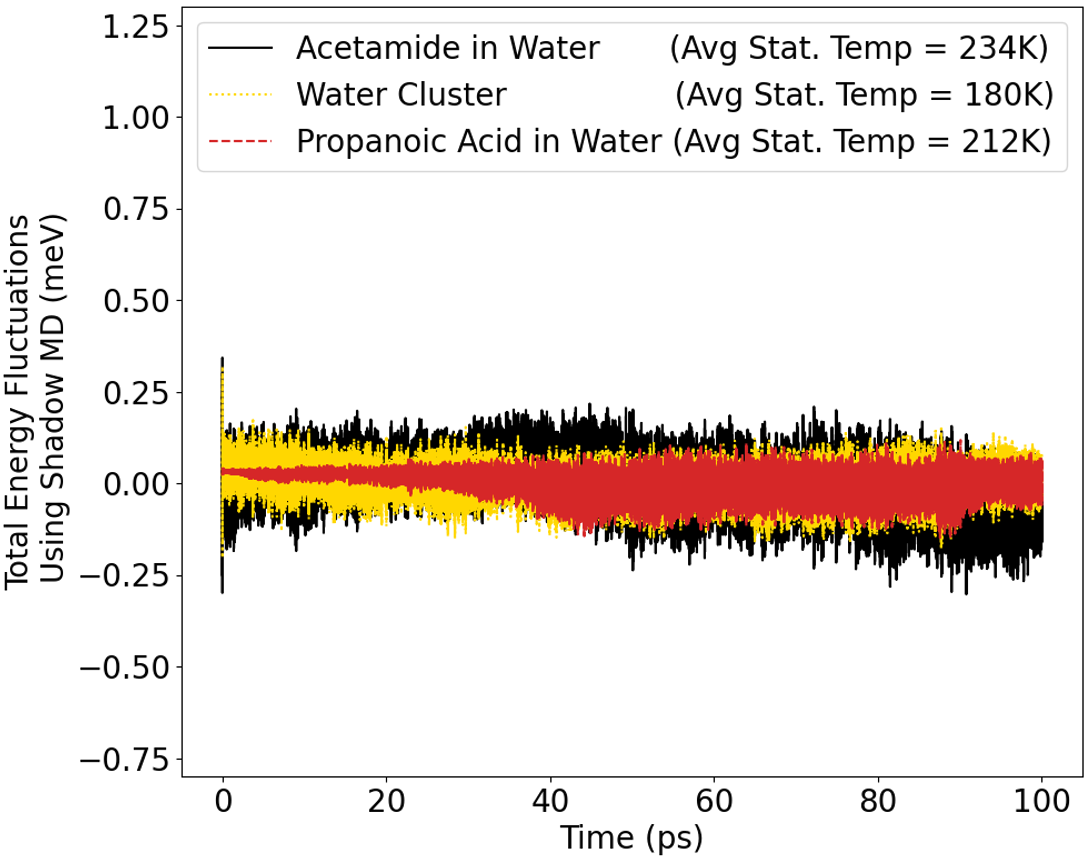}
    \caption{Fluctuations in total energy (kinetic energy + potential) for 100 ps shadow MD simulation of acetamide in water using an integration time step of $\delta t = 0.4$ fs with the fixed monopole/flexible dipole model. Average statistical temperatures across each simulation are given in the legend. The simulations were performed using the fixed monopole/flexible dipole model (Sec.\ \ref{fixedmono}) with Alg.\ \ref{alg: lrupdate-fc} which uses the conjugate gradient algorithm to approximate ${\bf \ddot d}$ with $k_{\rm max} = 4$ and a diagonal Jacobian preconditioner, $\boldsymbol{\Lambda}_{\rm S}^{-1}$.}
    \label{fig:dt4100ps-fc}
\end{figure}

The long-term behavior of the total energy (kinetic + potential) for our three test systems using the fixed monopole/flexible dipole shadow MD is demonstrated in Fig.\ \ref{fig:dt4100ps-fc}. All systems were simulated using Alg.\ \ref{alg: lrupdate-fc} with $k_{\rm max} = 4$ and the diagonal Jacobian preconditioner, $\boldsymbol{\Lambda}_{\rm S}^{-1}$. The three simulations demonstrate the long-term stability of shadow MD with no visible systematic drift in the total energy over the 100 ps of simulation time.

\subsection{Preconditioning}\label{Precond}
\begin{figure}
    \centering
    \includegraphics[width=\columnwidth]{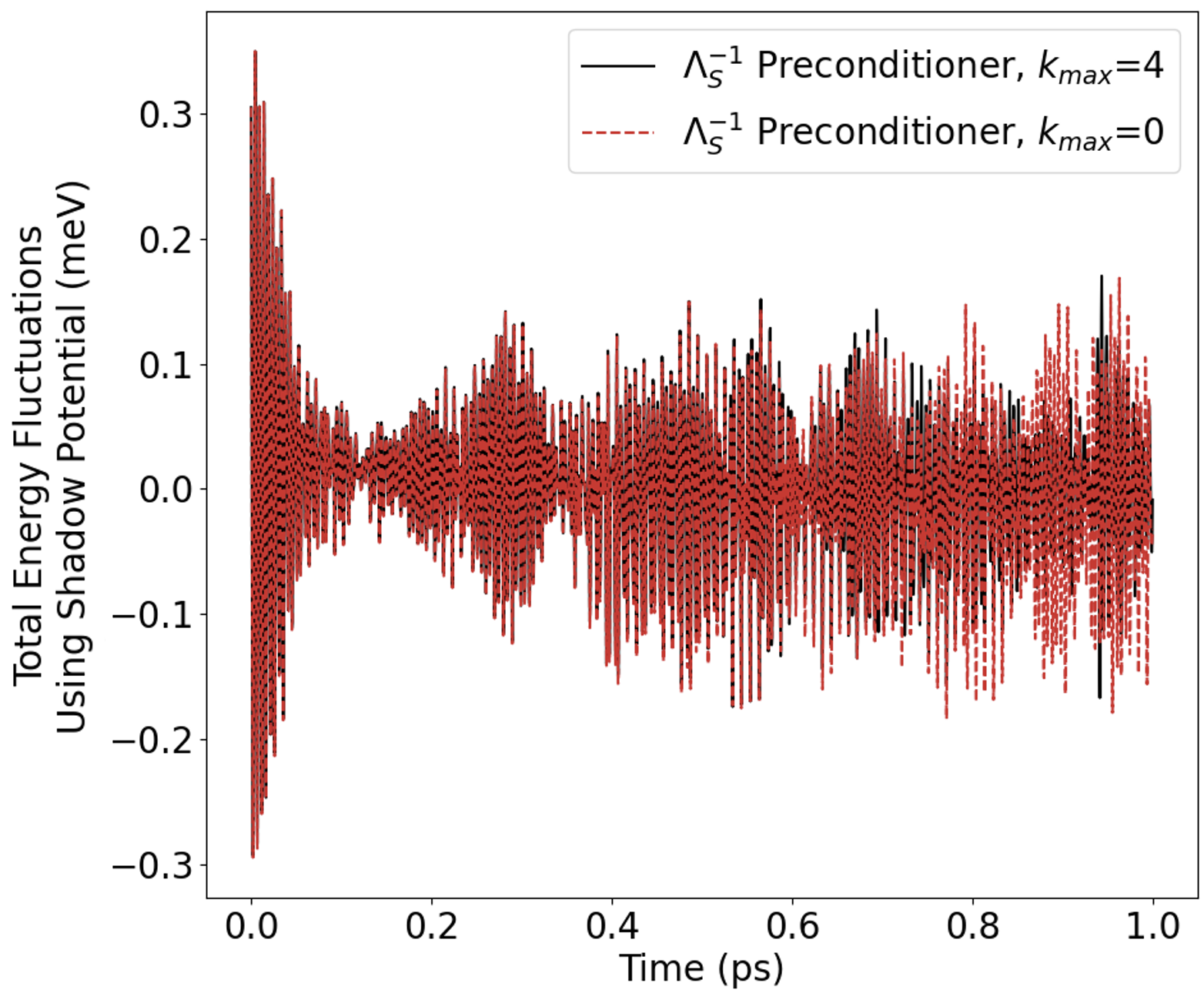}
    \caption{ The total energy fluctuations (kinetic + potential) across 1 ps of shadow MD simulations using the fixed monopole/flexible dipole model for the acetamide in water system with a timestep of $\delta t = 0.4$ fs. The shadow simulation (red dashed line) used only a diagonal preconditioner, $\boldsymbol{\Lambda}_{\rm s}^{-1}$, without any iterative conjugate gradient updates in the calculation of ${\bf \ddot d}$. The energy fluctuations are very similar to the more tightly converged reference calculation (black solid line), using a diagonal preconditioner with an adaptive maximum rank set to $k_{max}=4$ in the conjugate gradient algorithm in Alg.\ \ref{alg: lrupdate-fc}, though most updates used only 2-3 ranks.} 
    \label{fig:fc-precondComp}
\end{figure}

To further test the stability and computational efficiency of the shadow MD, we performed a simulation of the acetamide in water system using the fixed monopole/flexible dipole model using only a diagonal preconditioner for the propagation of ${\bf \ddot d}$, without any rank updates (or conjugated gradient steps), in comparison to a simulation using the conjugate gradient scheme in Alg.\ \ref{alg: lrupdate-fc} with up to rank-4 updates. Fig. \ref{fig:fc-precondComp} shows that both of the 1 ps simulations are closely aligned and stable, demonstrating that the diagonal ${\boldsymbol \Lambda}_{\rm S}^{-1}$ preconditioner alone can be sufficient to ensure high accuracy and stability without any iterative updates. This provides a significant speed up, because only a single Couloumb potential construction from the fixed monopoles and dipoles in necessary in each MD time step, which accelerates the calculations. 

\section{Summary and Conclusions}

We have developed shadow MD for flexible multipole models and fixed monopole/flexible dipole models within the framework of extended Lagrangian Born-Oppenheimer MD. In the Appendix we present detailed expressions for the shadow energy functions, potentials, and force terms, explicitly incorporating monopole-monopole, dipole-monopole, and dipole-dipole interaction terms. Additional information about our implementation is given in the Supplementary Information. In our formulations, the charge or multipole degrees of freedom are included as extended dynamical variables alongside the propagation of the nuclear coordinates and velocities. We demonstrate that introducing the additional dipole degrees of freedom preserves the stability and accuracy previously seen in monopole-only shadow molecular dynamics simulations. Our extended Lagrangian shadow MD provide a framework for stable, computationally efficient, and versatile molecular dynamics simulations involving long-range interactions among flexible multipoles. 

These developments are particularly relevant in the context of modern AI and ML techniques, which can be used to generate physics-informed and data-driven interatomic potentials for MD simulations. Such AI/ML-models aim to provide transferable, high-accuracy representations of interatomic interactions that are applicable across diverse sets of molecular systems, and therefore require accurate treatment of long-range charge interactions. Our shadow MD for multipoles is well-adapted to take advantage of these AI/ML methods.
However, the charge-independent potential, ${\bf V}({\bf R})$, along with the electronegativities, ${\boldsymbol{\chi}}$, chemical hardness, ${\bf u}$, and atomic polarizabilities, ${\boldsymbol \alpha}$, used in this work, were not AI/ML-optimized for our specific test systems. They were either approximated using charge-independent potentials from xTB or selected from tabulated values available in the literature.
To enable high-performance, high-fidelity MD simulations, we could instead employ modern AI/ML techniques to parameterize and predict the local, environment-dependent atomic properties of ${\boldsymbol \chi}$, ${\bf u}$, and ${\boldsymbol \alpha}$ as well as the charge-independent potential, $V({\bf R})$ \cite{SGoedecker15,TWKo20,Drautz2020, Rinaldi25, Goff23,LiNiklasson25}.

A key focus of future work is to combine our models with such AI/ML-based parameterizations trained on extensive first-principles datasets. The shadow multipole MD framework presented here is compatible with a wide range of AI/ML approaches, which can be applied by us or others, to develop general atomistic models that are applicable in MD simulations across diverse classes of complex materials.

\section{Appendices}

The purpose with the appendices is to provide the detailed expressions necessary for implementing the theory.
We first present the mono and multipole interaction matrices. Thereafter we present energy and force expressions before we describe how the kernel approximation can be constructed. At the end we present a short pseudo code for a generic MD scheme. 

\subsection{Interaction Matrices} \label{ap: interact}

For the position vector of an atom, $i$, we use the notation,
\begin{equation}\label{eq:Rposition}
    \mathbf{R}_i = 
    \begin{bmatrix}
        r_{ix} \\
        r_{iy} \\
        r_{iz}
    \end{bmatrix}.
\end{equation}
The vector between atoms $i$ and $j$ is defined as 
\begin{equation}\label{eq: rij}
    \mathbf{r}_{ij} = \mathbf{R}_i - \mathbf{R}_j,
\end{equation}
where the 2-norm of this vector is the interatomic distance, which is defined as
\begin{equation}\label{eq: norm}
    r_{ij} = \|\mathbf{r}_{ij}\|_2,
\end{equation}
and 
\begin{equation}\label{eq: rij_hat}
    \hat{\mathbf{r}}_{ij} = \dfrac{\mathbf{r}_{ij}}{r_{ij}}
\end{equation}
is a unit vector in the direction of $\mathbf{r}_{ij}$. The $\alpha = [x,y,z]$ components of $\mathbf{r}_{ij}$
are denoted by $ {{r}}^{\alpha}_{ij}$.

Here we give the explicit expressions for the matrix entries of the interaction matrices in Eq.\ (\ref{eq:dipole_energy}) for monopoles and dipoles, i.e.\ ${\bf C}$, ${\bf W}$ and ${\boldsymbol \Lambda}$. In the monopole-monopole matrix, $\mathbf{C}$, the matrix elements are given by
\begin{equation}\label{eq: Cmatfill}
    \begin{aligned}
        &C_{ij} = f(r_{ij}), ~~(i \ne j)\\ 
        &C_{ii} = u_i.
    \end{aligned}
\end{equation}
The monopole-dipole interactions matrices, $\mathbf{W}$ and $\mathbf{W}^{\rm T}$, have the matrix elements
\begin{equation}
    W_{ij} = W_{ji}^{\rm T} = f'(r_{ij}) \dfrac{\mathbf{r}_{ij}}{r_{ij}} = f'(r_{ij})\hat{\mathbf{r}}_{ij}.
\end{equation}
The dipole-dipole matrix, $\mathbf{\Lambda}$, has diagonal $3 \times 3$ atomic onsite blocks, 
\begin{equation}
    \boldsymbol{\Lambda}_{ii} = 
    \begin{bmatrix}
        \alpha_i^{-1} & 0 & 0\\
        0 & \alpha_i^{-1} & 0\\
        0 & 0 & \alpha_i^{-1}
    \end{bmatrix}
\end{equation}
and $3 \times 3$ off-diagonal blocks with matrix entries
    \begin{equation}
        \begin{aligned}
             \Lambda_{ij} = -1 \left( f''(r_{ij}) \dfrac{\mathbf{r}_{ij} \mathbf{r}_{ij}^{\rm T}}{r_{ij}^2} + \dfrac{f'(r_{ij})}{r_{ij}} \left( \mathbf{I} - \dfrac{\mathbf{r}_{ij} \mathbf{r}_{ij}^{\rm T}}{r_{ij}^2} \right) \right)\\
             = -1 \left( f''(r_{ij}) \hat{\mathbf{r}}_{ij} \hat{\mathbf{r}}_{ij}^{\rm T} + \dfrac{f'(r_{ij})}{r_{ij}} \left( \mathbf{I} - \hat{\mathbf{r}}_{ij}\hat{\mathbf{r}}_{ij}^{\rm T} \right) \right).
        \end{aligned}
    \end{equation}  
Here ${\bf I}$ is the $3 \times 3$ identity matrix.
Explicit examples of the interaction matrices for a 3-atom system are given in the Supplementary Information. 

The Coulomb interaction function, $f(r_{ij})$, is determined by the choice of the shape of the two overlapping charge distributions. The functional form and its derivatives used here are derived from the Coulomb energy of overlapping Gaussian charge distributions and are given by
    
\begin{equation}
    f(r_{ij}) = \dfrac{\operatorname{erf}(ar_{ij})}{r_{ij}} \label{eq:f0}
\end{equation}
\begin{equation}
    f'(r_{ij}) = u \dfrac{e^{-(ar_{ij})^2}}{r_{ij}} - \dfrac{\operatorname{erf(ar_{ij})}}{r_{ij}^2} \label{eq:f1}
\end{equation}
\begin{equation}
    \begin{aligned}
        &f''(r_{ij}) = -2ua^2 e^{-(ar_{ij})^2} - 2u \dfrac{e^{-(ar_{ij})^2}}{r_{ij}^2} \\&~~~~~~~~~~~~~~~+ 2 \dfrac{\operatorname{erf}(ar_{ij})}{r_{ij}^3} \label{eq:f2}
    \end{aligned}
\end{equation}
\begin{equation}
    \begin{aligned}
        &f'''(r_{ij}) = 4ua^4r_{ij}e^{-(ar_{ij})^2} + 4ua^2 \dfrac{e^{-(ar_{ij})^2}}{r_{ij}}  \\
    &~~~~~~~~~~~~+6u \dfrac{e^{-(ar_{ij})^2}}{r_{ij}^3} - 6 \dfrac{\operatorname{erf}(ar_{ij})}{r_{ij}^4} \label{eq:f3}
    \end{aligned}
\end{equation}
where
\begin{equation}
     a = \dfrac{\sqrt{\pi}}{2}u
\end{equation}
and with
\begin{equation}
  u = \dfrac{2u_iu_j}{(u_i + u_j)}.
\end{equation}
The function $f(r_{ij})$ decays as $1/r_{ij}$ at large distances and in the onsite limit, as $r_{ij} \rightarrow 0$, it reaches the value of $u$.

\subsection{Energy and Force Expressions}\label{ap: energyAndForce}

\subsubsection{Flexible Monopole Model}

The charge-dependent energy function for the regular Born-Oppenheimer monopole-only model is given by
\begin{equation}
    \begin{aligned}
        E(\mathbf{R,q}) = \sum_{i} q_i \,\chi_i + \frac{1}{2} \sum_{i} q_i^2 \, u_i \\
        + \frac{1}{2} \sum_{ij}^{i{\neq}j} q_i \, C_{ij}\,q_j .
    \end{aligned}
\end{equation}

The corresponding regular Born-Oppenheimer force with components $\alpha = [x, y, z]$ acting on atom $i$ for the flexible monopole-only model is
\begin{equation}
    \begin{aligned}
        &{F}_i^{\alpha}(\mathbf{R}, \mathbf{q}) =  -\frac{\partial U_{\rm BO}({\bf R})}{\partial R_i^\alpha}\\
        &~~~~~~~~~~~~~= -\frac{\partial V({\bf R})}{\partial R_i^\alpha}+ \sum_{ij}^{i \neq j} q_i \left(f'(r_{ij}) \dfrac{r^\alpha_{ij}}{r_{ij}} \right)q_j,
    \end{aligned}
\end{equation}
where $r^\alpha_{ij}$ is the $\alpha$-component of ${\bf r}_{ij}$. Additional force terms appear if ${\boldsymbol \chi}$ and ${\bf u}$ have been parameterized depending on their local environments that depend on ${\bf R}$. Here we simply assume that they are fixed ${\bf R}$-independent constants for each atomic type.

The shadow energy function for the monopole-only charge equilibration model is 
\begin{equation}\label{eq: suppMonoEnergy}
    \begin{aligned}
        \mathcal{E}(\mathbf{R,q,n}) = 
        \sum_{i} q_i[\mathbf{n}] \chi_i + \frac{1}{2} \sum_{i} (q_i[\mathbf{n}])^2 u_i \\+
        \frac{1}{2} \sum_{ij}^{i\neq j} 
        (2q_i[\mathbf{n}] - n_i) C_{ij} n_j,
    \end{aligned}
\end{equation}

~~
~~

\hspace{15pt}

~~

~~

which is an equivalent energy expression to Eq.\ (\ref{eq: shadowMono}), where ${\bf C}_{\rm S}$ is chosen as the diagonal matrix part of the Coulomb matrix with the chemical hardness parameters, $\{u_i\}$.

The corresponding force with components $\alpha = [x, y, z]$ acting on atom $i$ for the flexible monopole-only shadow model is then
\begin{equation}
    \begin{aligned}
        &{\cal F}_i^{\alpha}(\mathbf{R},  \mathbf{n}) =  -\frac{\partial{\cal  U}_{\rm BO}({\bf R},{\bf n})}{\partial R_i^\alpha}\\
        &~~~~~=-\frac{\partial V({\bf R})}{\partial R_i^\alpha}+\sum_{i,j}^ {(i \neq j)} (2q_i[{\bf n}] - n_i) \left(f'(r_{ij}) \dfrac{r^\alpha_{ij}}{r_{ij}} \right)n_j.
    \end{aligned}
\end{equation}
Once again, we here assume the chemical hardness terms, $\{u_i\}$, are ${\bf R}$-independent constants for each atomic type. If these terms are parameterized with respect to their environment we get additional force terms \cite{Goff23,LiNiklasson25}.


\subsubsection{Flexible Multipole Model}

The flexible multipole energy function for the regular Born-Oppenheimer model is defined by 
\begin{widetext}
\begin{equation}\label{eq: boEnergy-dipoles}
    \begin{aligned}
        &E({\bf R}, {\bf q}, {\bf p}) = \sum_i \chi_i q_i + \frac{1}{2}\sum_i  q_i^2 u_i + 
        \frac{1}{2} \sum_{i,j (i \ne j)}q_i f(r_{ij}) q_j \\
        &+ \sum_{i,j (i \ne j)} ({{\bf p}_{i}}^{\rm T} \hat{\mathbf{r}}_{ij}) f'(r_{ij}) q_j + \frac{1}{2} \sum{\bf p}_i^{\rm T}{\bf p}_i\alpha_i^{-1} \\
        &-\frac{1}{2}\sum_{i,j (i \neq j)} {\bf p}_i^{\rm T} \left(f''(r_{ij}) \hat{\mathbf{r}}_{ij}\hat{\mathbf{r}}_{ij}^{\rm T} + \frac{f'(r_{ij})}{r_{ij}}\left(\mathbf{I}-\hat{\mathbf{r}}_{ij} \hat{\mathbf{r}}_{ij}^{\rm T}\right) \right){\bf p}_j
    \end{aligned}
\end{equation}
\end{widetext}
where $\mathbf{p}_i$ and $\mathbf{p}_j$ are the dipole vectors for atoms $i$ and $j$, $\mathbf{I}$ is the $3\times 3$ identity matrix, and all other variables as defined above in Eqs.\ (\ref{eq:Rposition}-\ref{eq: rij_hat}, \ref{eq:f0}-\ref{eq:f2}, \ref{eq: suppMonoEnergy}). This equation is equivalent to Eq. (\ref{eq:dipole_energy})

The regular Born-Oppenheimer force with components ${\alpha = [x,y,z}]$ acting on atom $i$ for the flexible multipole model is then given by
\begin{widetext}
    \begin{equation}
    \begin{aligned}
    &{F}_i^{\alpha}(\mathbf{R}, \mathbf{q}, \mathbf{p}) =  -\frac{\partial U_{\rm BO}({\bf R})}{\partial R_i^\alpha}=-\frac{\partial V({\bf R})}{\partial R_i^\alpha}-\sum_{j (j \neq i)} q_i \left(f'(r_{ij}) \dfrac{r^\alpha_{ij}}{r_{ij}} \right)q_j\\
    &+ \mathbf{p}_i^{\rm T}\left[ \left( \dfrac{f'(r_{ij})}{r_{ij}} \right){\bf e}_\alpha +  \hat{r}^\alpha_{ij} \hat{\mathbf{r}}_{ij}  \left( f''(r_{ij}) - \dfrac{f'(r_{ij})}{r_{ij}} \right) \right]q_j\\
    &- q_i \left[ \left( \dfrac{f'(r_{ij})}{r_{ij}} \right){\bf e}_\alpha^{\rm T}  + \left( f''(r_{ij}) - \dfrac{f'(r_{ij})}{r_{ij}} \right) \hat{r}^\alpha_{ij} \hat{\mathbf{r}}_{ij}^{\rm T} \right] \mathbf{p}_j \\
    &+ \mathbf{p}_i^{\rm T}\left(  r^\alpha_{ij}\left(\dfrac{f'''(r_{ij})}{r_{ij}^3} - \dfrac{3f''(r_{ij})}{r_{ij}^4} + \dfrac{3f'(r_{ij})}{r_{ij}^5} \right) (\mathbf{r}_{ij}\mathbf{r}_{ij}^{\rm T}) 
    +  r^\alpha_{ij}\left( \dfrac{f''(r_{ij})}{r_{ij}^2} - \dfrac{f'(r_{ij})}{r_{ij}^3} \right)\mathbf{I} \right)\mathbf{p}_j \\
    &- {\bf p}_i^{\rm T}\left(\left( \dfrac{f''(r_{ij})}{r_{ij}^2} - \dfrac{f'(r_{ij})}{r_{ij}^3} \right) \left({\bf  r}_{ij}{\bf e}^{\rm T}_\alpha + {\bf e}_\alpha {\bf 
     r}_{ij}^{\rm T} \right)\right){\bf p}_j.
    \end{aligned}
\end{equation}
\end{widetext}
Here $\hat{r}_{ij}^\alpha$ is the $\alpha$-component of $\hat{\mathbf{r}}_{ij}$ and $\mathbf{e}_\alpha$ is a column vector of 
\begin{equation}
    \mathbf{e} = 
    \begin{bmatrix}
        1 & 0 & 0 \\
        0 & 1 & 0 \\
        0 & 0 & 1
    \end{bmatrix}
\end{equation}
such that
\begin{equation}
    {\bf e}_x = 
    \begin{bmatrix}
        1\\0\\0
    \end{bmatrix}, ~{\bf e}_y = 
    \begin{bmatrix}
        0\\1\\0
    \end{bmatrix}, {\bf e}_z = 
    \begin{bmatrix}
        0\\0\\1
    \end{bmatrix}.
\end{equation}

The corresponding shadow energy function for the flexible multipole model is defined as 
\begin{widetext}
\begin{equation}\label{eq: shEnergy-multipoles}
    \begin{aligned}
        &\mathcal E({\bf R}, {\bf q}, {\bf p}, {\bf n}, {\bf d}) = \sum_i \chi_i q_i + \frac{1}{2}\sum_i  q_i^2 u_i + 
        \frac{1}{2} \sum_{i,j (i \ne j)} (2q_i[{\bf n}, {\bf d}] - n_i) f(r_{ij}) n_j \\
        &+ \sum_{i,j (i \ne j)} ((2{{\bf p}[{\bf n}, {\bf d}]_{i}} - {\bf d}_i)^{\rm T} \hat{\mathbf{r}}_{ij}) f'(r_{ij}) n_j + \frac{1}{2} \sum{\bf p}_i^{\rm T}{\bf p}_i\alpha_i^{-1} \\
        &-\frac{1}{2}\sum_{i,j (i \neq j)} (2{\bf p}_i[{\bf n}, {\bf d}] - {\bf d}_i)^{\rm T} \left(f''(r_{ij}) \hat{\mathbf{r}}_{ij}\hat{\mathbf{r}}_{ij}^{\rm T} + \frac{f'(r_{ij})}{r_{ij}}\left(\mathbf{I}-\hat{\mathbf{r}}_{ij} \hat{\mathbf{r}}_{ij}^{\rm T}\right) \right){\bf d}_j.
    \end{aligned}
\end{equation}
\end{widetext}
which is equivalent to Eq.\ (\ref{eq:shadowFlexEnergy}) with a diagonal ${\bf G}_{\rm S}$. The corresponding shadow force terms with components ${\alpha = [x,y,z}]$ acting on atom $i$ for the flexible multipole model are then given by
\begin{widetext}
    \begin{equation} \label{eq: shForces-multipoles}
    \begin{aligned}
    &{\cal F}_i^{\alpha}(\mathbf{R},  \mathbf{n}, \mathbf{d}) =  -\frac{\partial{\cal  U}_{\rm BO}({\bf R},{\bf n},{\bf d})}{\partial R_i^\alpha}= -\frac{\partial V({\bf R})}{\partial R_i^\alpha}-\sum_{j (j \neq i)} (2q_i[{\bf n},{\bf d}] - n_i) \left(f'(r_{ij}) \dfrac{r^\alpha_{ij}}{r_{ij}} \right)n_j \\
    &+ \left(2\mathbf{p}_i[{\bf n},{\mathbf d}] - \mathbf{d}_i\right)^{\rm T} \left[ \left( \dfrac{f'(r_{ij})}{r_{ij}} \right){\bf e}_\alpha +  \hat{r}^\alpha_{ij} \hat{\mathbf{r}}_{ij}  \left( f''(r_{ij}) - \dfrac{f'(r_{ij})}{r_{ij}} \right) \right]n_j\\
    &- \left(2q_i[{\bf n},{\bf d}] - n_i \right) \left[ \left( \dfrac{f'(r_{ij})}{r_{ij}} \right){\bf e}_\alpha^{\rm T}  + \left( f''(r_{ij}) - \dfrac{f'(r_{ij})}{r_{ij}} \right) \hat{r}^\alpha_{ij} \hat{\mathbf{r}}_{ij}^{\rm T} \right] \mathbf{p}_j \\
    &+ \left(2\mathbf{p}_i[{\bf n},{\mathbf d}] - \mathbf{d}_i\right)^{\rm T}\left(  r^\alpha_{ij}\left(\dfrac{f'''(r_{ij})}{r_{ij}^3} - \dfrac{3f''(r_{ij})}{r_{ij}^4} + \dfrac{3f'(r_{ij})}{r_{ij}^5} \right) (\mathbf{r}_{ij}\mathbf{r}_{ij}^{\rm T}) 
    +  r^\alpha_{ij}\left( \dfrac{f''(r_{ij})}{r_{ij}^2} - \dfrac{f'(r_{ij})}{r_{ij}^3} \right)\mathbf{I} \right)\mathbf{d}_j \\
    &- \left(2\mathbf{p}_i[{\bf n},{\mathbf d}] - \mathbf{d}_i\right)^{\rm T} \left(\left( \dfrac{f''(r_{ij})}{r_{ij}^2} - \dfrac{f'(r_{ij})}{r_{ij}^3} \right) \left({\bf  r}_{ij}{\bf e}^{\rm T}_\alpha + {\bf e}_\alpha {\bf 
     r}_{ij}^{\rm T} \right)\right){\bf d}_j.
    \end{aligned}
\end{equation}
\end{widetext}

The energies and forces for the fixed-monopole/flexible dipole model are given by the same expressions as above, Eq.\ (\ref{eq: shEnergy-multipoles}) and Eq.\ (\ref{eq: shForces-multipoles}), for the flexible multipole models, but with fixed monopole charges, ${\bf q } = {\bf q}_0$, and ${\bf n} = {\bf q}_0$.

\subsection{Low-Rank Approximation of the Jacobian to Update the Kernel}\label{ap:lrupdate}

Computing the full kernel, $\mathbf {K} \in {\mathbb{R}}^{N \times N}$, at every time step during the integration of the multipole-dependent equations of motion in Eq.\ (\ref{eq:shadowMultipole_eqofmotion_2}), i.e.\
\begin{equation}
\ddot{\mathbf{x}} = -\omega^2 \mathbf{K} \left( \mathbf{c}[\mathbf{x}] - \mathbf{x} \right),
\end{equation}
is computationally expensive. To reduce this cost, we can instead apply a low-rank Krylov approximation of ${\bf K}$, acting on the residual function,
\begin{equation}
\mathbf{f}(\mathbf{x}) = \mathbf{c}[\mathbf{x}] - \mathbf{x}.
\end{equation}

The kernel $\mathbf{K}$ is defined as the inverse of the Jacobian,
\begin{equation}
\mathbf{K} = \mathbf{J}^{-1},
\end{equation}
of the residual function, $\mathbf{f}(\mathbf{x})$, with elements
\begin{equation}
J_{ij} = \frac{\partial f_i[\mathbf{x}]}{\partial x_j}.
\end{equation}
These elements are expensive to compute explicitly. To reduce the computational overhead, we can use a low-rank approximation. We can achieve this by first introducing a generalized definition of the Jacobian,
\begin{equation}
\mathbf{J} = \sum_{i,j}^N \mathbf{f}_{\mathbf{v}_i} \mathbf{L}_{ij} \mathbf{v}_i^{\mathrm{T}}.
\end{equation}
Here, $\mathbf{f}_{\mathbf{v}_i}$ denotes the directional derivative of the residual function,
\begin{equation}
\mathbf{f}_{\mathbf{v}_i}(\mathbf{x}) \equiv \left. \frac{d \mathbf{f}(\mathbf{x} + \lambda \mathbf{v}_i)}{d \lambda} \right|_{\lambda = 0} = \left. \frac{d \mathbf{c}[\mathbf{x} + \lambda \mathbf{v}_i]}{d \lambda} \right|_{\lambda=0} - \mathbf{v}_i,
\end{equation}
constructed using a complete set of linearly independent vectors, ${ \mathbf{v}_i }$. The matrix $\mathbf{L} = \mathbf{O}^{-1}$, where $\mathbf{O}$ is the overlap matrix with elements $O_{ij} = \mathbf{v}_i^{\mathrm{T}} \mathbf{v}_j$.

We then obtain a rank-$m$ approximation of the kernel, ${\bf K}_m = {\bf J}_m^{-1}$, through the Moore-Penrose pseudoinverse of the truncated rank-$m$ Jacobian,
\begin{equation}
\mathbf{K}_m = \sum_{i,j=1}^{m<N} \mathbf{v}_i \mathbf{M}_{ij} \mathbf{f}_{\mathbf{v}_i}^{\mathrm{T}},
\end{equation}
where $\mathbf{M} = \mathbf{S}^{-1}$ and $\mathbf{S}$ is the overlap matrix with elements $S_{ij} = \mathbf{f}_{\mathbf{v}_i}^{\mathrm{T}} \mathbf{f}_{\mathbf{v}_j}$.

The low-rank approximation, $\mathbf{K}_m$, can then be substituted into the equations of motion, i.e.\
\begin{equation} \label{eq:PrecondKryl}
\ddot{\mathbf{x}}  \approx -\omega^2 \left( \sum_{i,j=1}^{m<N} \mathbf{v}_i \mathbf{M}_{ij} \mathbf{f}_{\mathbf{v}_i}^{\mathrm{T}} \right) \left( \mathbf{c}[\mathbf{x}] - \mathbf{x} \right).
\end{equation}

The key challenge in constructing this low-rank approximation is selecting the set of $m$ vectors $\{ \mathbf{v}_i \}_{i = 1}^m$ that gives the smallest approximation error. The natural approach is to use a Krylov expansion, where only the orthogonal complement of each new Krylov vector is retained. This orthonormalized Krylov (or Arnoldi) subspace is generated using Alg.\ \ref{Kryl_Alg}. It is easy to see how this algorithm constructs a Krylov subspace, $\mathcal{K}_m(\mathbf{J})$, because $\mathbf{f}_{\mathbf{v}} = \mathbf{J} \mathbf{v}$.

\begin{algorithm}
\caption{Construction of orthonormal Krylov (or Arnoldi) subspace vectors}\label{Kryl_Alg}
    \SetAlgoNoLine
    \SetAlgoNoEnd
    \tcp{Initialize ${\bf v}_1$ and ${\bf f}_{{\bf v}_1}$} 
    ${\bf v}_1 = {\bf f}({\bf n})/||{\bf f}({\bf n})||$\\
    \vspace{1mm}
    ${\bf f}_{{\bf v}_1} = (\partial {\bf f}({\bf n} + \lambda {\bf v}_1))/(\partial \lambda) \Big |_{\lambda = 0} = {\bf J}{\bf v}_1$
    \vspace{3mm}

    \tcp{Build Up the Subspace}
    \For{$i = 2 ~~{\rm to}~~m$}{
        \vspace{1mm}
         ~~ \tcp{Gram-Schmidt Orthonormalization}
        ~~~~${\bf v}_i = {\bf f}_{{\bf v}_{i-1}}/||{\bf f}_{{\bf v}_{i-1}}||$\\ 
        \vspace{1mm}
        ~~~~\For{$j = 2 ~~{\rm to}~~i-1$}{
        ~~~~~~~~${\bf v}_i = {\bf v}_i - ({\bf v}_i^{\rm T}{\bf v}_j){\bf v}_j $
        }
        ~~~~$\mathbf{end}$\\
        \vspace{1mm}
        ~~~~${\bf v}_i = {\bf v}_i/||{\bf v}_i||$\\
        \vspace{1mm}
        ~~ \tcp{New Krylov Vector}
        ~~~~${\bf f}_{{\bf v}_i} = (\partial {\bf f}({\bf n} + \lambda {\bf v}_i))/(\partial \lambda) \Big |_{\lambda=0} = {\bf J}{\bf v}_i$
    }
    $\mathbf{end}$
\end{algorithm}

When an approximate Jacobian inverse $\mathbf{K}_0 \approx \mathbf{J}^{-1}$ is available, we can rewrite the kernel as,
\begin{equation}
\mathbf{K} = \left( \mathbf{K}_0 \mathbf{J} \right)^{-1} \mathbf{K}_0,
\end{equation}
and apply a low-rank approximation to $\left( \mathbf{K}_0 \mathbf{J} \right)^{-1}$ acting on the preconditioned residual, i.e.\ $\mathbf{K}_0 \left( \mathbf{c}[\mathbf{x}] - \mathbf{x} \right)$. Because $\left( \mathbf{K}_0 \mathbf{J} \right)^{-1} \approx \mathbf{I}$, a lower-rank can be used \cite{ANiklasson20}.

For this preconditioned Krylov approximation, we follow the same procedure as in Alg.\ \ref{Kryl_Alg} above, with the replacements
\begin{align}
\mathbf{f}(\mathbf{x}) &\rightarrow \mathbf{K}_0 \mathbf{f}(\mathbf{x}), ~\mbox{and}~
\mathbf{f}_{\mathbf{v}_i} \rightarrow \mathbf{K}_0 \mathbf{f}_{\mathbf{v}_i}.
\end{align}
This modification provides a more accurate approximation and a smaller subspace can in general be used. 

It is important to note that for the integration of the time-dependent dynamical variable ${\bf x}(t)$ in Eq.\ (\ref{eq:PrecondKryl}) we do not need to be highly accurate, as long as ${\bf x}(t)$ remains not too far away from the exact fully relaxed ground state. The main purpose of ${\bf x}(t)$ is simply to provide an approximate expansion point in the construction of the shadow energy function as was illustrated in Fig.\ \ref{fig:aceshifting}.

\subsection{Pseudocode}

To further facilitate and implementation of the shadow multipole extended Lagrangian Born-Oppenheimer MD scheme for the flexible multipole model, we can provide a simple algorithm outline. A key aspect to note is that the electronic degrees of freedom appear alongside the atomic positions and velocities as dynamical variables. 

\begin{algorithm}
\caption{Shadow Born-Oppenheimer Molecular Dynamics using a Velocity-Verlet Integration Scheme for the Flexible Multipoles Model. The coefficients, $\{c_k\}$, in the dissipative term for the modified Verlet schemes are given in Ref.\ \cite{ANiklasson09}} \label{alg:fullflexiblemulti}
    \SetAlgoNoLine
    \tcp{Input Coordinates and Parameters}
    ${\bf R} = {\bf R}_{\rm in}$, ~
    $\boldsymbol{\chi} = \boldsymbol{\chi}_{\rm in}$, ~
    ${\bf u} = {\bf u}_{\rm in}$, ~
    $\boldsymbol{\alpha} = \boldsymbol{\alpha}_{\rm in}$, ~
    ${\bf m} = {\bf m}_{\rm in}$
    \vspace{1mm}

    \tcp{Initialize Velocities (e.g., ${\bf v}_0 = {\bf 0}$)}
    ${\bf v}(t_0) = \bf{v}_0$ 
    \vspace{1mm}

    \tcp{Initialize ${\bf x}$ as Ground State Multipoles, ${\bf c}_0$}
    ${\bf x}(t_0) = {\bf c}_0(t_0)$ 
    \vspace{1mm}

    \tcp{Determine a Preconditioner}
    $\mathbf{K}_0 = \mathbf{J}^{-1}(t_0)$
    \vspace{1mm}

    \tcp{Initialize ${ \bf \ddot x}$}
    ${\bf \ddot x}(t_0) = {\bf 0}$
    \vspace{3mm}

    \tcp{Initialize Energy and Forces}
    Calculate using Eqs.\ (\ref{eq:Range_Separation} - \ref{eq:shadowMultipole_eqofmotion_1})
    \vspace{3mm}
    
    \tcp{Main MD Loop ($t = t_0$)}
    \While{$t \leq {\rm MaxTime}$}{
        \vspace{1mm}
        
        ~~~~\tcp{Update Velocities, First Half-Step} 
        ~~~~${\bf v}(t+\frac{1}{2}\delta t) = {\bf v}(t) + \frac{1}{2}\delta t{\bf \ddot{R}}(t)$ 
        \vspace{1mm}
        
        ~~~~\tcp{Position Update}
        ~~~~${\bf R}(t + \delta t) = {\bf R}(t) + \delta t  {\bf v}(t + \frac{1}{2} \delta t)$
        \vspace{1mm}

        ~~~~\tcp{Multipoles Integration using Verlet}
        ~~~~$\mathbf{x}(t+\delta t) = 2\mathbf{x}(t) - \mathbf{x}(t-\delta t) + \delta t^2\ddot{{\bf x}}(t)$
        \vspace{1mm}
        
        ~~~~\tcp{Add Weak Dissipative Term}
        ~~~~$\mathbf{x}(t+\delta t) = \mathbf{x}(t + \delta t) + \alpha \sum_{k=0}^{k_{\rm max}} c_k \mathbf{x}(t-k\delta t)$
        \vspace{1mm}

        ~~~~\tcp{Determine Relaxed Multipoles}
        ~~~~ ${\bf x} = {\bf c}[{\bf x}]$
        \vspace{3mm}
        
        ~~~~\tcp{Energy and Forces Update}
        ~~~~Calculated using Eqs.\ (\ref{eq:Range_Separation} - \ref{eq:shadowMultipole_eqofmotion_1})\\
        \vspace{3mm}

        ~~~~\tcp{Update ${\bf \ddot x}$ using Low-Rank Update}
        ~~~~\tcp{(Appendix \ref{ap:lrupdate}, Including ${\bf K}_0$)}
        ~~~~$ {\bf \ddot x} = -\omega^2 {\bf K} ({\bf c}[{\bf x}] - {\bf x})$\\
        \vspace{3mm}
        
        ~~~~\tcp{Update Velocities, Second Half-Step}
        ~~~~${\bf v}(t + \delta t) = {\bf v}(t + \frac{1}{2} \delta t) + \frac{1}{2}\delta t{\bf \ddot{R}}(t + \delta t)$
        \vspace{1mm}
        
        ~~~~\tcp{Increase Time}
        ~~~~$t = t + \delta t$
    }
\end{algorithm}

\section{Software Availability}
The Python and MATLAB prototype codes associated with the manuscript will be made available as part of open-source SEDACS package (https://github.com/lanl/sedacs).


\begin{acknowledgments}
This work is supported by the U.S. Department of Energy Office of Basic Energy Sciences (FWP LANLE8AN), the Los Alamos National Laboratory LDRD-ER program and by the U.S. Department of Energy through the Los Alamos National Laboratory. Los Alamos National Laboratory is operated by Triad National Security, LLC, for the National Nuclear Security Administration of the U.S. Department of Energy Contract No. 892333218NCA000001. This article has been approved for unlimited distribution with the LA-UR number: `LA-UR-25-29120'.
\end{acknowledgments}

\bibliography{references}

\begin{widetext}
    \section*{Supporting Information}

\newcommand{\beginsupplement}{%
        \setcounter{table}{0}
        \renewcommand{\thetable}{S\arabic{table}}%
        \setcounter{figure}{0}
        \renewcommand{\thefigure}{S\arabic{figure}}%
        \setcounter{equation}{0}
        \renewcommand{\theequation}{S\arabic{equation}}%
     }






\beginsupplement
\section*{Supporting Information for: Shadow Molecular Dynamics for Flexible Multipole Models}

Rae A. Corrigan Grove, Robert Stanton, Michael E. Wall, and Anders M. N. Niklasson

This article has been approved for unlimited distribution with the LA-UR number: `LA-UR-25-29120'.

\section*{Jacobian Examples}
The Jacobian matrix, ${\bf J}$, that defines the kernel, ${\bf K} = {\bf J}^{-1}$, for the monopole-only charge equilibration model is
\begin{equation}
    {\bf J} = {\bf A}^{-1} \frac{\partial {\bf b}}{\partial {\bf n}} - {\bf I}_{N \times N}
\end{equation}
or explicitly for three atoms (with $N=3$),
\begin{equation}\label{matrix: 3atomJ-si}
    \mathbf{J} = -
    \begin{bmatrix}
        \sum_{j}^N\{ \textbf{A}^{-1}\}_{1j} \frac{db_j}{dn_1} - \delta_{11} & \sum_{j}^N\{ \textbf{A}^{-1}\}_{1j} \frac{db_j}{dn_2} - \delta_{12} & \sum_{j}^N\{ \textbf{A}^{-1}\}_{1j} \frac{db_j}{dn_3} - \delta_{13} \\
        \sum_{j}^N\{ \textbf{A}^{-1}\}_{2j} \frac{db_j}{dn_1} - \delta_{21} & \sum_{j}^N\{ \textbf{A}^{-1}\}_{2j} \frac{db_j}{dn_2} - \delta_{22} & \sum_{j}^N\{ \textbf{A}^{-1}\}_{2j} \frac{db_j}{dn_3} - \delta_{23} \\
        \sum_{j}^N\{ \textbf{A}^{-1}\}_{3j} \frac{db_j}{dn_1} - \delta_{31} & \sum_{j}^N\{ \textbf{A}^{-1}\}_{3j} \frac{db_j}{dn_2} - \delta_{32} & \sum_{j}^N\{ \textbf{A}^{-1}\}_{3j} \frac{db_j}{dn_3} - \delta_{33} \\
    \end{bmatrix}.
\end{equation}
For example, the $J_{12}$ element of the Jacobian for the monopole-only model is
\begin{equation}\label{eq: jacobian12-si}
    \begin{aligned}
    &J_{12} = -(\{\textbf{A}^{-1}\}_{11} \frac{db_1}{dn_2} + \{\textbf{A}^{-1}\}_{12} \frac{db_2}{dn_2} + 
    \{\textbf{A}^{-1}\}_{13} \frac{db_3}{dn_2}) - \delta_{12} \\
    &~~~~~=-(\{\textbf{A}^{-1}\}_{11}(-\gamma_{12}) + \{\textbf{A}^{-1}\}_{12}(-\gamma_{22}) 
    + \{\textbf{A}^{-1}\}_{13} (-\gamma_{32})) - \delta_{12}
    \end{aligned}.
\end{equation}
In the above equations $\delta_{ii}=1$, $\delta_{ij}=0$ ($i \ne j)$, and the matrices used are defined as
\begin{equation}
    {\bf A}=
    \begin{bmatrix}
        u_1 & 0 & 0 & 1\\
        0 & u_2 & 0 & 1\\
        0 & 0 & u_3 & 1\\
        1 & 1 & 1 & 0
    \end{bmatrix},
    {\bf b} =
    \begin{bmatrix}
        -\chi_1 - \sum_j C_{1j}n_j\\
        -\chi_2 - \sum_j C_{2j}n_j\\
        -\chi_3 - \sum_j C_{3j}n_j\\
        Q_{\rm tot}
    \end{bmatrix},
    {\bf I} = 
    \begin{bmatrix}
        1 & 0 & 0 \\
        0 & 1 & 0 \\
        0 & 0 & 1
    \end{bmatrix}
\end{equation}
Calculation of the full Jacobian matrix for the flexible multipole model:
\begin{equation}
    {\bf J} = 
    - \begin{bmatrix}
        {\bf C}_{\rm S} & {\bf W}_{\rm S}^{\rm T} & {\bf 1}\\
        {\bf W}_{\rm S} & \boldsymbol{\Lambda}_{\rm S} & {\bf 0} \\
        {\bf 1}^{\rm T} & {\bf 0}^{\rm T} & 0
    \end{bmatrix}^{-1}_{4N \times 4N + 1}
    \begin{bmatrix}
        {\bf C}_{\rm L} & {\bf W}_{\rm L}^{\rm T}\\
        {\bf W}_{\rm L} & \boldsymbol{\Lambda}_{\rm L} \\
        {\bf 0}^{\rm T} & {\bf 0}^{\rm T}
    \end{bmatrix}
    -
    {\bf I}_{4N \times 4N}
\end{equation}
Calculation of the full Jacobian matrix for the fixed monopole/flexible dipole model:
\begin{equation}
    {\bf J} = -
    \boldsymbol{\Lambda}_{\rm S}^{-1} \boldsymbol{\Lambda}_{\rm L} - {\bf I}_{3N \times 3N}
\end{equation}
All ${\bf C}$, ${\bf W}$, and $\boldsymbol{\Lambda}$ matrices used in the equations above are given for three atoms below in the supplementary section titled ``Example Interaction Matrices for Three Atoms" and the ${\bf I}$ are identity matrices with the listed dimensions.

\section*{Example Interaction Matrices for Three Atoms}
The position vector for an atom, $i$, can be defined as 
\begin{equation}\label{eq:Rposition-si}
    \mathbf{R}_i = 
    \begin{bmatrix}
        r_{ix} \\
        r_{iy} \\
        r_{iz}
    \end{bmatrix}
\end{equation}
Additional variables can be defined, for example, where 
\begin{equation}\label{eq: rij-si}
    \mathbf{r}_{ij} = \mathbf{R}_i - \mathbf{R}_j
\end{equation}
is the distance vector between atoms $i$ and $j$ with the $r_{{ij}_\alpha}$ components, $[r_{{ij}_x},r_{{ij}_y}, r_{{ij}_z}]$, i.e.\ $[{{\bf r}_{ij}}_x,{{\bf r}_{ij}}_y, {{\bf r}_{ij}}_z]$. The norm of the interatomic distance is given by
\begin{equation}\label{eq: norm-si}
    r_{ij} = |\mathbf{r}_{ij}|
\end{equation}
 and 
\begin{equation}\label{eq: rij_hat-si}
    \hat{\mathbf{r}}_{ij} = \dfrac{\mathbf{r}_{ij}}{r_{ij}}
\end{equation}
is a unit vector in the direction of $\mathbf{r}_{ij}$.

The Coulomb interaction function, $f(r_{ij})$, is determined by the choice of the shape of the two overlapping charge distributions. The functional form and its derivatives used here are derived from the Coulomb energy of overlapping Gaussian charge distributions and are given by
\begin{equation}\label{eq: erf-derivatives-si}
    \begin{aligned}
        &f(r_{ij}) = \dfrac{\operatorname{erf}(ar_{ij})}{r_{ij}} \\
        &f'(r_{ij}) = u \dfrac{e^{-(ar_{ij})^2}}{r_{ij}} - \dfrac{\operatorname{erf(ar_{ij})}}{r_{ij}^2}\\
        &f''(r_{ij}) = -2ua^2 e^{-(ar_{ij})^2} - 2u \dfrac{e^{-(ar_{ij})^2}}{r_{ij}^2} + 2 \dfrac{\operatorname{erf}(ar_{ij})}{r_{ij}^3}\\
        &f'''(r_{ij}) = 4ua^4r_{ij}e^{-(ar_{ij})^2} + 4ua^2 \dfrac{e^{-(ar_{ij})^2}}{r_{ij}} +6u \dfrac{e^{-(ar_{ij})^2}}{r_{ij}^3} - 6 \dfrac{\operatorname{erf}(ar_{ij})}{r_{ij}^4}
    \end{aligned}
\end{equation}
where
\begin{equation}
    u = \dfrac{2u_iu_j}{(u_i + u_j)}
\end{equation}
and
\begin{equation}
    a = \dfrac{\sqrt{\pi}}{2}u
\end{equation}

Using these expressions,  we can determine the interaction matrices for monopoles and dipoles. These interaction matrices are split into several pieces: 1) the monopole-monopole matrix, $\mathbf{C}$ where
\begin{equation}\label{eq: Cmatfill-si}
    \begin{aligned}
        C_{ij} = f(r_{ij})\\
        C_{ii} = u_i,
    \end{aligned}
\end{equation}
2) the monopole-dipole matrices, $\mathbf{W}$ and $\mathbf{W}^{\rm T}$, where
\begin{equation}
    W_{ij} = W_{ji}^{\rm T} = f'(r_{ij}) \dfrac{\mathbf{r}_{ij}}{r_{ij}} = f'(r_{ij})\hat{\mathbf{r}}_{ij},
\end{equation}
and 3) the dipole-dipole matrix, $\mathbf{\Lambda}$, where the diagonal blocks are 
\begin{equation}
    \boldsymbol{\Lambda}_{ii} = 
    \begin{bmatrix}
        \alpha_i^{-1} & 0 & 0\\
        0 & \alpha_i^{-1} & 0\\
        0 & 0 & \alpha_i^{-1}
    \end{bmatrix}
\end{equation}
and with the off-diagonal blocks,
\begin{equation}
    \begin{aligned}
         \Lambda_{ij} = -1 \left( f''(r_{ij}) \dfrac{\mathbf{r}_{ij} \mathbf{r}_{ij}^{\rm T}}{r_{ij}^2} + \dfrac{f'(r_{ij})}{r_{ij}} \left( \mathbf{I} - \dfrac{\mathbf{r}_{ij} \mathbf{r}_{ij}^{\rm T}}{r_{ij}^2} \right) \right) \\
         = -1 \left( f''(r_{ij}) \hat{\mathbf{r}}_{ij} \hat{\mathbf{r}}_{ij}^{\rm T} + \dfrac{f'(r_{ij})}{r_{ij}} \left( \mathbf{I} - \hat{\mathbf{r}}_{ij}\hat{\mathbf{r}}_{ij}^{\rm T} \right) \right).
    \end{aligned}
\end{equation}

To better highlight the energy expression we also give some explicit examples for a three atom models system.
The Coulomb interaction matrix, $\bf{C}$, for monopole-monopole interactions between three atoms is
\begin{equation}
    {\bf C} = 
    \begin{bmatrix}
        u_1 & f(r_{12}) & f(r_{13})\\
        f(r_{21}) & u_2 & f(r_{23})\\
        f(r_{31}) & f(r_{32}) & u_3
    \end{bmatrix} = 
    \begin{bmatrix}
        u_1 & \dfrac{\operatorname{erf}(ar_{12})}{r_{12}} & \dfrac{\operatorname{erf}(ar_{13})}{r_{13}}\\
        \dfrac{\operatorname{erf}(ar_{21})}{r_{21}} & u_2 & \dfrac{\operatorname{erf}(ar_{23})}{r_{23}}\\
        \dfrac{\operatorname{erf}(ar_{31})}{r_{31}} & \dfrac{\operatorname{erf}(ar_{32})}{r_{32}} & u_3
    \end{bmatrix} .
\end{equation}

The ${\bf W}$ monopole-dipole interaction matrix for the three atom system is given by
\begin{equation}
    {\bf W} = 
    \begin{bmatrix}
        0 &f'(r_{12}){\hat r}_{12x} & f'(r_{13}){\hat r}_{13x}\\
        0 &f'(r_{12}){\hat r}_{12y} & f'(r_{13}){\hat r}_{13y}\\
        0 &f'(r_{12}){\hat r}_{12z} & f'(r_{13}){\hat r}_{13z}\\
        f'(r_{21}){\hat r}_{21x} & 0 & f'(r_{23}){\hat r}_{23x}\\
        f'(r_{21}){\hat r}_{21y} & 0 & f'(r_{23}){\hat r}_{23y}\\
        f'(r_{21}){\hat r}_{21z} & 0 & f'(r_{23}){\hat r}_{23z}\\
        f'(r_{31}){\hat r}_{31x} & f'(r_{32}){\hat r}_{32x} & 0\\
        f'(r_{31}){\hat r}_{31y} & f'(r_{32}){\hat r}_{32y} & 0\\
        f'(r_{31}){\hat r}_{31z} & f'(r_{32}){\hat r}_{32z} & 0\\
    \end{bmatrix}.
\end{equation}

The dipole-dipole interactions matrix, $\mathbf{\Lambda}$, for the three atom system can be divided into three separate parts.

The first $9 \times 3$ columns and rows of the $\mathbf{\Lambda}$ matrix for three atoms is
\begin{equation}
    \rotatebox{90}{$ {\mathbf {\Lambda}}_{1:9, 1:3} = \begin{matrix}
    \alpha_1^{-1} & 0 & 0 \\
    0 & \alpha_1^{-1} & 0 \\
    0 & 0 & \alpha_1^{-1} \\
    -\dfrac{f'(r_{21})}{r_{21}} - \left( f''(r_{21})-\dfrac{f'(r_{21})}{r_{21}} \right) \hat{r}_{21x}\hat{r}_{21x} & -\left( f''(r_{21})-\dfrac{f'(r_{21})}{r_{21}} \right) \hat{r} _{21x}\hat{r}_{21y} & -\left( f''(r_{21})-\dfrac{f'(r_{21})}{r_{21}} \right) \hat{r} _{21x}\hat{r}_{21z}\\
    -\left( f''(r_{21})-\dfrac{f'(r_{21})}{r_{21}} \right) \hat{r}_{21y}\hat{r}_{21x} & -\dfrac{f'(r_{21})}{r_{21}} - \left( f''(r_{21})-\dfrac{f'(r_{21})}{r_{21}} \right) \hat{r} _{21y}\hat{r}_{21y} & -\left( f''(r_{21})-\dfrac{f'(r_{21})}{r_{21}} \right) \hat{r} _{21y}\hat{r}_{21z}\\
    -\left( f''(r_{21})-\dfrac{f'(r_{21})}{r_{21}} \right) \hat{r}_{21z}\hat{r}_{21x} & -\left( f''(r_{21})-\dfrac{f'(r_{21})}{r_{21}} \right) \hat{r} _{21z}\hat{r}_{21y} & -\dfrac{f'(r_{21})}{r_{21}} - \left( f''(r_{21})-\dfrac{f'(r_{21})}{r_{21}} \right) \hat{r} _{21z}\hat{r}_{21z}\\
    -\dfrac{f'(r_{31})}{r_{31}} - \left( f''(r_{31})-\dfrac{f'(r_{31})}{r_{31}} \right) \hat{r}_{31x}\hat{r}_{31x} & -\left( f''(r_{31})-\dfrac{f'(r_{31})}{r_{31}} \right) \hat{r} _{31x}\hat{r}_{31y} & -\left( f''(r_{31})-\dfrac{f'(r_{31})}{r_{31}} \right) \hat{r} _{31x}\hat{r}_{31z}\\
    -\left( f''(r_{31})-\dfrac{f'(r_{31})}{r_{31}} \right) \hat{r}_{31y}\hat{r}_{31x} & -\dfrac{f'(r_{31})}{r_{31}} - \left( f''(r_{31})-\dfrac{f'(r_{31})}{r_{31}} \right) \hat{r} _{31y}\hat{r}_{31y} & -\left( f''(r_{31})-\dfrac{f'(r_{31})}{r_{31}} \right) \hat{r} _{31y}\hat{r}_{31z}\\
    -\left( f''(r_{31})-\dfrac{f'(r_{31})}{r_{31}} \right) \hat{r}_{31z}\hat{r}_{31x} & -\left( f''(r_{31})-\dfrac{f'(r_{31})}{r_{31}} \right) \hat{r} _{31z}\hat{r}_{31y} & -\dfrac{f'(r_{31})}{r_{31}} - \left( f''(r_{31})-\dfrac{f'(r_{31})}{r_{31}} \right) \hat{r} _{31z}\hat{r}_{31z}\\
    \end{matrix}$}
\end{equation}

The second  $9 \times 3$ columns and rows of $\mathbf{\Lambda}$ matrix for three atoms is 
\begin{equation}
    \rotatebox{90}{$ {\mathbf {\Lambda}}_{1:9, 4:6} = \begin{matrix}
    -\dfrac{f'(r_{12})}{r_{12}} - \left( f''(r_{12})-\dfrac{f'(r_{12})}{r_{12}} \right) \hat{r}_{12x}\hat{r}_{12x} & -\left( f''(r_{12})-\dfrac{f'(r_{12})}{r_{12}} \right) \hat{r} _{12x}\hat{r}_{12y} & -\left( f''(r_{12})-\dfrac{f'(r_{12})}{r_{12}} \right) \hat{r} _{12x}\hat{r}_{12z} \\
    
    -\left( f''(r_{12})-\dfrac{f'(r_{12})}{r_{12}} \right) \hat{r}_{12y}\hat{r}_{12x} & -\dfrac{f'(r_{12})}{r_{12}} - \left( f''(r_{12})-\dfrac{f'(r_{12})}{r_{12}} \right) \hat{r} _{12y}\hat{r}_{12y} & -\left( f''(r_{12})-\dfrac{f'(r_{12})}{r_{12}} \right) \hat{r} _{12y}\hat{r}_{12z} \\
    
    -\left( f''(r_{12})-\dfrac{f'(r_{12})}{r_{12}} \right) \hat{r}_{12z}\hat{r}_{12x} & -\left( f''(r_{12})-\dfrac{f'(r_{12})}{r_{12}} \right) \hat{r} _{12z}\hat{r}_{12y} & -\dfrac{f'(r_{12})}{r_{12}} - \left( f''(r_{12})-\dfrac{f'(r_{12})}{r_{12}} \right) \hat{r} _{12z}\hat{r}_{12z}\\
    
    \alpha_2^{-1} & 0 & 0  \\
    0 & \alpha_2^{-1} & 0\\
    0 & 0 & \alpha_2^{-1}\\
    
    -\dfrac{f'(r_{32})}{r_{32}} - \left( f''(r_{32})-\dfrac{f'(r_{32})}{r_{32}} \right) \hat{r}_{32x}\hat{r}_{32x} & -\left( f''(r_{32})-\dfrac{f'(r_{32})}{r_{32}} \right) \hat{r} _{32x}\hat{r}_{32y} & -\left( f''(r_{32})-\dfrac{f'(r_{32})}{r_{32}} \right) \hat{r} _{32x}\hat{r}_{32z}\\
    
    -\left( f''(r_{32})-\dfrac{f'(r_{32})}{r_{32}} \right) \hat{r}_{32y}\hat{r}_{32x} & -\dfrac{f'(r_{32})}{r_{32}} - \left( f''(r_{32})-\dfrac{f'(r_{32})}{r_{32}} \right) \hat{r} _{32y}\hat{r}_{32y} & -\left( f''(r_{32})-\dfrac{f'(r_{32})}{r_{32}} \right) \hat{r} _{32y}\hat{r}_{32z}\\
    
     -\left( f''(r_{32})-\dfrac{f'(r_{32})}{r_{32}} \right) \hat{r}_{32z}\hat{r}_{32x} & -\left( f''(r_{32})-\dfrac{f'(r_{32})}{r_{32}} \right) \hat{r} _{32z}\hat{r}_{32y} & -\dfrac{f'(r_{32})}{r_{32}} - \left( f''(r_{32})-\dfrac{f'(r_{32})}{r_{32}} \right) \hat{r} _{32z}\hat{r}_{32z}\\
    \end{matrix}$}
\end{equation}

The second  $9 \times 3$ columns and rows of $\mathbf{\Lambda}$ matrix for three atoms is 
\begin{equation}
    \rotatebox{90}{$ {\mathbf {\Lambda}}_{1:9, 7:9} = \begin{matrix}
     -\dfrac{f'(r_{13})}{r_{13}} - \left( f''(r_{13})-\dfrac{f'(r_{13})}{r_{13}} \right) \hat{r}_{13x}\hat{r}_{13x} & -\left( f''(r_{13})-\dfrac{f'(r_{13})}{r_{13}} \right) \hat{r} _{13x}\hat{r}_{13y} & -\left( f''(r_{13})-\dfrac{f'(r_{13})}{r_{13}} \right) \hat{r} _{13x}\hat{r}_{13z} \\
    
     -\left( f''(r_{13})-\dfrac{f'(r_{13})}{r_{13}} \right) \hat{r}_{13y}\hat{r}_{13x} & -\dfrac{f'(r_{13})}{r_{13}} - \left( f''(r_{13})-\dfrac{f'(r_{13})}{r_{13}} \right) \hat{r} _{13y}\hat{r}_{13y} & -\left( f''(r_{13})-\dfrac{f'(r_{13})}{r_{13}} \right) \hat{r} _{13y}\hat{r}_{13z}\\
    
     -\left( f''(r_{13})-\dfrac{f'(r_{13})}{r_{13}} \right) \hat{r}_{13z}\hat{r}_{13x} & -\left( f''(r_{13})-\dfrac{f'(r_{13})}{r_{13}} \right) \hat{r} _{13z}\hat{r}_{13y} & -\dfrac{f'(r_{13})}{r_{13}} - \left( f''(r_{13})-\dfrac{f'(r_{13})}{r_{13}} \right) \hat{r} _{13z}\hat{r}_{13z}\\
    
     -\dfrac{f'(r_{23})}{r_{23}} - \left( f''(r_{23})-\dfrac{f'(r_{23})}{r_{23}} \right) \hat{r}_{23x}\hat{r}_{23x} & -\left( f''(r_{23})-\dfrac{f'(r_{23})}{r_{23}} \right) \hat{r} _{23x}\hat{r}_{23y} & -\left( f''(r_{23})-\dfrac{f'(r_{23})}{r_{23}} \right) \hat{r} _{23x}\hat{r}_{23z} \\
    
     -\left( f''(r_{23})-\dfrac{f'(r_{23})}{r_{23}} \right) \hat{r}_{23y}\hat{r}_{23x} & -\dfrac{f'(r_{23})}{r_{23}} - \left( f''(r_{23})-\dfrac{f'(r_{23})}{r_{23}} \right) \hat{r} _{23y}\hat{r}_{23y} & -\left( f''(r_{23})-\dfrac{f'(r_{23})}{r_{23}} \right) \hat{r} _{23y}\hat{r}_{23z}\\
    
     -\left( f''(r_{23})-\dfrac{f'(r_{23})}{r_{23}} \right) \hat{r}_{23z}\hat{r}_{23x} & -\left( f''(r_{23})-\dfrac{f'(r_{23})}{r_{23}} \right) \hat{r} _{23z}\hat{r}_{23y} & -\dfrac{f'(r_{23})}{r_{23}} - \left( f''(r_{23})-\dfrac{f'(r_{23})}{r_{23}} \right) \hat{r} _{23z}\hat{r}_{23z}\\
    
    \alpha_3^{-1} & 0 & 0 \\
    0 & \alpha_3^{-1} & 0\\
    0 & 0 & \alpha_3^{-1}\\
    \end{matrix}$}
\end{equation}

The unscreened version of the $\boldsymbol{\Lambda}$ interaction matrix using the bare $1/r_{ij}$ interactions between point charges instead of the functional form determined by a finite Gaussian charge distribution, is less computationally intensive. However, this unscreened, bare form will most likely lead to serious stability problems in molecular dynamics simulations, because singularities may appear in the equations determining the relaxed flexible charges.

The unscreened, bare version of the $\boldsymbol{\Lambda}$ interaction matrix is given by

\begin{equation}
    \begin{aligned}
    &\gamma_{ij} = \frac{1}{r_{ij}}\\
    &\gamma_{ij}^2 = \frac{1}{r_{ij}^2}\\
    &\gamma_{ij}^3 = \frac{1}{r_{ij}^3}
    \end{aligned}
\end{equation}
where $r_{ij}$ is defined in Supplementary Equations \ref{eq:Rposition-si} - \ref{eq: norm-si}.

The first $9 \times 3$ columns and rows of the simplified $\mathbf{\Lambda}$ matrix for three atoms is
\begin{equation}
   {\mathbf {\Lambda}}_{1:9, 1:3} = \begin{bmatrix}
        1/\alpha_1 & 0 & 0 \\
        0 & 1/\alpha_1 & 0 \\
        0 & 0 & 1/\alpha_1 \\
        \gamma_{21}^3 - 3(\hat{r}_{21x})\gamma_{21}^3({\hat{r}}_{21x}) & - 3(\hat{r}_{21x})\gamma_{21}^3({\hat{r}}_{21y}) & - 3(\hat{r}_{21x})\gamma_{21}^3({\hat{r}}_{21z}) \\
        - 3(\hat{r}_{21y})\gamma_{21}^3({\hat{r}}_{21x}) & \gamma_{21}^3 - 3(\hat{r}_{21y})\gamma_{21}^3({\hat{r}}_{21y}) & - 3(\hat{r}_{21y})\gamma_{21}^3({\hat{r}}_{21z}) \\
        - 3(\hat{r}_{21z})\gamma_{21}^3({\hat{r}}_{21x}) & - 3(\hat{r}_{21z})\gamma_{21}^3({\hat{r}}_{21y}) & \gamma_{21}^3 - 3(\hat{r}_{21z})\gamma_{21}^3({\hat{r}}_{21z})\\
        \gamma_{31}^3 - 3(\hat{r}_{31x})\gamma_{31}^3({\hat{r}}_{31x}) & - 3(\hat{r}_{31x})\gamma_{31}^3({\hat{r}}_{31y}) & - 3(\hat{r}_{31x})\gamma_{31}^3({\hat{r}}_{31z}) \\
        - 3(\hat{r}_{31y})\gamma_{31}^3({\hat{r}}_{31x}) & \gamma_{31}^3 - 3(\hat{r}_{31y})\gamma_{31}^3({\hat{r}}_{31y}) & - 3(\hat{r}_{31y})\gamma_{31}^3({\hat{r}}_{31z}) \\
        - 3(\hat{r}_{31z})\gamma_{31}^3({\hat{r}}_{31x}) & - 3(\hat{r}_{31z})\gamma_{31}^3({\hat{r}}_{31y}) & \gamma_{31}^3 - 3(\hat{r}_{31z})\gamma_{31}^3({\hat{r}}_{31z})
    \end{bmatrix}
\end{equation}

The second $9 \times 3$ columns and rows of the simplified $\mathbf{\Lambda}$ matrix for three atoms is
\begin{equation}
     {\mathbf {\Lambda}}_{1:9, 4:6} = \begin{bmatrix}
        \gamma_{12}^3 - 3(\hat{r}_{12x})\gamma_{12}^3({\hat{r}}_{12x}) & - 3(\hat{r}_{12x})\gamma_{12}^3({\hat{r}}_{12y}) & - 3(\hat{r}_{12x})\gamma_{12}^3({\hat{r}}_{12z})\\
        - 3(\hat{r}_{12y})\gamma_{12}^3({\hat{r}}_{12x}) & \gamma_{12}^3 - 3(\hat{r}_{12y})\gamma_{12}^3({\hat{r}}_{12y}) & - 3(\hat{r}_{12y})\gamma_{12}^3({\hat{r}}_{12z})\\
        - 3(\hat{r}_{12z})\gamma_{12}^3({\hat{r}}_{12x}) & - 3(\hat{r}_{12z})\gamma_{12}^3({\hat{r}}_{12y}) & \gamma_{12}^3 - 3(\hat{r}_{12z})\gamma_{12}^3({\hat{r}}_{12z})\\
        1/\alpha_2 & 0 & 0 \\
        0 & 1/\alpha_2 & 0 \\
        0 & 0 & 1/\alpha_2 \\
        \gamma_{32}^3 - 3(\hat{r}_{32x})\gamma_{32}^3({\hat{r}}_{32x}) & - 3(\hat{r}_{32x})\gamma_{32}^3({\hat{r}}_{32y}) & - 3(\hat{r}_{32x})\gamma_{32}^3({\hat{r}}_{32z})\\
        - 3(\hat{r}_{32y})\gamma_{32}^3({\hat{r}}_{32x}) & \gamma_{32}^3 - 3(\hat{r}_{32y})\gamma_{32}^3({\hat{r}}_{32y}) & - 3(\hat{r}_{32y})\gamma_{32}^3({\hat{r}}_{32z}) \\
        - 3(\hat{r}_{32z})\gamma_{32}^3({\hat{r}}_{32x}) & - 3(\hat{r}_{32z})\gamma_{32}^3({\hat{r}}_{32y}) & \gamma_{32}^3 - 3(\hat{r}_{32z})\gamma_{32}^3({\hat{r}}_{32z})
    \end{bmatrix}
\end{equation}

The third $9 \times 3$ columns and rows of the simplified $\mathbf{\Lambda}$ matrix for three atoms is
\begin{equation}
     {\mathbf {\Lambda}}_{1:9, 7:9} = \begin{bmatrix}
        \gamma_{13}^3 - 3(\hat{r}_{13x})\gamma_{13}^3({\hat{r}}_{13x}) & - 3(\hat{r}_{13x})\gamma_{13}^3({\hat{r}}_{13y}) & - 3(\hat{r}_{13x})\gamma_{13}^3({\hat{r}}_{13z})\\
        - 3(\hat{r}_{13y})\gamma_{13}^3({\hat{r}}_{13x}) & \gamma_{13}^3 - 3(\hat{r}_{13y})\gamma_{13}^3({\hat{r}}_{13y}) & - 3(\hat{r}_{13y})\gamma_{13}^3({\hat{r}}_{13z})\\
        - 3(\hat{r}_{13z})\gamma_{13}^3({\hat{r}}_{13x}) & - 3(\hat{r}_{13z})\gamma_{13}^3({\hat{r}}_{13y}) & \gamma_{13}^3 - 3(\hat{r}_{13z})\gamma_{13}^3({\hat{r}}_{13z})\\
        \gamma_{23}^3 - 3(\hat{r}_{23x})\gamma_{23}^3({\hat{r}}_{23x}) & - 3(\hat{r}_{23x})\gamma_{23}^3({\hat{r}}_{23y}) & - 3(\hat{r}_{23x})\gamma_{23}^3({\hat{r}}_{23z})\\
        - 3(\hat{r}_{23y})\gamma_{23}^3({\hat{r}}_{23x}) & \gamma_{23}^3 - 3(\hat{r}_{23y})\gamma_{23}^3({\hat{r}}_{23y}) & - 3(\hat{r}_{23y})\gamma_{23}^3({\hat{r}}_{23z})\\
        - 3(\hat{r}_{23z})\gamma_{23}^3({\hat{r}}_{23x}) & - 3(\hat{r}_{23z})\gamma_{23}^3({\hat{r}}_{23y})& \gamma_{23}^3 - 3(\hat{r}_{23z})\gamma_{23}^3({\hat{r}}_{23z})\\
        1/\alpha_3 & 0 & 0\\
        0 & 1/\alpha_3 & 0\\
        0 & 0 & 1/\alpha_3
    \end{bmatrix}
\end{equation}

\end{widetext}

\end{document}